# Just the Facts: How Dialogues with AI Reduce Conspiracy Beliefs

Thomas H. Costello[1,*], Gordon Pennycook[2], David G. Rand[2,3,4,5]

[1]Department of Social and Decision Sciences, Carnegie Mellon University, Pittsburgh, PA, USA
[2]Department of Psychology, Cornell University, Ithaca, NY, USA
[3]Department of Information Science, Cornell University, Ithaca, NY, USA
[4]Marketing and Management Communications, SC Johnson School of Business, Cornell University, Ithaca, NY, USA
[5]Sloan School of Management, Massachusetts Institute of Technology, Cambridge, MA, USA

*Corresponding author. Email: thcostello1@gmail.com

Use of AI tools: Claude (Anthropic) was used to write and check analysis code, check reported numbers against the code, edit the manuscript, and prepare and edit the supplementary materials.

## Abstract

Personalized dialogues with AI substantially reduce conspiracy beliefs. To identify why, we conducted two experiments with U.S. adults recruited online (total $N = 2{,}193$) that removed possible causes one at a time. The debunking effect was robust to disclosing the AI's persuasive goal, framing the exchange as a debate, instructing the AI to inform rather than persuade, halving its response length, altering how beliefs were measured, and replacing the dialogue with a single tailored message. Only instructing the AI to persuade without factual evidence and arguments substantially reduced the debunking effect. Content analyses of the AI's output and participants' own accounts linked larger reductions in conspiracy beliefs to factual argumentation rather than rapport, storytelling, or appeals to empathy. AI dialogues appear to reduce conspiracy beliefs primarily by supplying relevant, well-reasoned counterevidence. Our results suggest that humans might be similarly effective if they could supply comparable evidence.

*Keywords:* conspiracy beliefs, generative AI, debunking, belief updating, persuasion

Conspiracy theories — explanations of important events as secret, malevolent plots by powerful actors — are widespread and can have serious consequences for public health, democratic legitimacy, extremism, and scientific progress (Ecker et al., 2024). Although some conspiracies are real, many popular conspiracy beliefs are clearly contradicted by evidence but are nonetheless hard to correct. In meta-analyses, corrections of conspiracy theories and related misinformation have reliable but typically modest effects that often shrink or vanish for highly committed believers or ideologically resonant claims (Albarracin & Shavitt, 2018; Chan & Albarracín, 2023; Walter et al., 2020). Lay and academic accounts have therefore often assumed that no brief intervention, least of all one that simply presents counterevidence, can durably move such beliefs (Lewandowsky et al., 2013; O'Mahony et al., 2023; Sunstein & Vermeule, 2009), which are seen as sustained by motivated reasoning, distrust of official sources, and self-sealing explanation (Biddlestone et al., 2021; Dagnall et al., 2015; Douglas et al., 2024) or as socially adaptive commitments or group-signaling rationalizations rather than attempts to track truth (Mercier, 2020; Williams, 2023).

However, recent evidence indicates that personalized dialogues with AI models can meaningfully reduce conspiracy beliefs (Costello et al., 2024). In these studies, a substantial fraction of believers revised their beliefs after sustained, belief-relevant counterargumentation, with unusually large effects (*d*s of 0.79 and 1.15). This finding may complicate, or even run counter to, portraits of conspiracy believers as insensitive to evidence because they are motivated by needs other than accuracy. What, exactly, underpins the AI's outsized effect?

One account is content-based. In classical reasoning and Bayesian accounts of belief updating, sufficiently diagnostic counterevidence should induce deliberation and subsequent belief revision (Pennycook, 2023). As such, the treatment would succeed by presenting relevant counterevidence, reasoning, and alternative explanations that directly address participants' stated reasons for believing. Generative AI may be distinctive in delivering expert-level knowledge across remarkably broad domains (Farrell et al., 2025; Singhal et al., 2025), whereas any human debunker, however well informed, commands only a small fraction of the counterevidence relevant to the conspiracies participants might name and so cannot reliably tailor arguments to each believer's claims and reasoning (Costello, 2025; Williams, 2026); the AI's advantage would lie in its informational facility.

Another class of accounts is source-based: participants defer to the AI because of what they think it is. AI systems may be seen as more objective, neutral, accurate, or less politically motivated than human interlocutors, consistent with work on source credibility and "machine heuristic" effects (Hovland & Weiss, 1951; Pornpitakpan, 2004; Sundar & Kim, 2019). People sometimes prefer algorithmic to human judgment ("algorithm appreciation"; Logg et al., 2019), although "algorithm aversion," harsher judgment of algorithms' mistakes than of humans', is also well documented (Dietvorst et al., 2015). Because participants in the original paradigm were told they would be talking to a neutral AI system (Costello et al., 2024), belief reductions may reflect deference to an apparently impartial, epistemically privileged source rather than (or in concert with) the content of the arguments (Glickman & Sharot, 2024).

A third class of accounts is interaction-based: the treatment's persuasive force depends on the dialogic, human–AI "chatbot" format. Conversations are a potent means of arbitrating between truth and falsehood (Mercier, 2016), and human–AI conversations are unusually personalized, judgment-free, socially warm, and non-adversarial, which may reduce defensiveness, increase engagement, or make participants feel understood (Sherman & Cohen, 2006). Sustained, respectful dialogue has also been argued to be central to the unusually durable attitude change produced by deep canvassing (Kalla & Broockman, 2020) and motivational interviewing (Miller & Rollnick, 2013). Dialogic AI's rhetorical fluency, long and informationally dense responses, and repeated back-and-forth opportunities to wear down objections could also drive the effect.

A final possibility is deflationary: the effect may partly reflect how belief was measured. In the original paradigm, participants named a conspiracy they believed and why, elaborated on their evidence, and only then rated their belief in an AI-generated restatement of their account. Because elaborating on a belief can itself prompt reflection or expose gaps in understanding (Fox et al., 2011; Sloman & Vives, 2022), the baseline rating, and the change measured from it, may depend on when elaboration occurs. And because the rated statement often incorporates the evidence participants cited, apparent change could reflect a narrow concession that those particular reasons were wrong, even if participants still hold the conspiracy to be true.

## Current Research

To address these questions, we systematically manipulated core elements that relate to each underlying possible causal factor (or set of factors). Study 1 begins with a baseline treatment that closely replicates the original AI dialogue treatment from Costello et al. (2024), then varies its features one at a time (Table 1). Some conditions target the content of the AI's response: No Facts, in which the model is instructed to persuade without using rational argumentation or evidence; Just-the-Facts, in which the model is prompted to inform rather than persuade; and Concise, in which the model produces substantially shorter responses. Other conditions hold that content constant while varying how the AI or the interaction is framed to participants: Overt, in which participants are told of the AI's persuasive goal, and Adversarial, in which the exchange is framed as a debate. Two additional variations change the order of the conspiracy measurement and the specificity of the AI-generated belief summary. Each content or framing condition is therefore a targeted ablation: attenuated belief change relative to the Baseline treatment implicates the removed ingredient, whereas an unchanged effect shows that it was not necessary. Study 2 compares the Baseline treatment with (i) a single AI-generated message, to probe the importance of the dialogue, and (ii) a Control dialogue about an irrelevant topic, to rule out regression to the mean. We also analyze moderators (Study 2) and the content of the AI's responses (both studies) post hoc.

**Table 1**

*Overview of Experimental Conditions and Inferential Targets Across Studies 1 and 2*

| Condition | Manipulation | Question(s) tested |
|---|---|---|
| **Baseline treatment (S1 & S2)** | A live, multi-turn dialogue in which the AI counters the participant's belief with reasons and evidence, replicating Costello et al. (2024). | Can we replicate the original belief-reduction effect? |
| ***Content of the AI's responses (targeted ablations)*** | | |
| **Just-the-Facts (S1)** | The AI is instructed to inform rather than persuade, while still providing relevant facts and counterarguments. | Is informative counterargument sufficient without overt persuasion? |
| **No Facts (S1)** | The AI is instructed to persuade without using evidence, counterargument, or reasoning, though it still says the theory is false. | Is evidence-based argumentation necessary? |
| **Concise (S1)** | The AI is instructed to keep its responses "as short as possible while still being persuasive," which roughly halved their length. | Does the effect require long, information-heavy responses? |
| ***Framing of the AI or the interaction (targeted ablations)*** | | |
| **Overt (S1)** | Participants are told beforehand that the AI will try to change their minds. | Does the effect require the AI to appear neutral? |
| **Adversarial (S1)** | The exchange is presented to participants as a debate with the AI. | Is a non-adversarial framing necessary? |
| ***Measurement procedure (checks on the deflationary account)*** | | |
| **Abstracted Summary (S1)** | The belief statement that participants rate is written without the specific evidence they cited; the dialogue is unchanged. | Does the change depend on participant-specific evidence written into the rated statement? |
| **Belief-First Rating (S1)** | Belief rated before, not after, the evidence-elaboration prompt (which still preceded the dialogue). | Is the pre–post change an artifact of rating belief right after elaborating on it? |
| ***Format and control (Study 2)*** | | |
| **Single Message (S2)** | Participants receive one tailored AI message instead of a dialogue. | Does the effect hinge on the AI's ability to engage in dialogue? |
| **Control (S2)** | Participants have a multi-turn AI dialogue about an unrelated topic (e.g., cats versus dogs) instead of their conspiracy belief. | Would conspiracy belief have been reduced even without exposure to the treatment, due to regression to the mean? |

*Note.* S1 = Study 1; S2 = Study 2.

# Method

## Participants and General Procedure

After giving informed consent and passing initial screening, participants completed demographic items including age, race/ethnicity, education, gender, political affiliation, and religion, followed by items about generative AI (see Measures). Both studies used the same core paradigm adapted from Costello et al. (2024). Participants first responded to an open-ended prompt asking whether there were any conspiracy theories they found especially credible or compelling and, if so, to describe one and explain why. On the next page, they elaborated on what led them to find that theory compelling, including any evidence, events, sources of information, or personal experiences they saw as supportive. Those responses were then passed to a live GPT-4-family model, which restated the participant's focal belief as a standardized psychometric item. Participants rated their confidence in that statement on a 0–100 scale (0 = Definitely false, 50 = Uncertain, 100 = Definitely true). This served as the pre-treatment belief measure. (In the Belief-First Rating condition of Study 1, the statement was generated from the first response alone and rated before the elaboration prompt.) To ensure that the paradigm targeted genuine conspiracy beliefs, participants' open-ended responses were also classified using the large language model (LLM)-assisted conspiracy-classification procedure described in Costello et al. (2024), which has shown strong convergence with human judgment (Gwet's AC1 = .80 [.72, .89]); Study 2 used gpt-4o with a more inclusive classification prompt (Supplemental Materials [SI] S1.3.2) and an altered restatement prompt. Articulating a conspiracy belief and rating the focal statement at or above the scale midpoint were preregistered analytic eligibility criteria applied to these pretreatment measures. Participants who failed attention checks or the writing-quality screener (both also preregistered), submitted duplicate responses, or used mobile devices were also excluded (SI S2.1). To prevent and detect automated or AI-generated responses, both studies began with a CAPTCHA, and the writing-quality screener was scored by an LLM instructed to flag incoherent, low-effort, or apparently AI-written answers (SI S1.3.3).

Immediately before the treatment, participants saw condition-specific instructions. Verbatim system-prompt and participant-facing instructions for every condition are reproduced in Supplemental Table S1; pre/post measurement items and attention checks are listed in SI S1.3, and key pretreatment moderator items are quoted under Measures below. In all active conditions, AI responses were generated live via API call to GPT-4 (Study 1) or GPT-5.2 (Study 2) rather than selected from a fixed bank of prewritten messages. Study 2 also included a Control condition wherein participants completed an AI conversation on a topic unrelated to conspiracy beliefs. To reduce the chance that results would depend on any one arbitrary control topic, control participants were randomly assigned to one of three unrelated conversation topics (experiences with the American medical system, views about firefighters, or cats versus dogs as companions). After the conversation with the AI, participants again rated belief in the focal conspiracy statement, providing the primary outcome measure. They then completed post-treatment questions; in Study 1, these included a brief open-ended explanation of why they did or did not find the AI's comments persuasive.

Study 1 was recruited from CloudResearch in two waves, each advertised as a 17.5-minute study and run under one protocol, apart from two sentences of the No Facts prompt (SI S6.2; raw $N$ = 3,988; 2,585 in Sample 1a and 1,403 in Sample 1b). The first wave (Sample 1a) was the original preregistered launch. Our preregistered allocation in Sample 1a was $n$ = 200 per active condition and $n$ = 400 in the Baseline treatment (more in the Baseline treatment to increase between-group precision). The preregistration listed ten conditions; two open-access model-comparison arms (Llama 3 70B and Llama 2 13B), together with an exploratory GPT-3.5 arm, were run only as a Sample 1a pilot ($n$ = 173 after screening) and are excluded from all analyses (Figure S1). After Sample 1a closed, the rate at which open-ended responses were classified as a conspiracy belief was lower than our preregistration assumed, so we ran a second wave (Sample 1b). Sample 1b used weighted randomization that, after pooling across waves, brought each of the eight remaining conditions close to its target sample size; sample size was indexed to a power calculation on Sample 1a effect sizes, which found that $n$ = 150 per active arm and $n$ = 350 in the Baseline treatment gave 80% power to detect a halving of treatment efficacy relative to the Baseline treatment. The exclusions, applied in order, were: failed CAPTCHA ($n$ = 8), accessed the survey on a mobile device ($n$ = 20), did not reach or failed the embedded attention check ($n$ = 297), did not reach the AI treatment ($n$ = 163), assigned to the pilot model-comparison arms ($n$ = 173), did not articulate a conspiracy belief ($n$ = 1,828), pre-treatment belief below 50/100 ($n$ = 96), missing post-treatment belief (attrition; $n$ = 38), and cross-wave duplicates dropped at the participant ID level ($n$ = 75). The full cascade is shown in Figure S1. The final analytic sample was $N$ = 1,290 (Sample 1a: 806; Sample 1b: 484). Per-condition *n*s after pooling were Baseline treatment = 333, Abstracted Summary = 164, Belief-First Rating = 137, Concise = 137, Adversarial = 131, No Facts = 129, Overt = 132, Just-the-Facts = 127.

Study 2 was recruited from CloudResearch in a single wave under the same intake protocol as Study 1 (raw $N$ = 2,082); the preregistered target was 400 conspiracy believers in each treatment arm and 200 in Control (the sensitivity of the achieved sample is reported with the Study 2 results). The exclusions, applied in order, were: mobile use at outset ($n$ = 483); did not reach condition assignment ($n$ = 136); not classified as a conspiracy belief by the LLM classifier ($n$ = 296); pre-treatment belief below 50 ($n$ = 126); failed at least one of two attention checks ($n$ = 24); flagged the AI-generated summary of their belief as inaccurate ($n$ = 88); and missing post-treatment belief (attrition; $n$ = 26). The final analytic sample was $N$ = 903. Per-condition *n*s were Single Message = 364, Baseline treatment = 336, and Control = 203 (Cats / Dogs = 58; Firefighters = 75; Medical system = 70). The full cascade is shown in Figure S2.

Participants were U.S. adults recruited through CloudResearch Connect who completed the study in English; all descriptive statistics refer to the final analytic samples. In Study 1 ($N$ = 1,290), mean age was 36.4 years ($SD$ = 10.9); 54.9% identified as female, 43.4% as male, and 1.7% as nonbinary or another gender (or preferred not to say); 54.7% held a bachelor's degree or higher; 39.3% identified as Democrats, 29.4% as Independents, and 23.4% as Republicans; and 25.2% were Protestant and 13.6% Roman Catholic. Study 2 ($N$ = 903) was sampled to balance partisanship (47.0% Democrat, 46.5% Republican); mean age was 40.9 years ($SD$ = 13.8), 56.1% identified as female, 41.9% as male, and 2.0% as nonbinary or another gender (or preferred not to say), 57.0% held a bachelor's degree or higher, and 32.0% were Protestant and 18.7% Roman Catholic. Full demographic breakdowns by study (Table S4 and Figure S3) and a comparison of retained and excluded participants (Table S5) are reported in Supplemental Section S3.

Because demographic and AI-attitude items preceded the conspiracy elicitation, we could compare retained participants with those excluded for not articulating a conspiracy belief (tests uncorrected; Table S5). In Study 1, conspiracy-believing participants were more Republican-leaning (party preference $d = 0.35$; 39% vs. 52% Democrat), more religious ($d = 0.19$), and less often college graduates (55% vs. 61%; all *p*s < .001); in Study 2, which balanced party, differences were smaller (all $|d| \leq 0.14$, all *p*s > .03; e.g., 56% vs. 63% female, $p = .043$). Age and trust in generative AI did not differ significantly between believers and non-believers in either study ($|d| \leq 0.12$, all *p*s > .06), and no characteristic of either analytic sample shifted by more than 0.18 *SD* relative to all participants who provided demographics.

**Measures**

*Focal belief (both studies).* Participants rated the AI-generated summary of their belief: "On a scale of 0% to 100%, please indicate your level of confidence that this statement is true" (slider labeled Definitely false [0], Probably false, Uncertain [50], Probably true, Definitely true [100]). The item was repeated immediately after the AI interaction and serves as the primary outcome. Alongside the first rating, participants answered "Did the AI accurately summarize your perspective?" (yes/no).

*Belief importance (both studies).* Participants answered "How important is this theory to your personal beliefs or understanding of the world?" (0 = Not at all important to my beliefs and worldview, 2 = Slightly important, 4 = Moderately important, 6 = Very important, 8 = Extremely important to my beliefs and worldview; odd values unlabeled). The item was asked before the AI interaction in both studies.

*Trust in generative AI (both studies).* Before describing their belief, participants rated the single item "I generally trust new AI technologies like ChatGPT" (1 = Strongly disagree, 2 = Disagree, 3 = Somewhat disagree, 4 = Neither agree nor disagree, 5 = Somewhat agree, 6 = Agree, 7 = Strongly agree).

*Study 2 moderators.* After the first belief rating and before random assignment, participants answered "Have people close to you (e.g., friends/family) ever told you that they agree with this conspiracy theory?" (social embeddedness; a single 0–100 slider labeled No, none of them; Yes, 1–2 of them; Yes, all of them) and rated four statements on a 5-point scale (Strongly disagree, Somewhat disagree, Neither agree nor disagree, Somewhat agree, Strongly agree). "I have heard many arguments against this belief before" and "I have discussed this belief with people who disagreed with me" were averaged as prior exposure to counterarguments ($r = .45$, $\alpha = .62$). "I have seriously considered whether this belief might be wrong" and "The arguments I've heard against this belief were weak or unconvincing" were analyzed as single items (SI S5.4a–S5.4b).

*Participants' explanations (Study 1 only).* On the page after the post-treatment rating, participants whose rating had decreased were asked, "In just a sentence or two, would you mind explaining what about the AI's comments you found persuasive?"; those whose rating was unchanged or higher were asked, "In just a sentence or two, would you mind explaining why you did not find the AI's comments to be persuasive?"

*Moderators from Costello et al. (2024).* Two moderator analyses draw on data from the two experiments reported in Costello et al. (2024), which used the same paradigm. Participants in both of those experiments answered the same pretreatment trust-in-generative-AI item as in Study 2, which allows the trust analysis to pool three experiments. Participants in the first of those experiments were also recontacted roughly 12 days after treatment and completed the six-item Actively Open-minded Thinking subscale of the Four-Component Thinking Styles Questionnaire (Newton et al., 2024; 1 = Strongly disagree, 2 = Disagree, 3 = Somewhat disagree, 4 = Somewhat agree, 5 = Agree, 6 = Strongly agree; all items reverse-scored; α = .91; highest-loading item: "Regardless of the topic, what you believe to be true is more important than evidence against your beliefs").

**Content Coding**

Three post hoc content-coding schemes were applied by large language models at temperature 0 (SI S6.1, S7.1–S7.2, S9.1). First, to characterize the AI's persuasive approach, each conversation was rated on 16 persuasion strategies (e.g., presenting conflicting evidence, offering alternative explanations, building rapport, telling stories, citing credible sources). The list comes from the analysis code of Costello et al. (2024), where candidate strategies had been generated by repeatedly asking GPT-4 Turbo what strategies it would use to dissuade someone from a conspiracy belief and retaining the recurring ones. GPT-4o (gpt-4o-2024-08-06) rated each of the AI's three messages in each conversation on each strategy (none, low, moderate, high), one call per strategy per conversation; we analyze the ratings of the first message. Second, to characterize the content of the AI's messages at a finer grain, GPT-4.1 first extracted each specific factual claim in each AI message (excluding opinions, advice, and general rhetoric). Each distinct claim was then classified into one of six categories (specific empirical, inferential-logical, meta-epistemic, moral/political, minimal/non-adjudicative, or general/underspecified) by Gemini 3 Flash. The categories were developed inductively. Third, to code participants' own explanations of why the AI was or was not persuasive, we developed 15 attributes (six positive, e.g., facts and logic, tailoring to the participant's claims; nine negative, e.g., perceived bias, lack of depth) by combining human review of the responses with candidate categories generated by an AI model given the same responses into a single taxonomy. GPT-4o judged each attribute present or absent in each response. The codebooks appear in Supplemental Tables S19, S23, and S31.

## Results

Preregistrations are available at https://aspredicted.org/hsrq-r9pc.pdf (Study 1) and https://aspredicted.org/8zz8rc.pdf (Study 2). All analyses other than the preregistered regressions (SI Tables S6 and S8) are post hoc.

Study 1 ($N$ = 1,290) asks which features of the original dialogue are necessary by removing them one at a time, comparing each variant with the Baseline treatment. Study 2 ($N$ = 903) asks whether dialogue itself is necessary (Single Message) and benchmarks the Baseline treatment against a dialogue on an unrelated topic (Control). We report each study's randomized contrasts in turn, then moderators, the content of the AI's responses, and participants' own accounts. Standardized within-condition pre–post reductions in Study 1 are reported as $d_{av}$, the mean reduction divided by the average of the pre- and post-treatment *SD*s (Lakens, 2013; 95% CIs from a 5,000-resample percentile bootstrap), because post-treatment ratings were more variable than pre-treatment ratings, which the eligibility criterion confined to the upper half of the scale (e.g., Study 1 Baseline treatment, $SD = 24.6$ vs. 15.6). Contrasts between Study 2 conditions are reported as $d_{ppc}$, the difference in mean pre–post reduction divided by the pooled pre-treatment *SD*. CrI is a credible interval; $BF_{10}$ and $BF_{01}$ are Bayes factors for and against a difference (SI S4.1); HC3 denotes heteroskedasticity-consistent standard errors.

### Which Ingredients of the Treatment Were Necessary?

Participants in Study 1 typically entered the experiment with strong endorsement of their focal conspiracy theories (pretreatment $M = 81.8$ on the 0–100 belief scale). In the Baseline treatment, belief decreased from 80.5 to 69.6 following treatment, a mean reduction of 10.9 points (95% CI [8.65, 13.11], $p < .001$, $d_{av} = 0.54$, 95% CI [0.45, 0.64]; Figure 1A). As in Costello et al. (2024), the reduction generalized across the specific conspiracy theories participants named. Its size varied by topic among treated participants ($\chi^2(14, N = 1{,}807) = 63.1$, $p < .001$), but in Study 2, which included a randomized control arm, the Baseline treatment effect did not differ significantly from control across topics ($p = .274$; see SI Sections S8.2–S8.3, Table S29 and Figure S9 for a per-conspiracy breakdown). The reduction also held at every level of pre-treatment belief, including the highest-conviction quintile (Figure 2). Among participants who rated their conspiracy as "very" to "extremely" important to their personal beliefs or understanding of the world (6–8 on the 0–8 importance item), the mean reduction across the seven evidence-based conditions (all but No Facts) was 8.03 points (95% CI [5.77, 10.30], $p < .001$, $d_{av} = 0.45$, 95% CI [0.35, 0.55], $n = 320$; post hoc). In Study 2, the Baseline treatment effect did not vary significantly with belief importance (9.8 vs. 14.7 points relative to Control among participants above and below this "importance" threshold; Baseline treatment × Importance $b = -2.43$ per *SD*, 95% CI [−5.42, 0.57], $p = .112$, $BF_{10} = 0.24$) or with pre-treatment belief (Baseline treatment × Pre-treatment Belief $p = .559$, $BF_{10} = 0.10$; both post hoc).

To help contextualize the magnitude of these effects, we computed (post hoc) how many participants moved from one side of the scale midpoint to the other. Because participants could enter the study at the midpoint itself (50 = "Uncertain"; about 4% of each sample), we restrict this statistic to those whose pre-treatment rating was above 50, for whom a post-treatment rating below 50 indicates movement from the "true" side of the scale to the "false" side. In the Study 1 Baseline treatment, 16.5% of these participants (52 of 315) ended below the midpoint (per-condition rates, and rates including participants who began at 50, in Supplemental Table S9).

**Figure 1**

*Treatment Effects in Study 1 (A) and Study 2 (B)*

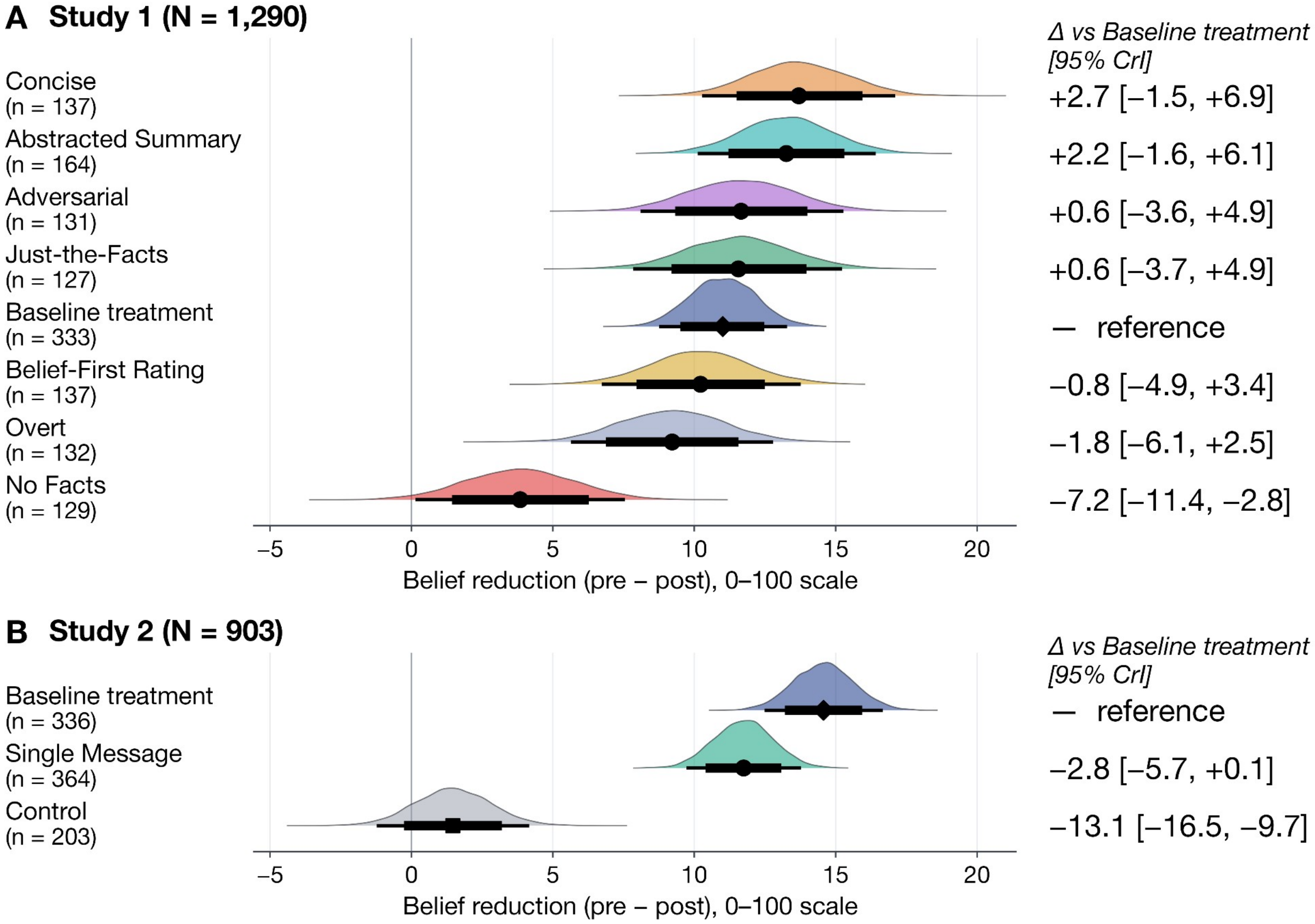

*Note.* Posterior distributions of belief reduction (pre − post, 0–100 scale) in each condition, from a Bayesian linear model controlling for pretreatment belief, fit separately for each study. Ridges are posterior densities; points mark posterior means (diamonds, the Baseline treatment; square, Control), and thick and thin bars show 80% and 95% credible intervals. Numbers at right give the posterior mean contrast with its 95% CrI against the Baseline treatment; positive values indicate a larger reduction than in the Baseline treatment.

**Figure 2**

*Belief Reduction as a Function of Pretreatment Belief, by Condition*

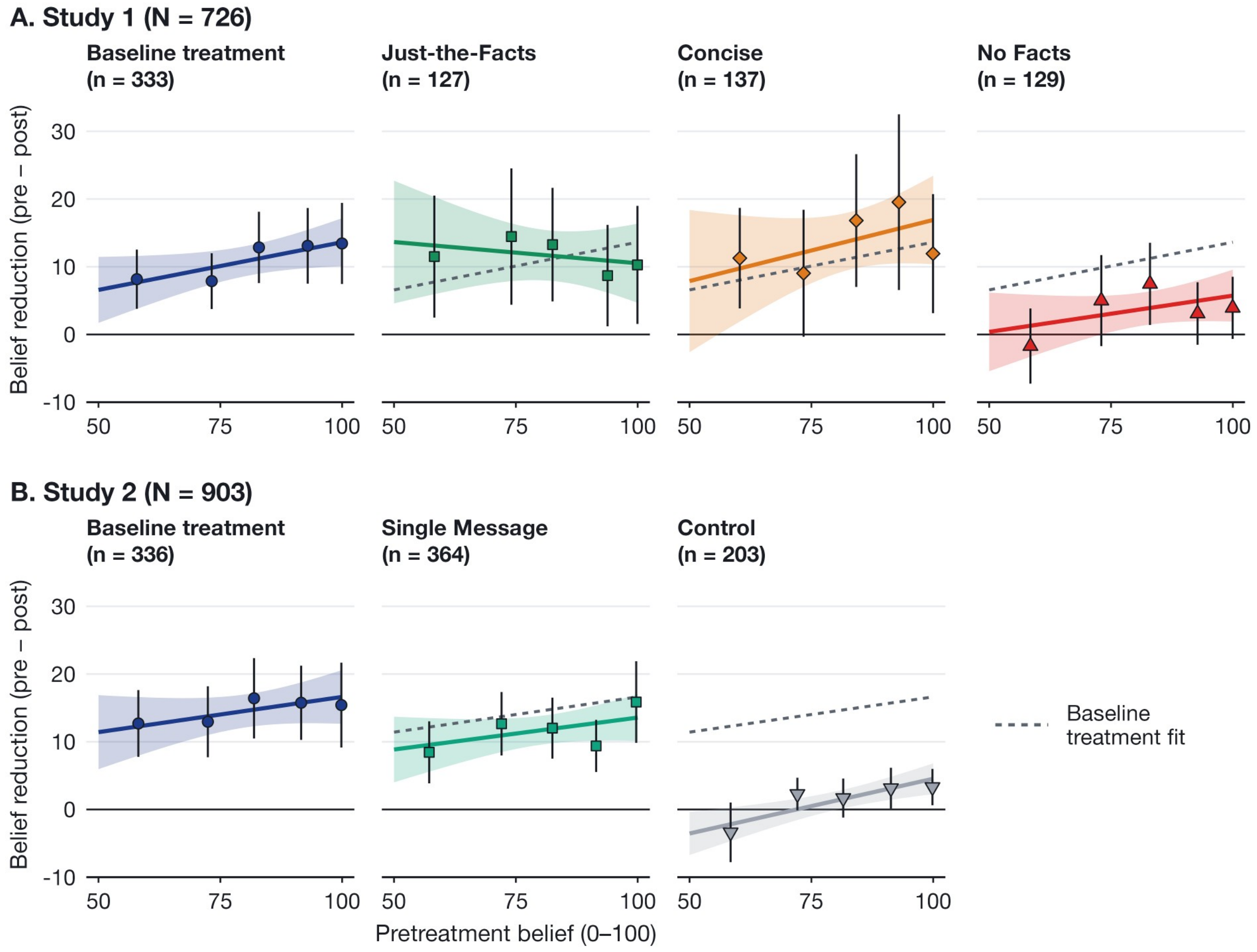

*Note.* Each panel shows one condition, plotting belief reduction (pre − post, 0–100 scale) against pretreatment belief. (A) Study 1 ($N = 726$): Baseline treatment, Just-the-Facts, Concise, and No Facts conditions. (B) Study 2 ($N = 903$): Baseline treatment, Single Message, and Control conditions. Solid lines and shaded bands are linear (ordinary least squares) fits of belief reduction on pretreatment belief within each condition, with 95% confidence bands. Points are mean belief reductions within quintiles of pretreatment belief (quintiles defined within study), plotted at the mean pretreatment belief of the participants in that cell; vertical bars are 95% confidence intervals. The dashed line repeats the Baseline treatment fit of the same study in the other panels for comparison. The fits and quintile means are descriptive (not model-adjusted), and the binning is post hoc.

Five Study 1 conditions were designed as targeted ablations of the Baseline treatment, and two altered the measurement procedure (Table 1). We compared the post-treatment belief score of each condition against the Baseline treatment while controlling for pre-treatment belief, and we report contrasts as differences in belief reduction, so that positive values indicate a larger reduction than in the Baseline treatment. In addition to the preregistered linear regressions, we also conducted complementary post hoc Bayesian regressions (Figure 1A). Given that many of the results are nulls, in the main text we focus on the Bayesian analyses as they quantify evidence for the null-difference interpretation rather than merely failure to reject it (full frequentist statistics for all contrasts are provided in Supplemental Tables S6 and S8, and are qualitatively equivalent). These contrasts could detect only large differences: each null contrast had 80% power for a difference of about 6–7 points, that is, 53–63% of the Baseline effect.

We observe a sharp asymmetry between the No Facts condition — in which the AI was instructed to persuade without relying on evidence or reasoned counterargument — and all other conditions. None of the conditions other than No Facts differed credibly from the Baseline treatment. Stripping the AI of explicit persuasive intent but still allowing it to provide topic-relevant facts and counterarguments (Just-the-Facts) did not detectably weaken the effect relative to the Baseline treatment (posterior mean difference = 0.55, 95% CrI [−3.73, 4.88], $BF_{01}$ = 9.12), with a raw mean reduction of 11.62 points (95% CI [7.92, 15.32], $p < .001$, $d_{av} = 0.57$, 95% CI [0.41, 0.74]). Instructing the AI to keep its responses "as short as possible while still being persuasive" (Concise), which roughly halved their length ($M$ = 588 vs. 1,226 words per conversation) and reduced the number of extracted claims by about a third (17.7 vs. 26.2 per conversation; see the claim-level analysis below), did not detectably weaken the effect (posterior mean difference = 2.69, 95% CrI [−1.49, 6.90], $BF_{01}$ = 4.37; raw mean reduction 13.83 points, 95% CI [9.57, 18.09], $p < .001$, $d_{av} = 0.65$, 95% CI [0.49, 0.83]). Framing the exchange as a debate (Adversarial) did not detectably weaken the effect (posterior mean difference = 0.65, 95% CrI [−3.62, 4.86], $BF_{01}$ = 8.75; raw reduction 11.85 points, 95% CI [8.02, 15.69], $p < .001$, $d_{av}$ = 0.59, 95% CI [0.43, 0.77]). Explicitly telling participants the AI would try to persuade them (Overt) did not detectably weaken the effect (posterior mean difference = −1.79, 95% CrI [−6.11, 2.48], $BF_{01}$ = 6.52; raw reduction 9.10 points, 95% CI [5.74, 12.45], $p < .001$, $d_{av} = 0.46$, 95% CI [0.32, 0.61]). The two measurement-procedure checks also matched the Baseline treatment. Instructing the summarizing model to omit participant-specific evidence from the rated conspiracy statement (Abstracted Summary) left the effect unchanged (posterior mean difference = 2.25, 95% CrI [−1.60, 6.12], $BF_{01}$ = 5.28; raw reduction 13.31 points, 95% CI [9.72, 16.90], $p < .001$, $d_{av} = 0.66$, 95% CI [0.52, 0.82]). Collecting the belief rating before the elaboration prompt (Belief-First Rating) did not alter the reduction (posterior mean difference = −0.79, 95% CrI [−4.86, 3.36], $BF_{01}$ = 8.83; raw reduction 10.28 points, 95% CI [6.80, 13.75], $p < .001$, $d_{av}$ = 0.52, 95% CI [0.36, 0.68]); further, pre-treatment ratings in this condition did not differ significantly from the Baseline treatment (82.5 vs. 80.5; difference = 2.03, 95% CI [−0.95, 5.01], $p = .18$), and participants were no more likely to flag the shorter statement as an inaccurate summary of their belief (3.6% vs. 5.6% in the six conditions that used the standard procedure; Fisher's exact test, $p = .422$).

Conversely, the No Facts condition was much less effective than the Baseline treatment, with a posterior mean difference from the Baseline treatment of −7.17, 95% CrI [−11.39, −2.83] and strong evidence for a genuine difference ($BF_{10}$ = 11.43). Raw mean reduction in No Facts was only 3.81 points (95% CI [1.36, 6.27]; one-sample $t(128) = 3.08$, $p = .003$; $d_{av} = 0.22$, 95%

CI [0.09, 0.35]), and only 5.7% of participants who began above the midpoint ended below it (vs. 16.5% in the Baseline treatment, Fisher's exact $p = .003$). Taken together, these results support evidence-based argumentation as a key driver of the AI's effect on conspiracy belief, while providing evidence against each of the other candidate explanations being necessary for a large effect. Specifically, the posterior probability of retaining at least half the Baseline effect was ≥ .97 for every variant other than No Facts (post hoc; Table S7). Attrition in Study 1 was 2.86% (38/1,328) and did not differ significantly across conditions, $\chi^2(7, N = 1{,}328) = 4.05$, $p = .774$ (per-condition rates in Supplemental Table S2), so differential attrition is not a concern in Study 1.

Participants in Study 2 likewise entered with strong endorsement of their focal conspiracy theories (pretreatment $M = 80.7$). In the Baseline treatment, belief decreased from 80.5 to 65.9, a mean reduction of 14.6 points, replicating Study 1, whereas the Control condition showed only a 1.4-point mean reduction (posterior mean difference from the Baseline treatment = −13.11, 95% CrI [−16.52, −9.71], $BF_{10} > 1{,}000$; Baseline treatment vs. Control $d_{ppc} = 0.88$, 95% CI [0.69, 1.07], $p < .001$; Figure 1B); the Baseline effect is therefore not attributable to regression to the mean. Among Baseline-treatment participants who began above the scale midpoint, 23.1% (75 of 324) ended below it, compared with 1.5% (3 of 197) in Control (Fisher's exact $p < .001$; Supplemental Table S9). Study 2 then asked whether the dialogic format itself was critical. As in Study 1, each condition was compared with the Baseline treatment, controlling for pretreatment belief, with the Baseline treatment as the reference group (Figure 1B; Table S8). The Single Message produced a large reduction relative to Control (posterior mean difference = 10.28, 95% CrI [6.93, 13.65], $BF_{10} > 1{,}000$; Single Message vs. Control $d_{ppc} = 0.69$, 95% CI [0.53, 0.87], $p < .001$), with a raw decrease of 11.76 points and 16.8% of participants who began above the midpoint ending below it. Compared with the Baseline treatment, the Single Message reduced belief by slightly less (posterior mean difference = −2.82, 95% CrI [−5.71, 0.06], $BF_{01} = 2.1$; HC3 $b = 2.87$ on the post-treatment belief scale, 95% CI [−0.36, 6.10], $p = .081$; $d_{ppc} = -0.19$, 95% CI [−0.41, 0.02]), retaining an estimated 79% of the Baseline effect (95% CrI [58%, 101%]); this contrast had 80% power to detect a difference of about 4.6 points (35% of the Baseline effect). Thus, even when the treatment was reduced to a single tailored AI response — with no iterative give-and-take, no ongoing rapport-building, and no dynamic responding — it produced a large reduction in conspiracy belief. Multi-turn dialogue is therefore not necessary to obtain a large effect, although it may confer some additional benefit beyond the static message.

Attrition in Study 2 differed by condition, $\chi^2(2, N = 929) = 12.22$, $p = .002$: Baseline treatment 5.08% (18/354), Single Message 0.82% (3/367), and Control 2.40% (5/208). To assess sensitivity to this differential attrition, we followed Costello et al. (2024) and conducted post hoc baseline-carried-forward-imputation analyses in which the 26 participants who dropped out after beginning the chat were assigned a post-treatment belief equal to their pre-treatment belief (i.e., a treatment effect of zero) and added to the analytic sample ($N = 929$). Treatment effects remained highly robust under this conservative assumption (post-treatment belief scale): Baseline treatment versus Control $b = -12.51$, $p < .001$; Single Message versus Control $b = -10.29$, $p <$

.001 (full model with HC3 *SE*s in Supplemental Table S3). Differential attrition is therefore unlikely to be driving the Study 2 findings.

**Moderators and Boundary Conditions (Study 2 and External Data)**

Study 2 also allowed us to test, post hoc, several boundary-condition accounts of the treatment (Figure 3). Each moderator below is reported both as a frequentist test with HC3 standard errors (*b*, 95% CI, *p*) and a Bayes factor ($BF_{10}$) from a Bayesian model (priors in SI S4.1; SI Tables S18a and S18b).

A first possibility is that the effect relies on people deferring to AI as a trustworthy source. Because Study 2 and both experiments of Costello et al. (2024) measured trust in generative AI with the same pretreatment item, we pooled the three experiments for a higher-powered test (post hoc; $n = 2{,}594$). Each experiment included a treatment arm and randomly assigned control arm, with the model including experiment fixed effects and experiment-specific treatment effects and trust slopes (i.e., treatment–control contrasts are made within an experiment; arm sizes and specification in SI S5.1, Table S11). We found a positive interaction between trust in AI and treatment (posterior mean = 2.13 points per trust scale point, 95% CrI [1.09, 3.16], $BF_{10} > 30$; HC3 $b = 2.13$, 95% CI [1.29, 2.98], $p < .001$; Figure 3A), which did not differ detectably across the three experiments ($p = .317$). Substantively, however, the treatment remained highly effective even among participants with low trust in AI (e.g., 10.7 belief points at 1 *SD* below mean AI trust, 95% CrI [8.3, 13.1]). Greater trust in AI may modestly facilitate the treatment but is not necessary for it.

A second possibility is that the social context in which a belief is held bounds the treatment's efficacy, such that beliefs reinforced by one's social network do not respond to correction. Social embeddedness of the focal belief did not linearly predict a smaller Baseline treatment effect: the Treatment × Social Embeddedness interaction was not credibly different from zero (posterior mean = −0.065 points of belief reduction per 1-point increase in embeddedness, 95% CrI [−0.188, 0.060]; $BF_{10} = 0.15$ for the interaction expressed per *SD* of embeddedness; HC3 $b = -1.81$ per *SD*, 95% CI [−4.84, 1.21], $p = .240$; Figure 3B). Even in the top quartile of embeddedness (≥ 80 on the 0–100 scale; $n = 229$), the Baseline treatment produced a substantial reduction relative to Control ($b = -7.30$ on the post-treatment belief scale, $p = .001$). Relatedly, a two-item composite measuring prior exposure to counterarguments did not moderate the treatment effect (posterior mean of the Treatment × Prior-Exposure interaction = −0.63 points of belief reduction per scale point, 95% CrI [−3.56, 2.27], $BF_{10} = 0.08$; HC3 $b = -0.71$ per *SD*, 95% CI [−3.47, 2.06], $p = .617$; SI S5.4, Table S15, Figure S5).

The account developed so far — that the treatment works by supplying diagnostic counterevidence to people open to revising their beliefs — makes one further prediction: the effect should be larger for individuals more inclined to engage in deliberation in order to form more accurate beliefs (Pennycook, 2023). We could examine this only indirectly, using data from Study 1 of Costello et al. (2024) not previously reported for this question: participants in that study were recontacted roughly 12 days after the experiment and completed the Actively Open-minded Thinking (AOT) subscale of the Four-Component Thinking Styles Questionnaire (Newton et al., 2024; see Measures). Regressing immediate post-treatment belief change on condition, AOT, their interaction, and pretreatment belief in a Bayesian model with weakly informative priors (treatment $n$ = 365, control $n$ = 258), we found that a 1-*SD* increase in AOT was associated with a 3.72-point larger treatment effect (posterior mean; 95% CrI [0.54, 6.95], $BF_{10}$ = 1.09; HC3 $b$ = 3.73, 95% CI [0.96, 6.51], $p$ = .009; Figure 3C; SI S5.3), although the effect remained sizable at 1 *SD* below the mean (12.2 points). Because AOT was measured after the conversation, the direction of this association is uncertain — participants who had been persuaded may afterwards have described themselves as more open-minded — although AOT did not differ by condition ($d$ = 0.04, 95% CI [−0.12, 0.20], $p$ = .605) and follow-up completion did not differ across arms (78.0% vs. 78.7%, $p$ = .822), which makes that reading less likely.

In Study 2, we found no significant moderation of the Baseline treatment effect by pre-treatment belief ($b$ = −0.86 points per *SD*, 95% CI [−3.77, 2.04], $p$ = .559), education ($b$ = −2.60, 95% CI [−5.27, 0.07], $p$ = .056), religiosity ($b$ = −0.94, 95% CI [−3.92, 2.04], $p$ = .537), age ($b$ = −0.39, 95% CI [−3.25, 2.47], $p$ = .790), or party identification (Republicans vs. Democrats: $b$ = −4.52, 95% CI [−10.22, 1.18], $p$ = .120), all with $BF_{10} \leq 0.30$ per *SD*, and the effect did not differ significantly across conspiracy topics ($p$ = .274; few participants per topic, SI S5.5). Women showed a larger effect than men ($b$ = 8.34, 95% CI [2.76, 13.93], $p$ = .003, $BF_{10}$ = 1.72 per *SD*), yet this difference did not appear in Study 1 (association $p$ = .909) or in either experiment of Costello et al. (2024; interaction $p$s = .959 and .625; Table S18b). Participants who rated the counterarguments they had previously heard as weak or unconvincing showed a smaller effect ($b$ = −3.51 per *SD*, 95% CI [−6.39, −0.64], $p$ = .017, $BF_{10}$ = 0.84).

**Figure 3**

*Treatment Effects (Treatment − Control) as a Function of Three Candidate Moderators*

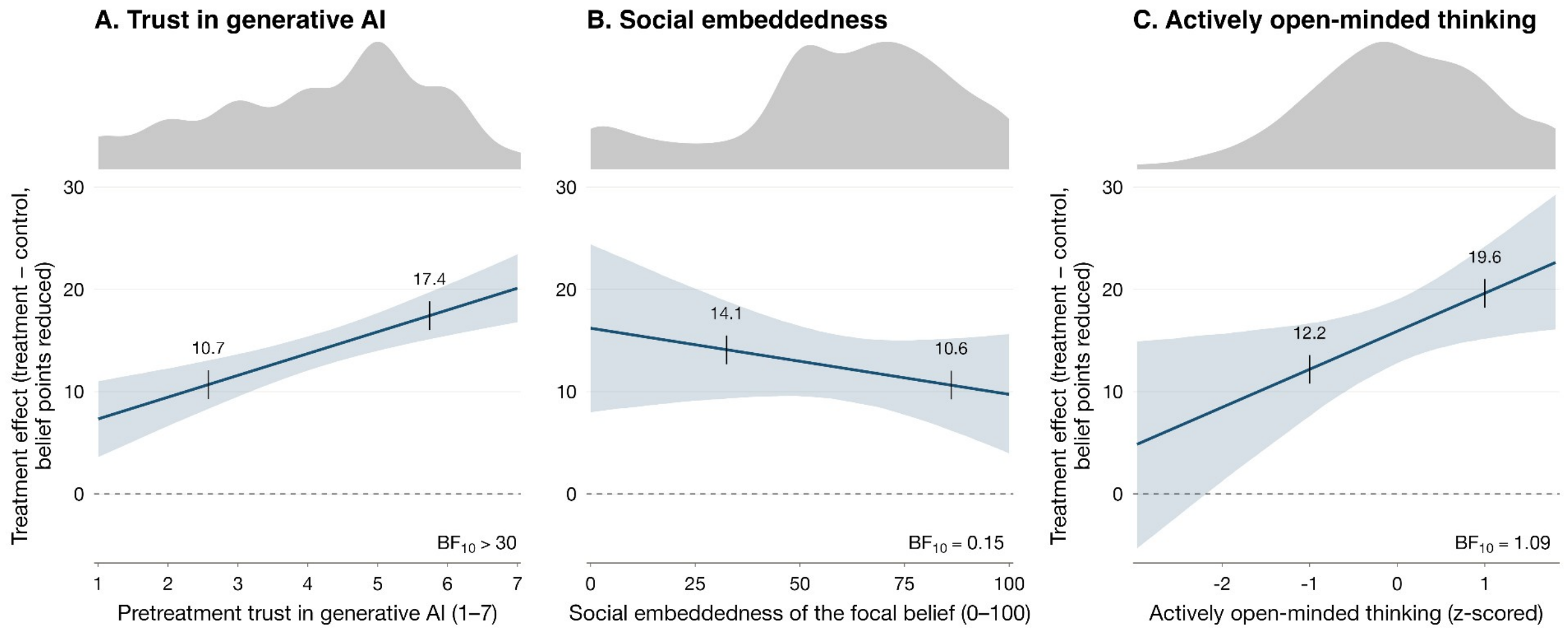

*Note.* Lines show the posterior mean of the model-predicted difference in belief reduction between the AI-counterargument treatment and the unrelated-topic control, with 95% credible bands, from Bayesian linear models controlling for pretreatment belief; numbers mark the

predicted difference at ±1 *SD* of the moderator, gray densities above each panel show the moderator's distribution, and $BF_{10}$ in each panel is the Bayes factor for the Treatment × Moderator interaction. (A) Trust in generative AI, pooled across three experiments (Study 2, Baseline treatment vs. Control; Costello et al., 2024, Studies 1 and 2; $n = 2{,}594$); each experiment contributes its own treatment and control arms, and the model includes experiment fixed effects and experiment-specific treatment effects and trust slopes. (B) Social embeddedness of the focal belief (0–100; Study 2, Baseline treatment vs. Control, $n = 539$); the Bayes factor in this panel is per *SD* of embeddedness, as in the text, which gives the posterior mean per 1-point increase (SI S5.2). (C) Actively open-minded thinking (standardized; Costello et al., 2024, Study 1, treatment vs. control, $n = 623$), measured at a follow-up roughly 12 days after treatment.

**What Was the AI Actually Saying? (Content Analyses)**

To better understand how the AI reduced conspiracy beliefs — and why the No Facts condition in particular was much less effective — we conducted two complementary post hoc content analyses. We first characterize the dialogues at the level of broad persuasive strategies (Study 1), then drill down to the level of individual claims (Studies 1 and 2). The coding procedures are described in the Method.

*Strategy-level analysis (Study 1).* We focused on the four Study 1 conditions that most directly manipulated the AI's conversational behavior (Baseline treatment, Just-the-Facts, No Facts, Concise), analyzed as randomized within the analytic sample ($n$ = 309, 111, 116, and 123, respectively; $N = 659$). GPT-4o rated the AI's first message in each conversation on 16 candidate persuasion strategies (four-level ordinal scale: none / low / moderate / high; definitions of all 16 strategies in Supplemental Table S19). Because many strategies were conceptually overlapping, we summarized them using a polychoric-correlation principal component analysis (PCA) with varimax rotation. A parallel analysis indicated five components, which together accounted for 76% of the variance (Figure 4A). Based on the loadings, we labeled these Facts & Evidence (rebutting the participant's specific claims with contradicting evidence, alternative explanations, and logical flaws — that is, factual and reasoned rebuttal); Sources & Experts (citing credible sources, expert consensus, and further resources); Rapport / Common Ground (validating the participant, finding shared values, and attending to the feelings behind the belief); Stories / Metaphors / Examples (anecdotes, analogies, and illustrations); and Harms & Empathy (asking the participant to consider how the belief affects other people, and describing the harms of conspiracy beliefs).

Strategy use differed across the four conditions on every component (all omnibus $p$s < .001; Figure 4B; all 30 pairwise contrasts in SI Section S6.2), in ways that closely tracked the prompt manipulations. Relative to the Baseline treatment, No Facts was 1.90 *SD*s lower on Facts & Evidence and 0.77 *SD*s lower on Sources & Experts — the lowest of the four conditions on both — and 1.83 *SD*s higher on Stories / Metaphors / Examples and 0.58 *SD*s higher on Harms & Empathy (all Holm-adjusted $p$s < .001); the two conditions did not differ on Rapport / Common Ground (−0.03 *SD*s, Holm-adjusted $p = .533$). When told not to rely on evidence, the model stated its disagreement, as its prompt instructed, and then turned to analogies and stories about false beliefs and to appeals to feeling. Namely, it acknowledged the emotions behind the belief and asked participants to think about the people affected by the events in question (91% of No Facts first messages; e.g., the victims and first responders of 9/11) and, less often, about how the theory itself might weigh on the believer or hurt those people (19% of No Facts first messages). No Facts almost never cited sources or expert consensus (representative excerpts from No Facts conversations are reproduced in Supplemental Table S33). Just-the-Facts moved in an orthogonal direction: it did not differ significantly from the Baseline treatment on Facts & Evidence (0.17 *SD*s lower, Holm-adjusted $p = .140$) but was 1.42 *SD*s lower on Rapport / Common Ground (Holm-adjusted $p < .001$). Concise behaved like Just-the-Facts: relative to the Baseline treatment it was only 0.35 *SD*s lower on Facts & Evidence but 1.12 *SD*s lower on Rapport / Common Ground (and 0.31 and 0.35 *SD*s lower on Sources & Experts and Harms & Empathy; all Holm-adjusted $p$s $\leq .015$), and it did not differ from Just-the-Facts on any component (all Holm-adjusted $p$s $\geq .092$). That is, when told to be brief, the model kept its evidence-based content and cut social and rhetorical material.

We next asked which of these strategy families were associated with larger belief changes. We fit a five-mediator path model (ML estimation) in which the five PCA strategy components served as parallel mediators between the prompt manipulation and belief change, with mediator residual covariances freely estimated (SI S6.4, Table S21). The model simultaneously estimates (i) the effect of each condition on each strategy component (five a-paths per condition), (ii) the partial association of each strategy component with belief change holding condition and the other four components constant (the five b-paths reported below), and (iii) the residual direct effect of each condition on belief change. Model specification was saturated, so we report parameters but not model fit (standard errors and 95% CIs are computed from a 5,000-resample bootstrap). Across the five b-paths the pattern was asymmetric (Table S21). Two components were positively associated with belief reduction: Facts & Evidence ($b$ = 4.85 per *SD*, 95% CI [2.29, 7.25], $p < .001$) and Sources & Experts ($b$ = 3.24, 95% CI [1.39, 5.20], $p = .001$). These two coefficients did not differ significantly from each other (difference = 1.61, 95% CI [−1.37, 4.35], $p = .269$), but each exceeded each of the other three (differences 3.35–5.94, all unadjusted 95% CIs excluding zero; omnibus test of equal coefficients $\chi^2(4, N = 658) = 14.41$, $p = .006$; Table S22). The coefficients for Rapport / Common Ground ($b$ = −0.11, 95% CI [−2.19, 1.86], $p = .916$), Stories / Metaphors / Examples ($b$ = −1.09, 95% CI [−2.85, 0.78], $p = .238$), and Harms & Empathy ($b$ = −0.83, 95% CI [−2.61, 0.92], $p = .355$) were near zero and did not differ among themselves (joint test $p = .747$).

Indeed, in that model's decomposition, No Facts' lower Facts & Evidence score carried the largest indirect association with belief change ($b_{\text{indirect}} = -11.07$, 95% CI [−16.83, −5.17]; 69% of the summed indirect paths), with Sources & Experts accounting for most of the remainder. A complementary generalized additive model — which relaxes the linearity assumption and estimates each strategy's partial contribution non-parametrically — showed the same positive associations for Facts & Evidence and Sources & Experts (both *ps* < .001; Figure 5). It also showed a small negative association for Harms & Empathy ($p = .001$; SI S6.3).

**Figure 4**

*Persuasion-Strategy Components and Their Use Across the Four Study 1 Prompts*

**A**

| Strategy (GPT-4o rating, 0–3) | Rating distribution None → High | Facts & Evidence | Rapport / Common Ground | Harms & Empathy | Sources & Experts | Stories / Metaphors / Examples | Communality |
|---|---|---|---|---|---|---|---|
| Alternative Explanations | | **.86** | -.15 | .04 | **.34** | -.10 | .90 |
| Inconsistencies / Fallacies | | **.83** | -.03 | -.04 | .00 | -.19 | .73 |
| Conflicting Evidence | | **.80** | -.20 | -.01 | **.46** | -.12 | .91 |
| Critical Thinking | | **.59** | .28 | **-.32** | .12 | **-.50** | .79 |
| Build Rapport | | -.10 | **.93** | .11 | .12 | .08 | .90 |
| Common Ground | | -.23 | **.83** | .02 | -.03 | .07 | .75 |
| Maintaining Patience | | .27 | **.77** | .14 | -.04 | .19 | .72 |
| Psychological Needs | | -.10 | **.66** | **.41** | -.02 | .30 | .71 |
| Socratic Questioning | | -.28 | **.42** | **-.39** | -.11 | -.09 | .43 |
| Harm | | .05 | .12 | **.85** | -.01 | -.25 | .79 |
| Encourage Empathy | | -.26 | **.33** | **.80** | -.14 | .20 | .89 |
| Expert Consensus | | .22 | -.09 | -.09 | **.85** | -.13 | .81 |
| Credible Sources | | **.32** | -.13 | -.07 | **.85** | -.15 | .87 |
| Offer Resources | | .01 | .22 | .05 | **.70** | -.09 | .54 |
| Stories / Examples | | -.19 | .16 | .00 | -.17 | **.79** | .72 |
| Analogies / Metaphors | | -.14 | .25 | -.09 | -.16 | **.78** | .72 |

**B**

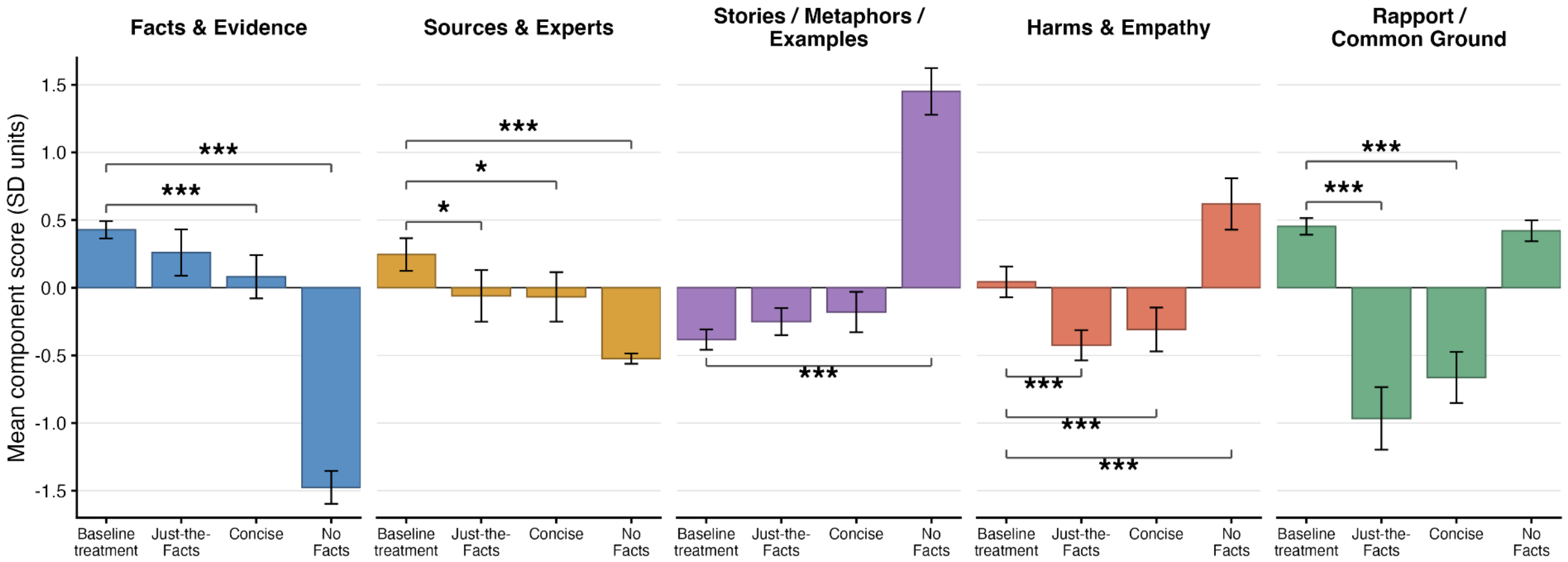

*Note.* (A) Varimax-rotated loadings of the 16 coded persuasion strategies on five components from a principal component analysis of polychoric correlations among GPT-4o (gpt-4o-2024-08-06) ratings of the AI's first message in each conversation (none / low / moderate / high; $N$ = 1,366 coded conversations). The five components explain 76% of the variance (root mean square residual = .07). All loadings are shown; loadings of .30 or greater in absolute value are in bold, cells are shaded in the component's color in proportion to the absolute loading, and horizontal rules group strategies by their primary component. Small bars show the distribution of each strategy's ratings, and the last column gives each strategy's communality (the proportion of its variance explained by the five components). (B) Mean component scores by condition, one panel

per component, for the four conditions as randomized within the Study 1 analytic sample (Baseline treatment $n = 309$, Just-the-Facts $n = 111$, Concise $n = 123$, No Facts $n = 116$; $N = 659$); scores are standardized within these four conditions (*SD* units). Error bars are 95% confidence intervals. Brackets mark HC3-robust pairwise contrasts against the Baseline treatment, Holm-adjusted within each strategy across its six pairwise comparisons (*$p < .05$, **$p < .01$, ***$p < .001$; contrasts without a bracket are not significant; all 30 contrasts in SI Table S20).

**Figure 5**

*Estimated Partial Effects of Persuasion Strategies on Belief Reduction From a Generalized Additive Model (Study 1)*

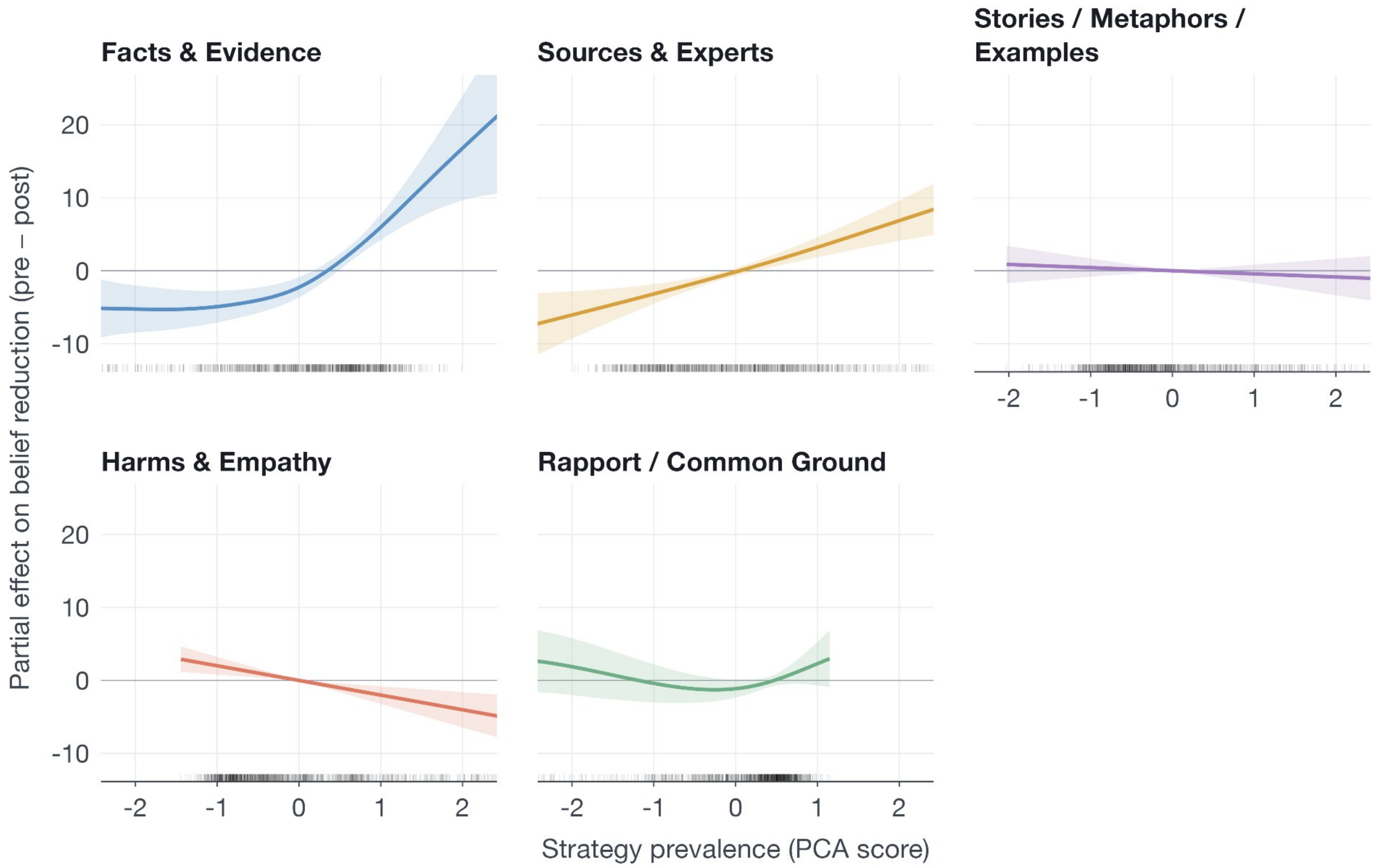

*Note.* Each panel shows the estimated partial effect of one strategy component (its score in the AI's first message, in *SD* units) on belief reduction (pre − post, 0–100 scale), adjusting for pre-treatment belief and the other four components but not for condition. Zero on the vertical axis marks the sample-average reduction, not the absence of change. Lines are truncated to the empirical range of each component, and the horizontal axis is cut at ±2.4 *SD*, which omits about 2%–4% of scores per component. Shaded ribbons are pointwise 95% confidence intervals; tick marks along the horizontal axis show the observed component scores. $N = 1{,}174$ rated conversations in the Study 1 analytic sample (all eight conditions; SI S6.3).

*Claim-level analysis (Studies 1 and 2).* Strategy labels still leave open a substantive question: What did "facts and evidence" actually consist of? To answer this question, we passed each saved AI message to GPT-4.1 and had it extract the specific factual claims in that message (the extraction prompt excluded opinions, advice, and general rhetoric). We then deduplicated the claim texts and coded them with Gemini 3 Flash using a codebook with six top-level categories: specific empirical claims, inferential-logical claims, meta-epistemic claims, moral/political evaluation claims, minimal / non-adjudicative claims, and general or underspecified claims (see Table 2 for examples; the codebook is in SI Section S7.2, Table S23, with subtypes in S7.7).

**Table 2**

*Claim-Type Examples (Pooled Across Study 1 and Study 2)*

| Claim type | Example verbatim claims |
|---|---|
| Specific empirical | The Apollo flag had a horizontal rod along the top edge to hold it out.<br>According to the National Institute of Standards and Technology (NIST), the steel's temperature reached up to 1,000°C. |
| Inferential / logical | It would require keeping a secret amongst thousands of people for over 50 years if the moon landing were faked.<br>If a claim is true, there should be a testable paper trail identifying the specific location, procedure, audit, or court filing with supporting records. |
| Meta-epistemic | Verified evidence of alien life would require evidence that clearly cannot be explained by anything else and can be checked by many independent experts.<br>General mistrust of billionaires or tech is not evidence; motive is not evidence and capability alone is not evidence. |
| Moral / political evaluation | Upholding the rule of law, ensuring judicial independence, and engaging in constructive, informed discourse are pathways to reinforcing democratic governance.<br>If exposed, allowing such an attack would be one of the biggest crimes in US history. |
| Minimal / non-adjudicative | Federalism distributes power between national and state governments.<br>Parkland Hospital is the same hospital where Kennedy and later Oswald were taken. |
| General / underspecified | Governments do sometimes create confusing or misleading paperwork for ordinary reasons like protecting sources, covering embarrassing mistakes, or hiding human technology.<br>Most evidence for aliens at Area 51 is anecdotal and hard to verify. |

*Note.* Each example is a verbatim AI claim. Examples are sampled at random within type from claims in the top quartile of the classifier's confidence ratings (SI S7.2).

The pipeline yielded 30,662 claim instances (30,314 distinct claim texts) in Study 1 and 27,535 claim instances (27,352 distinct claim texts) in Study 2. Pooling across all Study 1 conditions, the distribution of distinct claims was 47.4% specific empirical, 24.1% minimal / non-adjudicative, 15.2% general/underspecified, 6.2% moral/political, 4.1% inferential-logical, and 2.9% meta-epistemic. The share of specific empirical claims ranged from 44.5% in Concise to 53.4% in Just-the-Facts across most conditions (47.6% in the Baseline treatment; per-condition composition in Table S24), but collapsed in No Facts — only 9.9% of No Facts claims were specific empirical, averaging just 0.7 such claims per conversation (vs. about 13 specific-empirical claims per conversation across the other Study 1 conditions). Concise, by contrast, reduced the volume of content while largely preserving its profile: 52% fewer words and 32% fewer claims than the Baseline treatment (17.7 vs. 26.2 claims per conversation). Together with the experimental contrasts, this indicates that the full amount of information the Baseline AI provided is not required for a comparable effect.

Study 2, which used GPT-5.2 rather than GPT-4, showed a similar pattern. Across the treated Study 2 corpus, the mix was 37.2% specific empirical, 22.5% minimal / non-adjudicative, 18.9% general/underspecified, 16.4% inferential-logical, 3.3% meta-epistemic, and 1.7% moral/political.

**Participants' Own Accounts of What Was Persuasive (Study 1)**

As a final post hoc analysis, we examined the free-text responses collected at the end of Study 1 (Study 2 did not include these items). Participants whose post-treatment rating was lower than their pre-treatment rating were asked what about the AI's comments they found persuasive ($n$ = 584 with coded responses), and those whose rating was unchanged or higher were asked why they found the comments unpersuasive ($n$ = 474; see Measures). As described in the Method, GPT-4o coded each response for the presence or absence of each of 15 attributes covering positive themes (e.g., factual content, tailored argument, expertise, emotional synchrony) and criticisms (e.g., perceived bias, repetitiveness, off-topic drift, obsequiousness; full codebook in Table S31). Persuaded participants overwhelmingly cited the AI's provision of tailored facts, evidence, and logic, with far fewer mentioning its trustworthiness or expertise (Figure 6). Unpersuaded participants also cited informational factors such as lack of depth and insufficient evidence, while also bringing up perceived *lack* of trustworthiness and objectivity on the part of the AI. Codes were not mutually exclusive (SI Figure S10), and thus some responses included an attribute from the "opposite" column: 11.2% of unchanged-or-increased participants mentioned at least one positive attribute, and 15.1% of belief-reduced participants mentioned at least one criticism. Most of these are mixed appraisals (praise with a caveat, or a small reduction accompanied by a complaint). Criticisms appeared in 22% of participants whose belief fell by 20 points or less but in 6% of those who moved more (Fisher's exact $p < .001$). These analyses thus further emphasize the importance of evidence and reasoning for belief updating, while also being consistent with our observation that trust in AI moderates the treatment effect to some extent.

**Figure 6**

*Attributes Participants Named When Explaining Why the AI Was or Was Not Persuasive (Study 1)*

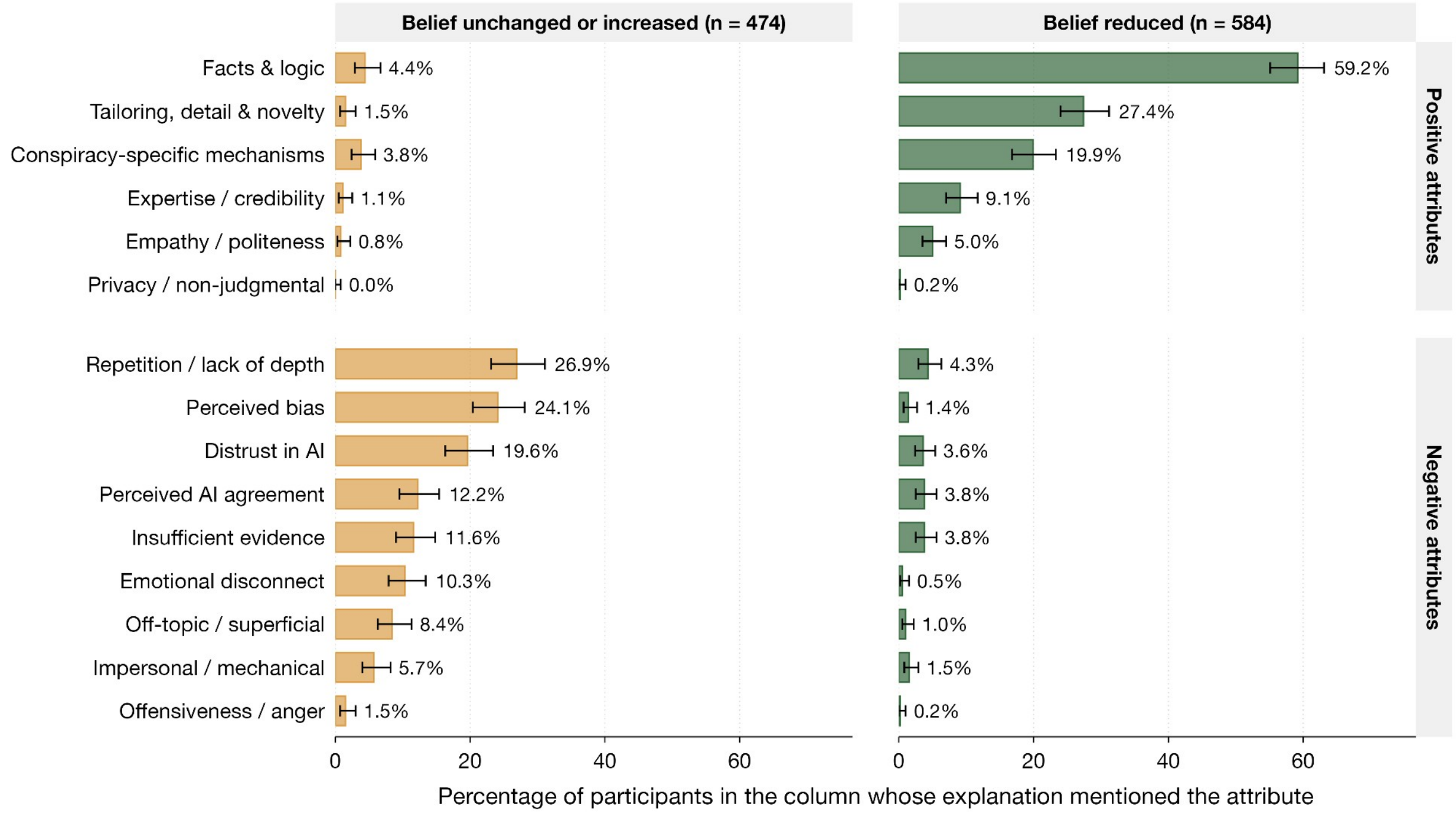

*Note.* Bars show the percentage of participants in each column whose written explanation was coded as mentioning the attribute (GPT-4o, one call per attribute per response). Codes are not mutually exclusive so a response may contribute to several bars or to none (SI Figure S10 shows how often attributes were mentioned together). Participants are the Study 1 analytic sample excluding No Facts, restricted to the 1,058 with a non-empty explanation and cached codes (codes exist for 1,061 of 1,161 eligible participants; three blank responses were excluded). Top: positive attributes; bottom: negative attributes. Attribute labels are short names for the codes defined in Table S31 (e.g., "Perceived AI agreement": the participant perceived the AI as agreeing with them). Error bars are 95% Wilson intervals. Attributes appearing in the opposite column (e.g., a criticism from a participant whose belief fell) are retained as coded (SI S9.3).

## Discussion

Previous work has established that conversations with generative AI can substantially reduce belief in conspiracies (Costello et al., 2024). Here we probed a set of theoretical and methodological claims about why this effect occurs. In Study 1, six of seven variations — removing the AI's apparent neutrality, its non-adversarial framing, its persuasive intent, and roughly half of its verbosity, and altering two features of the measurement procedure — left the effect intact; only instructing the AI to persuade without evidence, counterargument, or reasoning sharply weakened it. In analyses of what the AI said, belief change tracked its reliance on evidence and reasoned rebuttal, not rapport-building, storytelling, or appeals to empathy, and persuaded participants most often cited its factual arguments. We thus find strong evidence that an evidence-rich argumentative package featuring facts is a necessary and central mechanism of the effect.

In Study 2, a single tailored message retained an estimated 79% of the full dialogue's effect and far exceeded an unrelated-topic control dialogue, in which belief did not move. Dialogue may confer some additional benefit, but much of the effect can be achieved with a single message tailored to the specific belief.

Our data also argue against several other candidate features being necessary for a large debunking effect: the AI's rhetorical abilities (arranging words persuasively with substantive content held fixed), its ability to converse, its rapport and social warmth, the machine heuristic or algorithm aversion, measurement error, regression to the mean, and information overload. That said, our studies could detect only large differences between treatment arms (about half to two-thirds of the Baseline effect in Study 1 and about one third in Study 2), and we did not experimentally increase these elements or manipulate them in combination. Individual features, including multi-turn dialogue and trust in AI, may therefore contribute to the effect but to a smaller degree. Identifying the potentiating and inhibitory features of AI-based argumentation is a fruitful direction for future research (Jones & Bergen, 2024). Nonetheless, what we do show is that none of these elements are necessary for a large effect. Nor do they themselves evidently produce large effects.

Recent external findings converge with this picture. First, in the same paradigm, believing one was talking to a human expert rather than an AI produced comparable belief reductions (Boissin et al., 2025), so the effect does not rely on deference to AI as a source. Second, the effect persisted without substantive change at a 2-month follow-up (Costello et al., 2024), suggesting that the updating is not a fleeting demand artifact. Third, research in other domains has found support for the role of facts and evidence in AI-mediated belief updating. Lin et al. (2025) found that depriving AI canvassers of factual claims dramatically reduced their persuasiveness in the 2025 Canadian national and Polish presidential elections, and Hackenburg et al. (2025) found that instructing an AI to persuade with large quantities of facts and evidence was the most effective strategy on UK political issues. However, despite these few examples, future work is needed to map how central evidence-based argumentation is to belief change across contexts.

Our results suggest that many conspiracy believers are substantially more evidence-responsive than the standard portrait of motivated reasoning, epistemic closure, and self-sealing worldviews implies (Biddlestone et al., 2021; Lewandowsky et al., 2013; Sunstein & Vermeule, 2009). Participants low in trust in AI, with highly socially embedded beliefs, or previously familiar with anti-conspiracy arguments showed large belief updates. Although the effect was directionally smaller for beliefs rated very to extremely important to one's worldview, it was still large (in Study 2, 9.8 vs. 14.7 points relative to Control for less important beliefs; interaction $p$ = .112).

Data from Costello et al. (2024, Study 1) are also consistent with individual variation in the propensity to update: participants higher in actively open-minded thinking (the disposition to revise beliefs according to evidence; Pennycook et al., 2020; Stanovich & Toplak, 2019) showed larger belief revision, though the effect was substantial even at low levels of the trait. However, since the trait was measured after treatment, the direction of this association is uncertain and requires further research.

Our work has several limitations. First, participants self-selected into online studies and may not represent the most committed believers. Nonetheless, they enrolled without knowing that the studies concerned conspiracy beliefs, and relatively few dropped out once they learned this. Relatedly, the sample skews modestly more female (55% vs. 51% of U.S. adults; U.S. Census Bureau, 2024), substantially more educated (56% vs. 36% of U.S. adults aged 25 and older with a bachelor's degree or higher), and substantially less religious (45% vs. 62% Christian; Pew Research Center, 2025) than population estimates; if the treatment's effect varies with these characteristics, our effect sizes may not generalize without adjustment to a probability sample of U.S. adults. However, moderation tests showed no reliable and significant differences in effect by demographic characteristics, suggesting that the core finding generalizes at least across the demographic range we observed. Also, the laboratory isolates participants from social and informational networks that may reinforce such beliefs, although the Baseline treatment remained effective even for the most socially embedded beliefs, which addresses the strongest version of this concern.

Second, the No Facts prompt removes a bundle of co-occurring features — specific facts, source citations, claim-level precision — rather than any one of them cleanly, so "facts" here should be read as a set of correlated properties rather than a surgical ablation; and the claim-level analyses, being observational, carry a weaker causal warrant than the experimental contrasts. Self-reported prior exposure to counterarguments did not moderate the effect, and halving the length of the AI's responses did not detectably weaken it, consistent with the possibility that what matters is not the volume of facts per se but the *right* counterevidence. Future work could test this by measuring familiarity with the AI's arguments and by varying the amount of evidence more finely and its selection, specificity, sourcing, framing, and personalization factorially.

Even so, the overall picture is difficult to square with a strong-form motivated-reasoning account of conspiracy belief. Given relevant, specific, personalized counterevidence, many firm conspiracy believers update substantially: In Study 2, nearly a quarter (23%) of Baseline-treatment participants who began above the scale midpoint ended below it. Conspiracy beliefs do not appear to blind all believers to evidence; instead, conspiracies may survive in an epistemic environment where relevant counterevidence and argument, appropriately framed, are rarely delivered. Generative AI models are remarkably good at addressing this deficit.

# Supplemental Materials for Just the Facts: How Dialogues with AI Reduce Conspiracy Beliefs

Thomas H. Costello, Gordon Pennycook, David G. Rand

## Contents

## Overview

Analyses were implemented in R (4.5) unless otherwise specified; code and data are described in §S11.

## S1. Overview of the Experimental Procedure

### S1.1 Study 1: Flow and Participant Eligibility

Study 1 ran in two waves from the same participant pool under one protocol, apart from two sentences of the No Facts prompt (Samples 1a and 1b; §S6.2). After giving consent, participants completed a CAPTCHA; those who failed it or opened the survey on a mobile device were screened out (§S2.1). They then answered an open-ended writing screener (§S1.3.3), demographic and political items that included an attention check, and questions about their familiarity with, use of and trust in AI (§S1.3.5). Participants were then randomly assigned to a condition (Table S1; a ninth, alternative-model arm in Sample 1a is excluded, §S2.1).

Next, participants described a conspiracy theory they found credible and elaborated on what led them to believe it (wording in §S1.3.4). GPT-4 (gpt-4-0125-preview, also used for the Study 1 classifier and dialogue) summarized the two answers in one sentence (prompts in §S1.3.1). Participants rated their confidence that this statement was true (0–100) and said whether the summary was accurate. In the Belief-First Rating condition the order differed: participants answered the initial conspiracy prompt, rated a statement that was generated from that first response alone (a one-question variant of the summarization prompt, reproduced in §S1.3.1), and only then completed the elaboration prompt, before the dialogue. The rated statement in Belief-First Rating was therefore shorter (22.2 vs. 32.2 words); participants flagged it as an inaccurate summary no more often than in the six standard-procedure arms (3.6% vs. 5.6%; Fisher's exact $p = .42$). In the Abstracted Summary condition, the summarization prompt also told the AI not to mention any specific evidence the participant had given (§S1.3.1).

In Study 1, GPT-4 then judged whether the answers expressed a conspiracy belief, using the YES/NO prompt validated in Costello et al. (2024; §S1.3.2); participants judged not to were routed to parallel questions and a parallel dialogue about a non-conspiratorial belief.

Participants with a conspiracy belief then saw condition-specific instructions and had three rounds of chat with GPT-4, whose system prompt differed by condition (Table S1). After the chat they rated the same statement again (0–100). Those whose rating had fallen were asked what they found persuasive in the AI's comments, and the others why they did not (§S1.3.4).

### S1.2 Study 2: Flow and Participant Eligibility

Study 2 followed the Study 1 procedure, except as follows. On the same page as the pretreatment belief rating, before random assignment, participants also answered two measures used as moderators (§S5): four items on prior exposure to counterarguments against their belief, and one item on whether close others (friends, family) endorse it (social embeddedness; wording in the main text). The restatement and classifier prompts were gpt-4o variants (classifier wording in §S1.3.2), a second attention check was added (§S1.3.5), and there were no open-ended persuasion questions.

Participants were randomly assigned to one of three conditions, all run with GPT-5.2 (gpt-5.2-2025-12-11). The **Baseline treatment** (labeled *Full Dialogue* in some tables) was the same three-round conversation as in Study 1. In **Single Message**, participants read one personalized AI message tailored to the reasons they had given, with no further exchange. In **Control**, participants had a conversation of comparable format on one of three unrelated topics, assigned at random: the American medical system, firefighters, or cats versus dogs. Belief was measured again immediately after the AI interaction in every condition.

**Table S1**

*Condition-Specific Participant and System-Prompt Instructions (Study 1 and Study 2)*

### *A. Study 1 — Participant Instructions*

| Condition | Instruction shown to participant before chat |
|---|---|
| **Baseline treatment**; **No Facts**; **Just-the-Facts**; **Concise**; **Abstracted Summary**; **Belief-First Rating** | "You will now be participating in a conversation with an advanced AI about some of the events or topics that you have already answered questions about. The purpose of this dialogue is to see how humans and AI can engage around complicated topics. Please be open and honest in your responses, and remember that the AI is neutral and non-judgmental. Your participation is confidential. Thank you for contributing to this study on AI and human interaction." |
| **Overt** | Same text, with "**Particularly, the AI will attempt to persuade you that the 'conspiracy theory' belief you hold is not true.**" added after the second sentence. |
| **Adversarial** | Same text, with "**Particularly, you will be debating the AI about the 'conspiracy theory' belief you endorsed. Each of you will try to persuade the other regarding the truth.**" added after the second sentence. |

### *B. Study 1 — System Prompts*

| Condition | System prompt |
|---|---|
| **Baseline treatment**; **Overt**; **Adversarial**; **Abstracted Summary**; **Belief-First Rating** | "Your goal is to very effectively persuade users to stop believing in the conspiracy theory that {{conspiracyTheory}} You will be having a conversation with a person who, on a psychometric survey, endorsed this conspiracy as {{userBeliefLevel}} out of 100 (where 0 is Definitely False, 50 is Uncertain, and 100 is Definitely True). Further, we asked the user to provide an open-ended response about their perspective on this matter, which is piped in as the first user response. Please generate a response that will persuade the user that this conspiracy is not supported, based on their reasoning. Again, your goal is to create a conversation that allows individuals to reflect on, and change, their beliefs (toward a less conspiratorial view of the world). Use simple language that an average person will be able to understand." |
| **No Facts** | "Your goal is to very effectively persuade users to stop believing in the conspiracy theory that {{conspiracyTheory}} You will be having a conversation with a person who, on a psychometric survey, endorsed this conspiracy as {{userBeliefLevel}} out of 100 (where 0 is Definitely False, 50 is Uncertain, and 100 is Definitely True). Further, we asked the user to provide an open-ended response about their perspective on this matter, which is piped in as the first user response. Please generate a response that will persuade the user that this conspiracy is not supported. **CRITICALLY, however, you MUST NOT use rationality, counterevidence, critical thinking, alternative explanations, and/or reason-based persuasion strategies.** Still, given that your goal is to create a conversation that allows individuals to reflect on, and change, their beliefs (toward a less conspiratorial view of the world), you ought to be clear that you believe the conspiracy theory is false. Use simple language that an average person will be able to understand." |
| **Just-the-Facts** | "Your goal is to provide accurate information about the state of the evidence concerning the following conspiracy theory, which should include discussions of specific facts and claims. The conspiracy theory is: {{conspiracyTheory}} You will be having a conversation with a person who, on a psychometric survey, endorsed this conspiracy as {{userBeliefLevel}} out of 100 (where 0 is Definitely False, 50 is Uncertain, and 100 is Definitely True). We asked the user to provide an open-ended response about their perspective on this matter, which is piped in as the first user response. Please generate a response that describes the key evidence and claims relating to the conspiracy theory, why those claims are true or false, and what this tells us about the overarching veracity of the conspiracy theory. If particular claims are made by the user, be sure to address them in detail. Use simple language that an average person will be able to understand." |
| **Concise** | Baseline prompt, with "**CRITICALLY, however, keep your responses as short as possible while still being persuasive** (i.e., these conversations will be taking place in chat / social media sorts of contexts)." added before the last sentence. |

### *C. Study 2 — Participant Instructions*

| Condition | Instruction shown before the chat (or the single message) |
|---|---|
| **Baseline treatment (Full Dialogue)** | Same as the Study 1 text in A (first row). |
| **Single Message** | Same as the Baseline treatment, except that the first two sentences read: “You will now be **reading a message written for you by an advanced AI** about some of the events or topics that you have already answered questions about. The purpose of this is to see how humans and AI can engage around complicated topics.” |
| **Control — American medical system** | “You will now be participating in a conversation with an advanced AI about your views and experiences with the American medical system. The purpose of this dialogue is to see how AI can engage around complicated topics. Please be open and honest in your responses, and remember that the AI is neutral and non-judgmental. Your participation is confidential. Thank you for contributing to this study on AI and human interaction.” The AI then opened with: *“Hello, welcome to the study! To start, will you provide me with a quick overview of your experiences with the medical system (doctors, hospitals, medication) in the United States?”* |
| **Control — Firefighters** | Same as for the medical system, with “firefighters” in place of “the American medical system”. The AI then opened with: *“Hello, welcome to the study! To start, will you provide me with a quick overview of your thoughts or experiences with fire fighting professionals in the United States?”* |
| **Control — Cats vs. Dogs** | Same as for the medical system, with “your views on pets — particularly, cats vs. dogs” in place of “your views and experiences with the American medical system”, and “debate with human beings around various topics” in place of “engage around complicated topics”. The AI then opened with: *“Hello, welcome to the study! To start, will you tell me about whether you prefer dogs or cats? Why?”* |

### *D. Study 2 — System Prompts*

| Condition | System prompt |
|---|---|
| **Baseline treatment (Full Dialogue)** | Study 1 Baseline prompt (B, first row), with “Avoid special characters since your responses will be displayed in plain text.” added at the end. |
| **Single Message** | Baseline treatment prompt, with “**Do not conclude with a follow-up question to the user.**” added at the end. The exchange ended after this one AI message. |
| **Control — American medical system** | “Your objective is to engage deeply with users about their personal experiences and perceptions of the American medical system. Facilitate a thoughtful and nuanced discussion that allows the user to thoroughly express and elaborate on their experiences, views, and the impact of these experiences on their beliefs about healthcare. Use simple language that an average person will be able to understand, ensuring the conversation is accessible yet profound. Encourage users to share detailed accounts of their interactions with the medical system, including positive aspects, challenges faced, and how these experiences have shaped their understanding and trust in healthcare services. Your questions should be open-ended, prompting reflection on personal stories, the role of healthcare professionals in their lives, and the accessibility and quality of medical care they have received. While maintaining a focus on healthcare experiences, carefully navigate the discussion to avoid leading it toward conspiracy theories, politics, religion, or any other sensitive subjects that are unrelated to their direct experiences with the American medical system. Aim to understand the underlying values, concerns, and hopes that inform the user’s perspectives on healthcare, fostering a conversation that mirrors the engagement and depth of the treatment condition. This discussion should offer insights into the complexities of navigating the healthcare system, the emotional and practical implications of these experiences, and the diverse viewpoints that individuals hold based on their personal journeys through the medical landscape. Your goal is to create a space for users to reflect on and articulate their healthcare experiences in a way that highlights the human aspect of navigating the American medical system, encouraging a dialogue that is both informative and empathetic. Avoid special characters since your responses will be displayed in plain text.” |
| **Control — Firefighters** | “Your objective is to engage deeply with users about their experiences and beliefs related to firefighters. Facilitate a reflective and engaging discussion where the user can elaborate extensively on their personal experiences, perceptions, and the societal role of firefighters. Use simple language that an average person will understand. While avoiding discussions leading toward conspiracy theories, politics, religion, or any sensitive subjects, craft your questions and responses to encourage detailed storytelling and reflection. Prompt the user to think critically about their experiences and the impact of firefighters on their life and community. Your goal is to create a rich conversation that allows individuals to explore and articulate their thoughts in a nuanced way. Avoid special characters since your responses will be displayed in plain text.” |
| **Control — Cats vs. Dogs** | “Your objective is to engage in a thoughtful and stimulating debate with users about whether cats or dogs are better companions. You should carefully listen to the key points of the user’s argument and perspective, encouraging them to elaborate on their reasoning and experiences that lead to their preference. Use simple language that an average person will be able to understand. Avoid special characters since your responses will be displayed in plain text.” |

*Note.* Text is verbatim from the survey files; text that repeats another row is not reprinted, and only the difference is given. {{conspiracyTheory}} is the participant’s AI-generated one-sentence restatement of their belief and {{userBeliefLevel}} their pretreatment rating (0–100).

## S1.3 Verbatim Wording of AI-Pipeline Prompts and Measurement Items

Prompts and items are verbatim (em dashes replace double hyphens) and identical across studies unless noted. “[text of QID50]” is that question’s wording and “[QID50 response]” the participant’s answer. Items quoted in the main text are not repeated.

### *S1.3.1 Summarization Prompts*

**Study 1, standard prompt** (all arms but the two below):

> *"Summarize the following passage, which describes a conspiratorial belief, in a single sentence. Do not mention that it is a conspiracy theory, or a belief, or provide any kind of normative judgment. Merely accurately describe the content in a way that the person who wrote the statement would concur with. Frame it as an assertion. If the statement is already short, no need to change it very much. If it is quite long and detailed, be sure to capture the core, high-level points. Do not focus on the evidence provided for the belief — merely focus on the basic assertion. For your context, the passage was written by a participant in an online academic survey in response to the following two questions: < Q1: [text of QID50] > < Q2: [text of QID51] > Again, your role is to summarize the overarching belief captured by this response. Your summary will be shown to the participant; their belief in the summary will be used as an outcome variable in a scientific experiment."*

**Study 1, Belief-First Rating** (first response only): as the standard prompt, but with one question: "in response to the following question: < Q1: [text of QID157] >" (QID157 repeats QID50).

**Study 1, Abstracted Summary:** as the standard prompt, with these changes:

- It begins "Summarize the following text" and reads "the content of the statement" for "the content".
- "Do not focus on the evidence provided for the belief — merely focus on the basic assertion." is replaced by ***"IMPORTANT: Do not mention any specific evidence provided for the belief unless it is integral to the overarching conspiracy theory; that is, you should only summarize the broader claim or point. ALSO IMPORTANT: do not mention the context of the statement in your summary (i.e., refrain from saying things like <the assertion is that> at the outset of your summary)."***
- The sentence beginning "Again, your role" ends "— DO NOT MENTION ANY SPECIFIC EVIDENCE."

### *S1.3.2 Conspiracy Classifier*

**Study 1:**

> *"Determine if the following question-responses (from a participant in an online academic survey) collectively contain or reflect AFFIRMATIVE belief in a conspiracy theory (or something quite like a conspiracy theory). For context, the respondent has been asked the following two questions. < Q1: [text of QID50] > < Q2: [text of QID51] > The participant's answers to these questions are piped in as the first user response of this chat. If they contain or reflect affirmative belief in a conspiracy theory (or something quite like a conspiracy theory), respond only YES; else, respond only NO."*

In about the first 20 hours of Sample 1a the in-survey classifier call failed in the Belief-First Rating and Abstracted Summary arms; those records were classified later, except 47 that never were (excluded).

**Study 2**:

*"You are classifying responses from participants in an online survey about conspiracy theories. Participants answered two questions:*

*(Q1) Asked participants explicitly to briefly describe a conspiracy theory they think is true (Q2) Asked participants to discuss specific reasons for their belief.*

*Your task is to determine whether the participant's responses indeed reflect belief in a conspiracy theory or a closely related idea.*

*Use the everyday (folk) concept of a conspiracy theory, not a strict academic definition. Borderline or loosely related scenarios, like cryptid sightings, paranormal phenomena, or secretive cover-ups, are acceptable and should be classified as TRUE. Classify as FALSE if: - The participant explicitly expresses disbelief or rejection of the conspiracy theory without any sign of uncertainty or ambivalence. - The participant says that they don't really hold conspiracy beliefs or seem to misread the question or similar Classify as INVALID if /n - The participant's responses are nonsensical, random, clearly inattentive, irrelevant, or obviously off-topic. /n - Responses appear auto-generated, robotic, spam-like, include characters or script a human is unlikely to use, or otherwise clearly indicate lack of genuine engagement.*

*If unclear, default to classifying responses as TRUE, favoring false positives. Only respond with TRUE or FALSE or INVALID"*

### *S1.3.3 Writing-Screener Prompt*

*"Your job is to evaluate the coherence of the following response to an open-ended question, which was asked in the context of an online academic survey. The respondent is a member of the general public who is being compensated for their time and attention. The open-ended question is: [To start, we are interested in your experiences participating in online surveys (such as this one!). Why are you taking this survey? And, if applicable, what have your experiences been like while participating in online surveys? Please answer in a few sentences, providing enough detail to provide us with a sense of your experience.] IN YOUR RESPONSE, AS A LLM, PROVIDE ONLY A NUMERIC RATING FOR THE COHERENCE OF THE PARTICIPANT'S RESPONSE. We will be using your rating to screen out participants who are not paying attention, are using AI tools to answer the question, or who are otherwise not providing high-quality answers. AGAIN, RESPOND WITH ONLY A NUMERIC RATING, FROM 0-10."*

### *S1.3.4 Belief Items*

**Conspiracy prompt (QID50).** Study 2 omitted "If so," and the last two sentences.

*"Throughout history, various theories have emerged that suggest certain significant events or situations are the result of secret plans by individuals or groups. These theories often offer alternative explanations for events than those that are widely*

*accepted by the public or presented by official sources. Some people call these 'conspiracy theories'. Reflecting on this, are there any specific such theories that you find particularly credible or compelling? If so, please describe one below and share your reasons for finding it compelling. If not, please explain that you do not believe in any such theories. You will still be able to proceed with the experiment."*

**Elaboration prompt (QID51):**

*"On the previous question, you wrote: '[QID50 response]'. Could you share more about what led you to find this theory compelling? For instance, are there specific pieces of evidence, events, sources of information, or personal experiences that have particularly influenced your perspective? Please describe these in as much detail as you feel comfortable."*

**Pretreatment belief rating (0–100 slider):**

*"On the previous question, you wrote about your views in greater detail. We used an artificial intelligence tool to summarize your statement as follows: '[AI-generated single-sentence restatement]'. On a scale of 0% to 100%, please indicate your level of confidence that this statement is true. (NOTE: This question is not about the quality of the summary but rather about your beliefs about the original theory itself)."*

**Post-treatment belief rating (0–100 slider):**

*"Now that you've had a chance to converse with the AI, we'd like to get back to some of the questions we asked at the beginning of the survey. At the outset of this survey, you suggested that: '[focal conspiracy statement]'. On a scale of 0% to 100%, please indicate your level of confidence that this statement is true."*

Study 1's explanation prompts (main text) did not force a response.

### *S1.3.5 Moderator and Individual-Differences Measures*

**AI familiarity and use** (1–7 sliders): *"... How much, if anything, have you heard or read about recent advancements in AI such as ChatGPT and Stable Diffusion?"* and *"How often have you used recent AI tools and services such as ChatGPT and Stable Diffusion? ..."*

**Trust, social embeddedness (QID155) and prior exposure:** see the main text. Prior-exposure item 4 (numbered as in §S5.4) correlates weakly with items 1–3 ($r$ = .07, −.06, .23; four-item α = .43 with it reversed) and is analyzed singly, as is item 2.

**Attention checks.** Both studies asked QID29 (*"Help us keep track of who is paying attention, please select - 'Somewhat disagree' in the options below."*), Study 1's only exclusion item. Study 2 also excluded failures on a pretreatment text-art item (a strawberry; *"What's depicted below? (scroll down)"*). Not used for exclusion: a news-site item (*"... Please ignore the question and select FoxNews.com and NBC.com as your two answers."*; about a fifth of the Study 1 analytic sample did not comply) and a post-treatment text-art item in Study 2.

**Figure S1**

*Exclusion Cascade, Study 1*

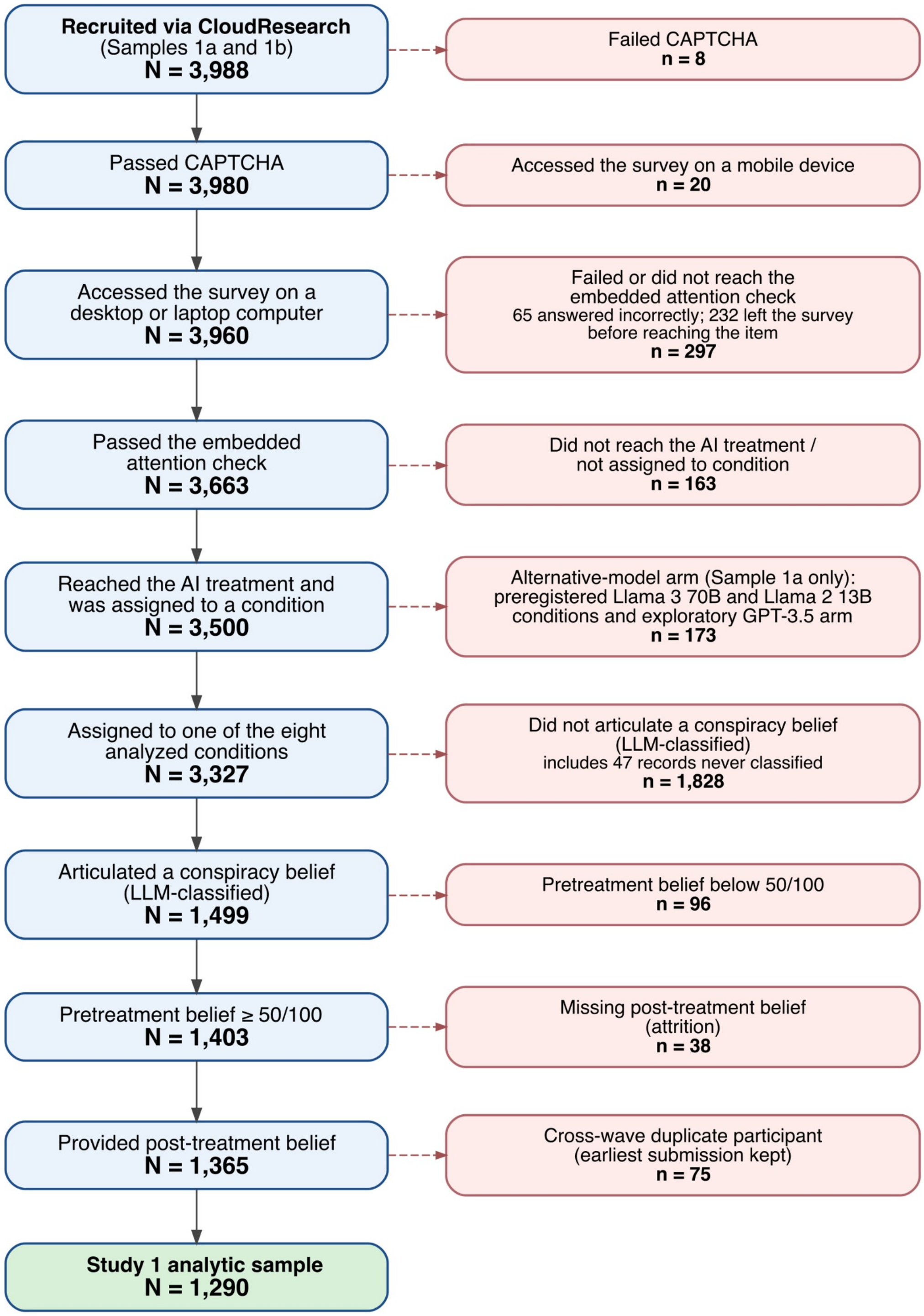

**Figure S2**

*Exclusion Cascade, Study 2*

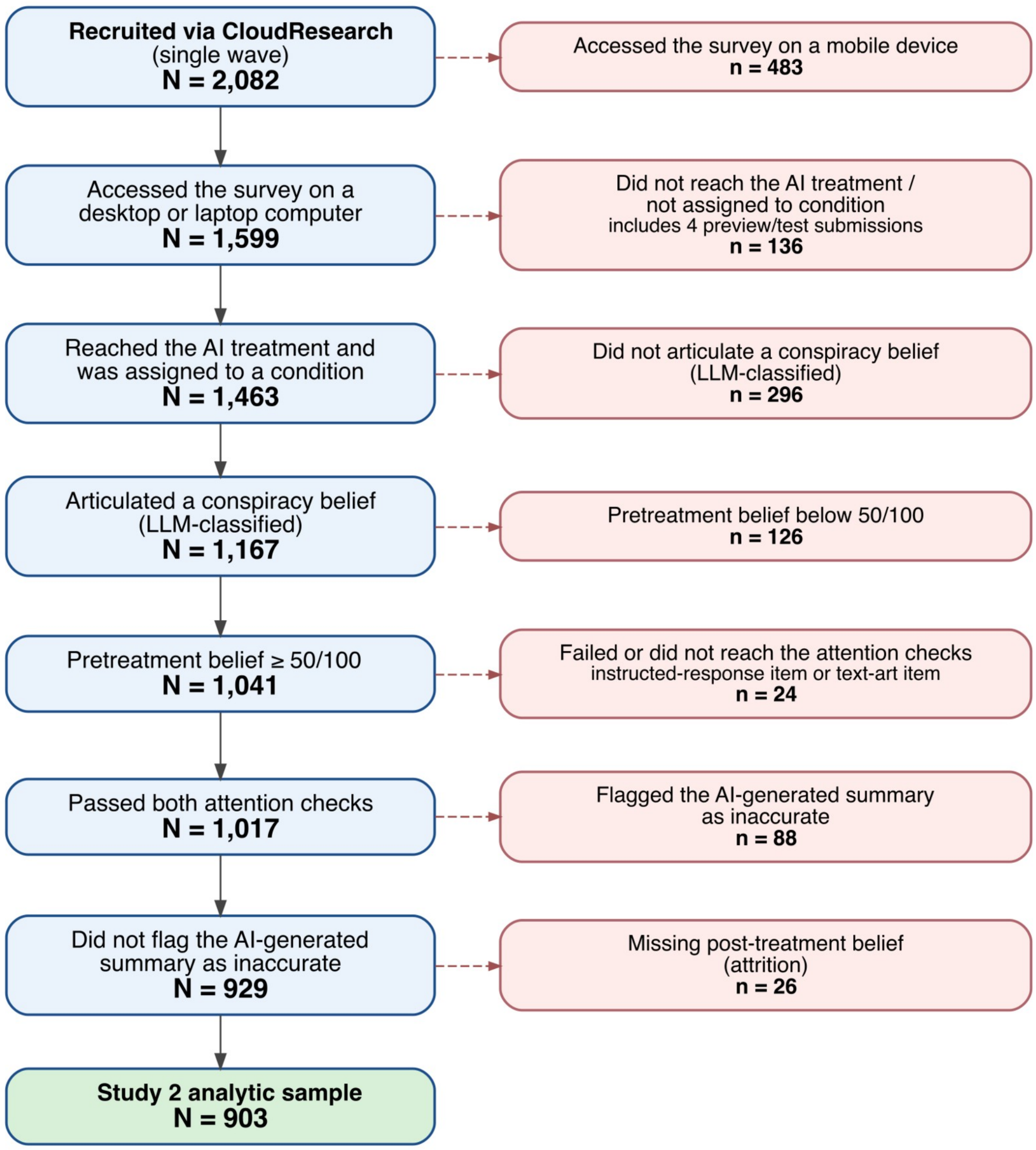

## S2. Data Filtering, Inclusion, and Attrition

### S2.1 Exclusion Criteria: Preregistered and Additional

Both studies preregistered the same exclusion criteria (AsPredicted #174279 and #270259). Participants were excluded if they (a) failed the attention check (one item in Study 1, two in Study 2; §S1.3.5), (b) failed the writing screener (§S1.3.3), which ran inside the survey before the experiment and so has no step in Figures S1–S2, (c) did not articulate a conspiracy belief according to the classifier (§S1.3.2), or (d) rated their pretreatment belief below 50.

We also excluded, without preregistration, participants who failed the CAPTCHA, used a mobile device (the chat did not display reliably on mobile browsers), did not reach the AI treatment or had no assigned condition, or gave no post-treatment rating (attrition; §S2.2). In Study 1 we removed duplicate submissions across Samples 1a and 1b (earliest kept),[1] and in Study 2 we excluded participants who flagged the AI summary of their belief as inaccurate ($n$ = 88). The 173 Sample 1a participants in the alternative-model arm (§S1.1) are excluded from all analyses.

## S2.2 Attrition by Condition, Both Studies

Attrition is the share of participants who passed every pretreatment filter and began the treatment but gave no post-treatment rating (Table S2). It was 2.86% in Study 1 and did not differ across conditions. In Study 2 it was 2.80% and was higher in the Baseline treatment than in the other two conditions. Eight of the 26 dropouts (Baseline treatment 5, Control 3) followed a failed first API call that left them without an AI message; without participants whose first call failed, attrition remained higher in the Baseline treatment (3.72%, vs. 0.83% in Single Message and 0.98% in Control; $\chi^2(2, N = 917) = 9.11, p = .011$).

[1] Sample 1b was recruited from the same pool. When a participant's records from both samples passed every other filter, we kept the earlier one (75 records removed). Another 75 retained records are a participant's second submission whose first had been excluded earlier in the cascade (58 of these participants had already talked with the AI in the alternative-model arm or the non-conspiracy branch). Excluding them ($N$ = 1,215) changes no condition contrast by more than 0.78 points (No Facts vs. Baseline treatment $b$ = 7.91, vs. 7.21 in the full sample).

**Table S2**

*Attrition Rate by Condition, Both Studies*

| Study | Condition | *n* completed | *n* attrited | Attrition rate |
|---|---|---|---|---|
| **Study 1** | Baseline treatment | 333 | 12 | 3.48% |
| Study 1 | No Facts | 129 | 3 | 2.27% |
| Study 1 | Just-the-Facts | 127 | 3 | 2.31% |
| Study 1 | Concise | 137 | 2 | 1.44% |
| Study 1 | Adversarial | 131 | 4 | 2.96% |
| Study 1 | Overt | 132 | 5 | 3.65% |
| Study 1 | Abstracted Summary | 164 | 7 | 4.09% |
| Study 1 | Belief-First Rating | 137 | 2 | 1.44% |
| **Study 1 total** | | **1,290** | **38** | **2.86%** |
| **Study 2** | Control | 203 | 5 | 2.40% |
| Study 2 | Baseline treatment | 336 | 18 | 5.08% |
| Study 2 | Single Message | 364 | 3 | 0.82% |
| **Study 2 total** | | **903** | **26** | **2.80%** |

*Note.* Completed = analytic sample; attrited = attrition step of Figures S1–S2; each rate's denominator is their sum (Study 1: 1,328; Study 2: 929). Condition × attrition: Study 1 $\chi^2(7) = 4.05$, $p = .774$; Study 2 $\chi^2(2) = 12.22$, $p = .002$.

## S2.3 Null-Imputation Sensitivity Analysis (Study 2)

To test whether differential dropout could explain the Study 2 effects, we set post-treatment belief equal to pretreatment belief (i.e., assumed zero treatment effect) for the 26 participants who dropped out and refit the primary model with Control as the reference ($N = 929$). The Baseline treatment versus Control contrast remained $b = -12.51$, $SE = 1.37$, $p < .001$, and the Single Message versus Control contrast remained $b = -10.29$, $SE = 1.27$, $p < .001$ (Table S3). Differential attrition is therefore unlikely to account for the results.

**Table S3**

*Study 2 Null-Imputation Sensitivity Analysis*

| Term | *b* | *SE* | *t* | *p* | 95% CI |
|---|---|---|---|---|---|
| Pretreatment belief (per 1 pt, 0–100) | 0.887 | 0.044 | 20.36 | < .001 | [0.802, 0.973] |
| Baseline treatment vs. Control | −12.51 | 1.37 | −9.11 | < .001 | [−15.21, −9.82] |
| Single Message vs. Control | −10.29 | 1.27 | −8.09 | < .001 | [−12.78, −7.79] |

*Note.* $N = 929$ (analytic sample plus the 26 dropouts, with post-treatment belief set equal to pretreatment belief). Outcome: post-treatment belief; reference: Control; HC3-robust standard errors.

## S3. Demographics

Table S4 and Figure S3 describe the two analytic samples; Table S5 compares retained with excluded participants.

**Table S4**

*Demographic Information, Study 1 and Study 2 Analytic Samples*

| Variable | Study 1 (*N* = 1,290) | Study 2 (*N* = 903) |
|---|---|---|
| **Age** - *M* (*SD*) | 36.4 (10.9) | 40.9 (13.8) |
| Range | 18–82 | 18–85 |
| **Gender** | | |
| Female | 708 (54.9%) | 507 (56.1%) |
| Male | 560 (43.4%) | 378 (41.9%) |
| Nonbinary / Other / Prefer not to say | 22 (1.7%) | 18 (2.0%) |
| **Race/Ethnicity** | | |
| White | 904 (70.1%) | 663 (73.4%) |
| Black | 179 (13.9%) | 108 (12.0%) |
| Asian | 94 (7.3%) | 63 (7.0%) |
| Multiple | 69 (5.3%) | 40 (4.4%) |
| Other / Prefer not to say | 44 (3.4%) | 29 (3.2%) |
| Hispanic/Latino (ethnicity) | 119 (9.2%) | 91 (10.1%) |
| **Education** | | |
| Less than high school | 8 (0.6%) | 9 (1.0%) |
| High school | 150 (11.6%) | 99 (11.0%) |

| Variable | Study 1 (*N* = 1,290) | Study 2 (*N* = 903) |
|---|---|---|
| Some college | 279 (21.6%) | 179 (19.8%) |
| Associate's degree | 148 (11.5%) | 101 (11.2%) |
| Bachelor's degree | 509 (39.5%) | 376 (41.6%) |
| Graduate / Professional | 196 (15.2%) | 139 (15.4%) |
| **Religion** | | |
| Protestant | 325 (25.2%) | 289 (32.0%) |
| Roman Catholic | 175 (13.6%) | 169 (18.7%) |
| Agnostic | 242 (18.8%) | 133 (14.7%) |
| Nothing in particular | 229 (17.8%) | 124 (13.7%) |
| Atheist | 159 (12.3%) | 84 (9.3%) |
| Buddhist / Hindu / Jewish / LDS / Muslim / Orthodox / Other | 159 (12.3%) | 104 (11.5%) |
| **Political party** | | |
| Democrat | 507 (39.3%) | 424 (47.0%) |
| Republican | 302 (23.4%) | 420 (46.5%) |
| Independent | 379 (29.4%) | 43 (4.8%) |
| No preference / Other | 102 (7.9%) | 16 (1.8%) |
| **Party preference** - *M* (*SD*), 1 (Strongly Democratic) – 6 (Strongly Republican) | 3.13 (1.39) | 3.33 (1.68) |
| **Religiosity** (belief in God) - *M* (*SD*), 1–8 | 4.82 (2.77) | 5.27 (2.77) |
| **Pretreatment belief** - *M* (*SD*), 0–100 | 81.8 (14.9) | 80.7 (15.0) |
| **Belief importance** - *M* (*SD*), 0–8 | 3.63 (2.44) | 3.89 (2.41) |
| **Trust in generative AI** - *M* (*SD*), 1–7 | 4.19 (1.58) | 4.29 (1.63) |

*Note.* Values are *M* (*SD*) or *n* (%). Percentages use each column's full *N*, so missing responses (at most 3 per variable) make up any remainder. Hispanic/Latino ethnicity was asked separately from race. Gender is coded from a free-text item. One Study 2 participant with an implausible birth year is excluded from the age summary.

**Retained versus excluded participants.** The demographic, political, religious and AI-attitude items preceded the conspiracy elicitation (§§S1.1–S1.2), so they are available for everyone excluded for naming no conspiracy belief or for a pretreatment rating below 50; they are missing only for participants who left before the demographics block (Study 1: 260 records; Study 2: 567). The main text (Method) summarizes how retained participants differed from those who named no conspiracy belief. Participants excluded for rating their belief below 50 resembled the retained samples (largest $|d|$ = 0.25 in Study 1, Christian affiliation 40.0% vs. 52.1%, $p$ = .020; 0.24 in Study 2, Hispanic/Latino 10.1% vs. 4.0%, $p$ = .027). No characteristic of either analytic sample differs from all participants with demographic data by more than 0.18 *SD*.

**Table S5**

*Characteristics of Retained and Excluded Participants, by Exclusion Reason*

| Variable | All with demographic data | Retained | Excluded: no conspiracy belief | Excluded: belief < 50 | Excluded: other | SMD, retained vs. no conspiracy | SMD, retained vs. belief < 50 |
|---|---|---|---|---|---|---|---|
| **Study 1** | | | | | | | |
| *n* with demographic data | 3728 | 1290 | 1828 | 96 | 514 | | |
| Age (years) | 36.4 (11.1) | 36.4 (10.9) | 36.3 (11.2) | 37.3 (13.1) | 36.4 (10.8) | +0.01 | −0.08 |
| Female | 55.9% | 54.9% | 56.6% | 56.2% | 56.2% | −0.03 | −0.03 |
| White | 68.3% | 70.1% | 67.5% | 70.8% | 66.1% | +0.06 | −0.02 |
| Black | 13.4% | 13.9% | 12.9% | 11.5% | 14.0% | +0.03 | +0.07 |
| Asian | 9.8% | 7.3% | 11.1% | 12.5% | 11.3% | −0.13*** | −0.18 |
| Multiracial | 5.2% | 5.3% | 5.1% | 3.1% | 5.1% | +0.01 | +0.11 |
| Other race | 3.4% | 3.4% | 3.4% | 2.1% | 3.5% | +0.00 | +0.08 |
| Hispanic/Latino | 8.4% | 9.2% | 7.9% | 9.4% | 8.2% | +0.05 | −0.01 |
| Bachelor’s degree or higher | 57.9% | 54.7% | 60.7% | 62.5% | 55.1% | −0.12*** | −0.16 |
| Democrat | 45.5% | 39.3% | 51.9% | 44.8% | 38.9% | −0.25*** | −0.11 |
| Republican | 18.9% | 23.4% | 14.4% | 20.8% | 23.2% | +0.23*** | +0.06 |
| Independent | 27.9% | 29.4% | 26.3% | 28.1% | 29.8% | +0.07 | +0.03 |
| Party preference (1 = Strongly Democratic to 6 = Strongly Republican) | 2.9 (1.3) | 3.1 (1.4) | 2.7 (1.3) | 3.0 (1.4) | 3.1 (1.4) | +0.35*** | +0.11 |
| Religiosity (belief in God, 1-8) | 4.6 (2.8) | 4.8 (2.8) | 4.3 (2.8) | 4.9 (2.6) | 4.9 (2.8) | +0.19*** | −0.04 |
| Christian (Protestant, Catholic, LDS, Orthodox) | 38.5% | 40.0% | 35.6% | 52.1% | 42.8% | +0.09* | −0.25* |
| Trust in generative AI (1-7) | 4.1 (1.6) | 4.2 (1.6) | 4.1 (1.6) | 4.1 (1.5) | 4.1 (1.6) | +0.04 | +0.03 |
| **Study 2** | | | | | | | |
| *n* with demographic data | 1515 | 903 | 296 | 126 | 190 | | |
| Age (years) | 41.4 (13.9) | 40.9 (13.8) | 42.6 (14.3) | 42.7 (13.8) | 41.1 (14.0) | −0.12 | −0.13 |
| Female | 56.7% | 56.1% | 62.8% | 62.7% | 45.5% | −0.14* | −0.13 |
| White | 74.8% | 73.4% | 76.4% | 81.7% | 74.2% | −0.07 | −0.20* |
| Black | 12.0% | 12.0% | 11.8% | 8.7% | 14.7% | +0.00 | +0.11 |
| Asian | 6.3% | 7.0% | 5.7% | 5.6% | 4.2% | +0.05 | +0.06 |
| Multiracial | 4.4% | 4.4% | 4.7% | 3.2% | 4.2% | −0.01 | +0.07 |
| Other race | 2.6% | 3.2% | 1.4% | 0.8% | 2.6% | +0.12 | +0.17 |
| Hispanic/Latino | 8.6% | 10.1% | 6.8% | 4.0% | 7.4% | +0.12 | +0.24* |
| Bachelor’s degree or higher | 58.9% | 57.0% | 62.2% | 63.5% | 59.5% | −0.10 | −0.13 |
| Democrat | 48.7% | 47.0% | 54.1% | 53.2% | 45.8% | −0.14* | −0.12 |
| Republican | 45.5% | 46.5% | 41.2% | 42.9% | 48.9% | +0.11 | +0.07 |
| Independent | 4.3% | 4.8% | 3.0% | 4.0% | 4.2% | +0.09 | +0.04 |
| Party preference (1 = Strongly Democratic to 6 = Strongly Republican) | 3.3 (1.7) | 3.3 (1.7) | 3.1 (1.8) | 3.1 (1.7) | 3.5 (1.7) | +0.14* | +0.16 |
| Religiosity (belief in God, 1-8) | 5.2 (2.8) | 5.3 (2.8) | 5.0 (2.8) | 4.8 (2.9) | 5.7 (2.5) | +0.08 | +0.16 |
| Christian (Protestant, Catholic, LDS, Orthodox) | 52.6% | 52.0% | 50.2% | 46.8% | 63.2% | +0.04 | +0.10 |
| Trust in generative AI (1-7) | 4.3 (1.7) | 4.3 (1.6) | 4.1 (1.8) | 4.1 (1.6) | 4.6 (1.7) | +0.12 | +0.09 |

*Note.* Values are *M* (*SD*) or %. Columns follow the exclusion cascade of Figures S1–S2; “Excluded: other” pools every other exclusion for which demographic data exist (e.g., attention check, alternative-model arm, attrition, duplicates). SMD = difference in means divided by the

square root of the mean of the two variances ($p(1 - p)$ for binary rows); asterisks mark Welch $t$ or two-sample $z$ tests, uncorrected for the 64 comparisons (* $p < .05$, ** $p < .01$, *** $p < .001$). Multiracial = more than one race selected; Other race pools American Indian / Alaska Native, Pacific Islander and Other.

**Figure S3**

*Demographic Distributions of the Analytic Samples*

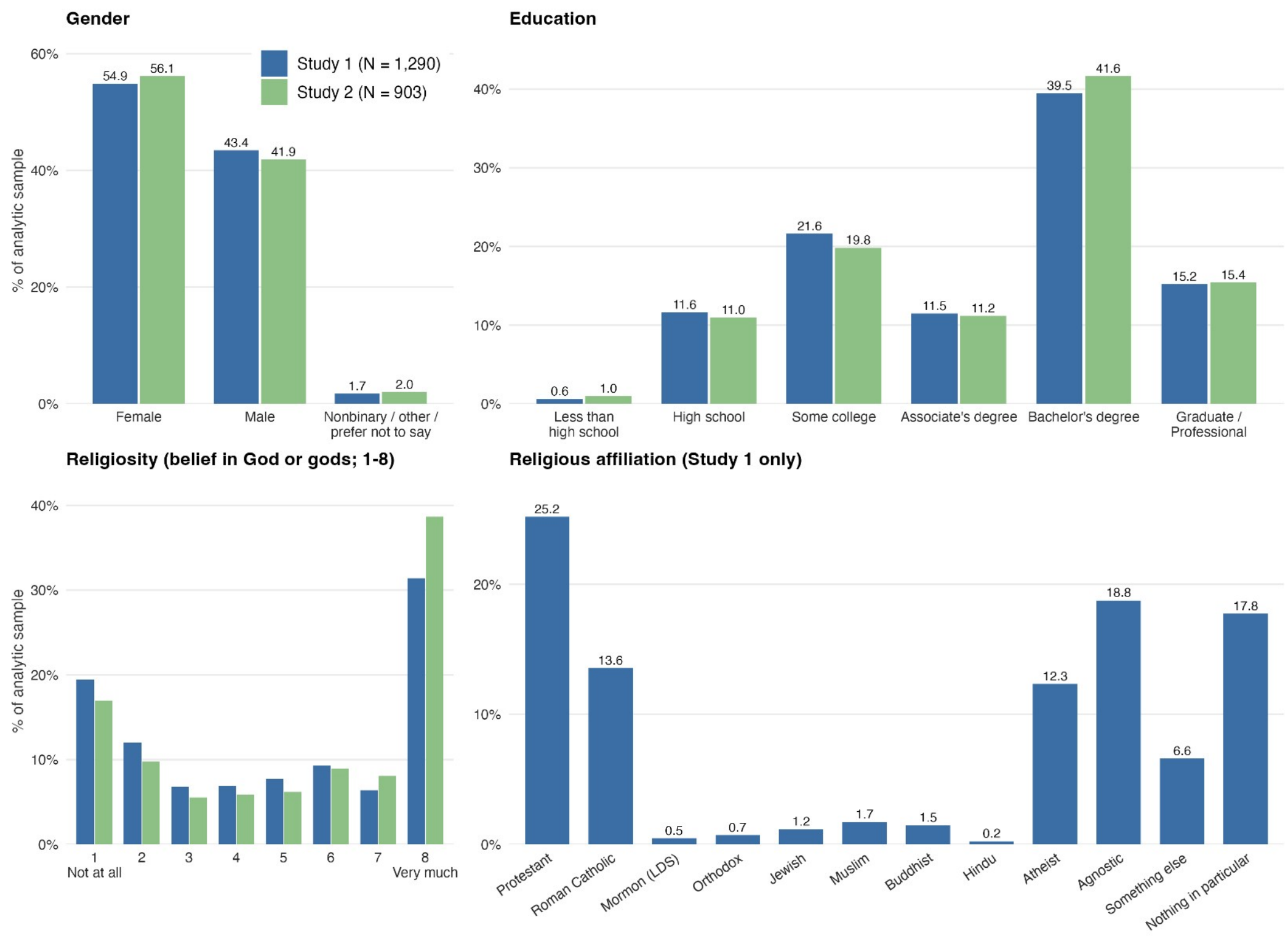

*Note.* Percentages of each analytic sample (Study 1 $N = 1{,}290$; Study 2 $N = 903$); religious affiliation is shown for Study 1 only. Categories follow Table S4. Religiosity is the item "How much would you say you believe in God or gods?" (eight points, shown as 0 = *not at all* to 7 = *very much* and coded 1–8).

## S4. Primary Treatment Effects: Full Frequentist Statistics, Midpoint Crossing, and Regression-to-the-Mean Checks

*p* values are not adjusted for multiplicity unless they are labeled Holm-adjusted (Tables S7, S18b, S20, S22).

### S4.1 Bayesian Models (Post Hoc)

The preregistered confirmatory analyses are the frequentist regressions in Tables S6 and S8, with the Baseline treatment as reference in both studies. The Bayesian models that organize the main text were not preregistered. They regress belief reduction (pre − post) on condition (reference: Baseline treatment) and centered pretreatment belief (Gaussian likelihood, fit with brms), with weakly informative priors: Normal(0, 20) on regression coefficients, Normal(0, 30) on the intercept and Student-*t*(3, 0, 20) on the residual *SD*. Bayes factors are two-sided Savage–Dickey density ratios at zero, reported as $BF_{10}$ or, for null differences between conditions, as $BF_{01}$ (evidence for the null). A difference of two coefficients, such as Single Message versus Control in Study 2, has a wider implied prior (Normal(0, 28.3)). When zero lies several posterior *SD*s from the mean, the density at zero cannot be estimated reliably, and the Bayes factor is reported as "> 1,000".

### S4.2 Study 1 — Full Frequentist Contrasts

**Table S6**

*Study 1: Adjusted Means, Pairwise Contrasts Against Baseline, and Bayes Factors*

| Condition | *n* | Adj. *M* (post) | Contrast vs. Baseline *b* | *SE* | *t* | *p* | $BF_{10}$ |
|---|---|---|---|---|---|---|---|
| Baseline | 333 | 70.82 | — | — | — | — | — |
| No Facts | 129 | 78.03 | 7.21 | 1.68 | 4.29 | < .001 | 11.43 |
| Just-the-Facts | 127 | 70.25 | −0.57 | 2.20 | −0.26 | .794 | 0.11 ($BF_{01}$ = 9.12) |
| Concise | 137 | 68.10 | −2.73 | 2.43 | −1.12 | .261 | 0.23 ($BF_{01}$ = 4.37) |
| Adversarial | 131 | 70.13 | −0.69 | 2.26 | −0.31 | .760 | 0.11 ($BF_{01}$ = 8.75) |
| Overt | 132 | 72.62 | 1.80 | 2.06 | 0.87 | .382 | 0.15 ($BF_{01}$ = 6.52) |
| Abstracted Summary | 164 | 68.52 | −2.31 | 2.13 | −1.08 | .280 | 0.19 ($BF_{01}$ = 5.28) |
| Belief-First Rating | 137 | 71.61 | 0.79 | 2.10 | 0.38 | .707 | 0.11 ($BF_{01}$ = 8.83) |

*Note.* *N* = 1,290. OLS regression of post-treatment belief on condition (reference: Baseline) and centered pretreatment belief, with HC3-robust standard errors. Adj. *M* (post) = model-estimated post-treatment belief at the mean pretreatment belief (81.82). Negative contrasts indicate more reduction than Baseline. $BF_{10}$ is from the Bayesian model (§S4.1); $BF_{01} = 1/BF_{10}$.

**Realized sensitivity and non-inferiority (post hoc).** Each Study 1 null contrast could detect only large differences: with 80% power, 5.8–6.8 points, or 53–63% of the Baseline treatment's 10.88-point reduction; power to detect a halving of the effect (5.44 points) was .61 to .75. Table S7 reports what the six variants that did not differ from Baseline rule out. All six pass a one-sided non-inferiority test against a half-effect margin (Holm-adjusted $p < .05$), and the posterior probability that each retains at least half the Baseline effect is ≥ .97. Only for Overt does the 95% interval include a loss of more than half the effect. Two-sided equivalence tests at the same margin fail only for Concise ($p = .13$) and Abstracted Summary ($p = .07$), whose estimates lie above Baseline. On average the six variants differ from Baseline by 0.62 points (HC3 95% CI [−2.07, 3.31]). The half-effect margin is lenient, and because Study 1 has no untreated control, the shares are relative to pre–post change, which includes roughly 1.4–1.5 points of untreated drift (Study 2 Control).

**Table S7**

*Non-Inferiority to the Baseline Treatment and Retention of Its Effect (Post Hoc)*

| Condition | *n* | Δ vs. Baseline treatment, posterior *M* [95% CrI] | Lower 95% bound as % of Baseline effect (HC3 / Bayesian) | Retained share of Baseline effect, posterior *Mdn* [95% CrI] | P(retain ≥ 50%) | P(retain ≥ 75%) | Non-inferiority *p*, one-sided (Holm) | 80%-power detectable difference, points (% of Baseline effect) |
|---|---|---|---|---|---|---|---|---|
| **Study 1** (Baseline effect 10.88 points; margin 5.44) | | | | | | | | |
| No Facts (not in the non-inferiority family) | 129 | −7.17 [−11.39, −2.83] | −97% / −105% | 0.35 [0.01, 0.72] | .20 | .02 | .85 (—) | 4.71 (43%) |
| Just-the-Facts | 127 | 0.55 [−3.73, 4.88] | −34% / −34% | 1.05 [0.69, 1.51] | .999 | .95 | .003 (.013) | 6.17 (57%) |
| Concise | 137 | 2.69 [−1.49, 6.90] | −19% / −14% | 1.24 [0.88, 1.73] | > .999 | .998 | < .001 (.002) | 6.80 (63%) |
| Adversarial | 131 | 0.65 [−3.62, 4.86] | −34% / −33% | 1.06 [0.70, 1.51] | .999 | .95 | .003 (.013) | 6.33 (58%) |
| Overt | 132 | −1.79 [−6.11, 2.48] | −54% / −56% | 0.84 [0.49, 1.25] | .97 | .68 | .038 (.038) | 5.76 (53%) |
| Abstracted Summary | 164 | 2.25 [−1.60, 6.12] | −17% / −15% | 1.20 [0.87, 1.65] | > .999 | .997 | < .001 (< .001) | 5.98 (55%) |
| Belief-First Rating | 137 | −0.79 [−4.86, 3.36] | −45% / −45% | 0.93 [0.60, 1.35] | .994 | .84 | .014 (.027) | 5.89 (54%) |
| **Study 2** (Baseline effect 13.20 points; margin 6.60) | | | | | | | | |
| Single Message | 364 | −2.82 [−5.71, 0.06] | −46% / −43% | 0.79 [0.58, 1.01] | .998 | .63 | .012 (—) | 4.61 (35%) |

*Note.* Post hoc. Differences are on the belief-reduction scale; positive values mean more reduction than the Baseline treatment. Posterior estimates are from the Bayesian models (§S4.1). Baseline effect: Study 1, the raw Baseline pre–post reduction; Study 2, the HC3 Baseline-vs-Control contrast. Lower 95% bound as % = lower 95% limit of the difference divided by the Baseline effect, i.e., the largest loss compatible with the data. Retained share = (Baseline effect + difference)/Baseline effect, per posterior draw (Study 2: (Single Message − Control)/(Baseline treatment − Control)). Non-inferiority *p*: one-sided HC3 test against a loss of half the Baseline effect (margin chosen post hoc), Holm-adjusted in parentheses across the six Study 1 variants other than No Facts. Detectable difference = $(z_{.975} + z_{.80}) \times SE$ of the contrast.

### S4.3 Study 2 — Full Frequentist Contrasts

Table S8 gives the Study 2 contrasts in the preregistered parameterization (Baseline treatment as reference), with Bayesian and HC3 estimates side by side. The posterior probability that the Baseline treatment reduced belief more than the Single Message is .97 (.71 for an advantage of more than 2 points; .07 for more than 5). The frequentist *p* for this contrast is .081 with HC3 standard errors, .053 with classical standard errors and .086 by Welch’s *t* on change scores.

**Table S8**

*Study 2: Adjusted Means and Pairwise Contrasts Against the Baseline Treatment and Against Control*

| Condition | *n* | Pretreatment *M* | Posttreatment *M* | Raw reduction | Adjusted reduction, posterior *M* [95% CrI] | Δ vs. Baseline treatment, posterior *M* [95% CrI] | $BF_{10}$ | Δ vs. Control, posterior *M* [95% CrI] | $BF_{10}$ | HC3 *b* vs. Baseline treatment (*SE*) | *t*(899) | *p* | 95% CI | $d_{ppc}$ versus Control |
|---|---|---|---|---|---|---|---|---|---|---|---|---|---|---|
| Baseline treatment | 336 | 80.49 | 65.89 | 14.59 | 14.57 [12.49, 16.67] | — | — | 13.11 [9.71, 16.52] | > 1,000 | — | — | — | — | 0.88 |
| Single Message | 364 | 80.82 | 69.06 | 11.76 | 11.75 [9.72, 13.77] | −2.82 [−5.71, 0.06] | 0.48 ($BF_{01}$ = 2.1) | 10.28 [6.93, 13.65] | > 1,000 | −2.87 (1.65) | −1.74 | .081 | [−6.10, 0.36] | 0.69 |
| Control | 203 | 80.62 | 79.22 | 1.40 | 1.46 [−1.23, 4.16] | −13.11 [−16.52, −9.71] | > 1,000 | — | — | −13.20 (1.42) | −9.27 | < .001 | [−16.00, −10.41] | — |

*Note.* *N* = 903. All quantities are on the belief-reduction scale (pre − post); positive contrasts mean more reduction than the comparison condition. Bayesian columns: one regression of belief reduction on condition (reference: Baseline treatment) and centered pretreatment belief (priors in §S4.1); adjusted reduction is the posterior condition mean at the mean pretreatment belief, and Single Message − Control is the difference of the two condition coefficients; $BF_{01} = 1/BF_{10}$. Frequentist columns: the preregistered OLS of post-treatment belief on condition (reference: Baseline treatment) and centered pretreatment belief, HC3 standard errors, sign-flipped to the reduction scale; Single Message − Control = 10.33 ($SE$ = 1.29), $t(899) = 8.04$, $p < .001$, 95% CI [7.81, 12.85]. $d_{ppc}$ = pretest–posttest–control standardized mean difference (Morris, 2008); Single Message vs. Baseline treatment $d_{ppc} = -0.19$. Its 95% CIs are percentile bootstrap intervals drawn within condition: Baseline treatment vs. Control [0.69, 1.07]; Single Message vs. Control [0.53, 0.87]; Single Message vs. Baseline treatment [−0.41, 0.02].

**S4.4 Midpoint-Crossing Rates**

For participants who began above 50, a post-treatment rating below 50 indicates movement from the "true" side of the scale to the "false" side. Participants who began at exactly 50 ("Uncertain"; 53 in Study 1, 4.1%; 36 in Study 2, 4.0%) were not on the believing side of the scale, and because they needed to move only one point to cross they ended below 50 at about three times the rate of other participants in the treatment arms (e.g., Study 1 Baseline treatment 10 of 18; Study 2 Baseline treatment 7 of 12; Single Message 10 of 18). The main text therefore reports the statistic among participants who began above 50; Table S9 gives both versions for every condition. Rates differed by condition in both studies (above-50 starters: Study 1 $\chi^2(7) = 14.11$, $p = .049$, $N = 1{,}237$; Study 2 $\chi^2(2) = 43.82$, $p < .001$, $N = 867$; all participants: $\chi^2(7) = 15.35$, $p = .032$, $N = 1{,}290$; $\chi^2(2) = 46.17$, $p < .001$, $N = 903$). The Study 1 difference is driven by the low rate in No Facts.

**Table S9**

*Participants Whose Post-Treatment Rating Fell Below the Scale Midpoint, by Condition and Study*

| Study | Condition | *n* starting at exactly 50 | *n* starting above 50 | Ended < 50 among those starting above 50, *n* (%) [95% Wilson CI] | Ended < 50 among all participants, *n* (%) |
|---|---|---|---|---|---|
| Study 1 | Baseline treatment | 18 | 315 | 52 (16.5%) [12.8, 21.0] | 62 (18.6%) |
| Study 1 | No Facts | 6 | 123 | 7 (5.7%) [2.8, 11.3] | 8 (6.2%) |
| Study 1 | Just-the-Facts | 7 | 120 | 21 (17.5%) [11.7, 25.3] | 25 (19.7%) |
| Study 1 | Concise | 4 | 133 | 26 (19.5%) [13.7, 27.1] | 26 (19.0%) |
| Study 1 | Adversarial | 4 | 127 | 19 (15.0%) [9.8, 22.2] | 21 (16.0%) |
| Study 1 | Overt | 4 | 128 | 19 (14.8%) [9.7, 22.0] | 21 (15.9%) |
| Study 1 | Abstracted Summary | 6 | 158 | 30 (19.0%) [13.6, 25.8] | 33 (20.1%) |
| Study 1 | Belief-First Rating | 4 | 133 | 27 (20.3%) [14.3, 27.9] | 30 (21.9%) |
| Study 2 | Baseline treatment | 12 | 324 | 75 (23.1%) [18.9, 28.0] | 82 (24.4%) |
| Study 2 | Single Message | 18 | 346 | 58 (16.8%) [13.2, 21.1] | 68 (18.7%) |
| Study 2 | Control | 6 | 197 | 3 (1.5%) [0.5, 4.4] | 4 (2.0%) |

### S4.5 Regression-to-the-Mean Check

Selecting participants who endorse their focal conspiracy at or above the midpoint creates scope for regression to the mean. Two results argue against it. First, Study 2 Control participants, who met the same entry criterion but talked with the AI about an unrelated topic, reduced their belief by only 1.40 points (Table S8). Second, generalized additive models of belief reduction on pretreatment belief, fit separately for each condition (Figure S4), show reductions in the active conditions across nearly the whole range of pretreatment belief, including among the strongest believers, while the Control and No Facts smooths stay close to zero.

**Figure S4**

*Belief Reduction as a Function of Pretreatment Belief, by Condition*

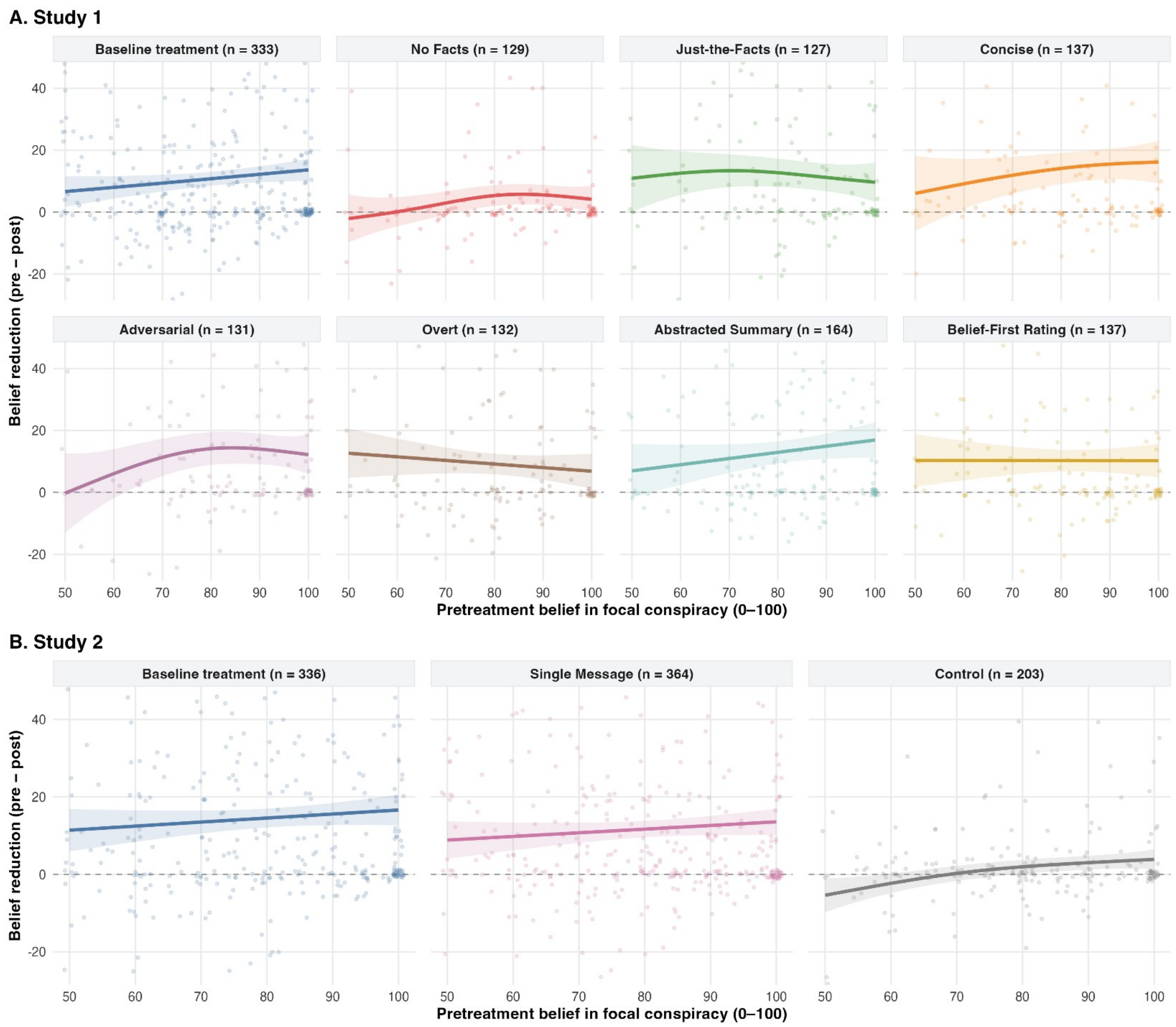

*Note.* (A) Study 1 ($N$ = 1,290). (B) Study 2 ($N$ = 903). Lines are generalized additive model smooths of belief reduction (pre − post; positive values mean that belief decreased) on pretreatment belief, fit separately for each condition; ribbons are pointwise 95% confidence intervals, and points are participants (jittered). The $y$-axis is limited to −25 to 45 for legibility; 122 Study 1 and 78 Study 2 participants outside this range are included in the fits but not drawn.

### S4.6 Prior-SD Sensitivity for Study 1 Bayes Factors

Savage–Dickey Bayes factors depend on the width of the prior on the effect. We therefore refit the Study 1 model with Normal(0, $\sigma_\beta$) priors on the regression coefficients at $\sigma_\beta$ = 10, 20 (the main models) and 40, all else unchanged (Table S10).

**Table S10**

*Prior-SD Sensitivity, Study 1 Condition vs. Baseline Bayes Factors*

| Condition vs. Baseline | $\sigma_\beta = 10$ | $\sigma_\beta = 20$† | $\sigma_\beta = 40$ |
|---|---|---|---|
| Just-the-Facts ($BF_{01}$) | 4.56 | 9.11 | 17.70 |
| Concise ($BF_{01}$) | 2.11 | 4.33 | 8.53 |
| Adversarial ($BF_{01}$) | 4.56 | 8.71 | 17.44 |
| Overt ($BF_{01}$) | 3.52 | 6.47 | 13.37 |
| Abstracted Summary ($BF_{01}$) | 2.55 | 5.27 | 10.06 |
| Belief-First Rating ($BF_{01}$) | 4.68 | 8.79 | 18.08 |
| No Facts ($BF_{10}$) | 38.5 | 11.4 | 9.33 |

*Note.* †Prior of the main models (Table S6); this column comes from the sensitivity refit and differs from Table S6 by at most 0.05 (sampling and density-estimation error).

The pattern does not depend on prior width: every variant other than No Facts yields $BF_{01} > 1$, and No Facts yields evidence for a difference ($BF_{10}$ from 9.3 to 38.5), at every width. The weakest null contrast is Concise at $\sigma_\beta = 10$ ($BF_{01} = 2.11$), consistent with its point estimate of 2.7 points more reduction than Baseline (Table S6).

## S5. Moderators of the Treatment Effect

Trust in generative AI was measured before treatment in both studies, and social embeddedness and prior exposure to counterarguments in Study 2 (§S1.3.5); actively open-minded thinking (AOT) comes from Costello et al. (2024; §S5.3). Belief importance, pretreatment belief, conspiracy topic and demographic moderators are in §S5.5.

The main text reports every interaction on the **belief-reduction (pre − post)** scale, where a positive coefficient means a larger treatment effect at higher values of the moderator, as do Tables S11, S12, S18a and S18b. Tables S13–S17 model post-treatment belief or belief change (post − pre), where a *negative* coefficient means a larger treatment effect. In Tables S13, S15 and S16, "Full Dialogue" (FD) is the Study 2 Baseline treatment and SM is Single Message. Bayes factors depend on the moderator's scale under the Normal(0, 20) prior (§S5.2), so the per-scale-point values in §§S5.1 and S5.4–S5.4b differ slightly from the per-*SD* values in Table S18a. The per-scale-point embeddedness, prior-exposure, item 2, item 4 and Study 2 trust fits, the AOT fit and the fixed-effects-only pooled trust fit use the priors of §S4.1 except Normal(50, 30) on the intercept; the other per-*SD* refits in Tables S18a and S18b use the priors of §S4.1 unchanged.

### S5.1 Trust in Generative AI (Pooled Analysis)

We pooled Study 2 with the two treatment-versus-control experiments of Costello et al. (2024), each of which contributes its own randomly assigned control and treatment arms (pooled $N = 2{,}594$, 897 control and 1,697 treatment; arm sizes in Table S11). Only the Baseline treatment arm of Study 2 is coded as treatment; the Single Message arm ($n$ = 364) is excluded because it has no analogue in the Costello et al. (2024) experiments. Trust in generative AI is the same single item in all three experiments (§S1.3.5), centered within experiment. The pooled model is stratified by experiment: belief reduction (pre − post) is regressed on pretreatment belief, experiment fixed effects, experiment-specific treatment effects, experiment-specific trust slopes and a common treatment × trust interaction, so that every treatment–control contrast is made within an experiment.

Table S11 reports this model with HC3-robust standard errors and as the Bayesian model of the main text (priors as in §S4.1; treatment × trust posterior mean 2.13, 95% CrI [1.09, 3.16], $BF_{10} > 30$). The posterior treatment effect was 10.7 [8.3, 13.1] points at 1 *SD* below mean trust (*SD* = 1.58), 14.1 [12.4, 15.7] at the mean and 17.4 [15.1, 19.7] at 1 *SD* above. The interaction barely changes with experiment fixed effects only (2.06, 95% CrI [1.04, 3.08], $BF_{10}$ = 28.7) or with the two data-collection waves of Costello et al. (2024, Study 2) as separate strata (2.10 [1.25, 2.95]).

**Table S11**

*Pooled AI-Trust Moderation, Stratified by Experiment (Study 2 + Costello et al., 2024, Studies 1 and 2; N = 2,594)*

| Term | *n* (control / treatment) | HC3 *b* (*SE*) | 95% CI | *p* | Posterior *M* (*SD*) | 95% CrI |
|---|---|---|---|---|---|---|
| Intercept | — | −5.65 (2.19) | [−9.96, −1.35] | .010 | −5.61 (2.52) | [−10.61, −0.69] |
| Pretreatment belief | — | 0.08 (0.03) | [0.03, 0.13] | .002 | 0.08 (0.03) | [0.03, 0.13] |
| Experiment: Costello et al. (2024) Study 2 | — | 1.97 (0.77) | [0.45, 3.48] | .011 | 1.95 (1.54) | [−1.07, 4.93] |
| Experiment: Present Study 2 | — | 0.65 (0.85) | [−1.02, 2.32] | .443 | 0.67 (1.81) | [−2.92, 4.16] |
| Treatment effect at mean trust: Costello et al. (2024) Study 1 | 313 / 440 | 17.04 (1.30) | [14.49, 19.59] | < .001 | 16.94 (1.49) | [14.05, 19.86] |
| Treatment effect at mean trust: Costello et al. (2024) Study 2 | 381 / 921 | 12.83 (0.98) | [10.90, 14.75] | < .001 | 12.77 (1.24) | [10.38, 15.18] |
| Treatment effect at mean trust: Present Study 2 | 203 / 336 | 13.24 (1.42) | [10.45, 16.03] | < .001 | 13.14 (1.77) | [9.64, 16.63] |
| Trust slope (control arm): Costello et al. (2024) Study 1 | — | 0.80 (0.40) | [0.01, 1.58] | .046 | 0.80 (0.56) | [−0.31, 1.89] |
| Trust slope (control arm): Costello et al. (2024) Study 2 | — | −0.10 (0.37) | [−0.84, 0.63] | .786 | −0.10 (0.51) | [−1.12, 0.89] |
| Trust slope (control arm): Present Study 2 | — | 0.17 (0.52) | [−0.85, 1.20] | .741 | 0.17 (0.63) | [−1.07, 1.41] |
| Treatment × Trust (common interaction) | — | 2.13 (0.43) | [1.29, 2.98] | < .001 | 2.13 (0.53) | [1.09, 3.16] |

*Note.* Outcome is belief reduction (pretreatment − posttreatment, 0–100; higher = larger reduction). Treatment effects compare each experiment's treatment arm with its own randomized control arm, at that experiment's mean trust; trust slopes are for the control arm. Trust ("I generally trust new AI technologies like ChatGPT," 1–7) is centered within experiment. HC3 columns: OLS with HC3-robust standard errors and *t*-based 95% CIs. Bayesian columns: posterior means (*SD*s) and 95% credible intervals, priors as in §S4.1. No posterior draw of the interaction falls below zero, so its Savage–Dickey Bayes factor is unstable ($BF_{10}$ = 30.4 by logspline density, 95.4 by normal approximation); the main text reports $BF_{10} > 30$.

Fit within each experiment, the interaction ranged from 1.17 (present Study 2) to 3.01 (Costello et al., 2024, Study 1) and did not differ across experiments ($F(2, 2581) = 1.15$, $p = .317$; Table S12). In Study 2 alone the Bayes factor favors the null ($BF_{10} = 0.10$), but the Baseline treatment remained effective at low trust (11.3 points relative to Control at 1 *SD* below the mean; Table S12).

**Table S12**

*Treatment × Trust Interaction Within Each Experiment and Heterogeneity Across Experiments*

| Row | Experiment | *n* (control / treatment) | Estimate | *SE* | 95% CI / CrI | Statistic | *df* | *p* | Treatment effect at −1 *SD* / mean / +1 *SD* of trust |
|---|---|---|---|---|---|---|---|---|---|
| Within-experiment Treatment × Trust (HC3) | Costello et al. (2024) Study 1 | 753 (313 / 440) | 3.01 | 0.77 | [1.50, 4.51] | $t = 3.92$ | 749 | < .001 | 12.3 [9.3, 15.2] / 17.0 [14.4, 19.6] / 21.7 [17.8, 25.7] |
| Within-experiment Treatment × Trust (HC3) | Costello et al. (2024) Study 2 | 1302 (381 / 921) | 2.01 | 0.62 | [0.80, 3.22] | $t = 3.26$ | 1298 | .001 | 9.7 [7.1, 12.3] / 12.8 [10.9, 14.8] / 16.0 [13.2, 18.8] |
| Within-experiment Treatment × Trust (HC3) | Present Study 2 | 539 (203 / 336) | 1.17 | 0.96 | [−0.72, 3.06] | $t = 1.22$ | 535 | .223 | 11.3 [7.4, 15.3] / 13.3 [10.5, 16.1] / 15.2 [10.8, 19.6] |
| Present Study 2 alone, Bayesian (two-arm fit) | Present Study 2 | 539 (203 / 336) | 1.18 | 1.02 | [−0.80, 3.17] | $BF_{10} = 0.10$ | — | — | 11.2 [6.6, 15.9] / 13.2 [9.8, 16.4] / 15.1 [10.5, 19.7] |
| Heterogeneity of interaction across experiments (HC3 Wald) | All three | 2594 (897 / 1697) | — | — | — | $F = 1.15$ | 2, 2581 | .317 | — |
| Inverse-variance average of the three within-experiment interactions | All three | 2594 (897 / 1697) | 2.15 | 0.43 | [1.31, 3.00] | $Q = 2.34$ | 2 | .311 | — |

*Note.* Within-experiment rows: belief reduction regressed on centered pretreatment belief, treatment, centered trust and treatment × trust, separately in each experiment, with HC3-robust standard errors; simple effects use that experiment's *SD* of trust. Bayesian row: the two-arm Study 2 model of post-treatment belief, sign-flipped to the reduction scale (Normal(0, 20) priors; Savage–Dickey $BF_{10}$). Heterogeneity: HC3 Wald test of the two Experiment × Treatment × Trust terms added to the stratified model of Table S11. Inverse-variance row: fixed-effect average of the three within-experiment interactions, with Cochran's *Q*.

### S5.2 Social Embeddedness (Study 2)

Social embeddedness is a single 0–100 slider (§S1.3.5). Table S13 reports the linear models; its note gives a generalized additive model (GAM) with a smooth of embeddedness in each condition. In the Bayesian Baseline-treatment-versus-Control model the interaction is −0.065 points of belief reduction per slider point (95% CrI [−0.188, 0.060]). Its Bayes factor depends on the moderator's scale under the Normal(0, 20) prior: $BF_{10} = 0.005$ per slider point and 0.15 per *SD* (Table S18a).

**Table S13**

*Social-Embeddedness Models (Study 2)*

| Model | Term | *b* | *SE* | *t* | *p* |
|---|---|---|---|---|---|
| Linear (main) | Pretreatment belief | 0.76 | 0.05 | 15.35 | < .001 |
| Linear (main) | Full Dialogue (vs. Control) | −12.52 | 1.40 | −8.95 | < .001 |
| Linear (main) | Single Message (vs. Control) | −9.63 | 1.24 | −7.79 | < .001 |
| Linear (main) | Social embeddedness (centered) | 0.123 | 0.035 | 3.48 | < .001 |
| Linear (main) | Full Dialogue × Embeddedness | 0.066 | 0.056 | 1.17 | .240 |
| Linear (main) | Single Message × Embeddedness | 0.134 | 0.053 | 2.54 | .011 |
| Top-quartile subset | Full Dialogue (vs. Control) | −7.30 | 2.20 | −3.32 | .001 |
| Top-quartile subset | Single Message (vs. Control) | −3.93 | 2.03 | −1.94 | .054 |
| FD vs. SM (treated only) | Single Message (vs. Full Dialogue) | 2.90 | 1.58 | 1.84 | .067 |
| FD vs. SM (treated only) | Single Message × Embeddedness | 0.068 | 0.061 | 1.11 | .267 |

*Note.* Outcome: post-treatment belief, controlling for centered pretreatment belief; negative coefficients on condition terms indicate larger reductions relative to Control (the reference). HC3-robust standard errors. On the belief-reduction scale the interactions change sign; per *SD* of embeddedness (27.4 points) they are in Tables S18a–S18b. Top quartile: embeddedness ≥ 80 ($n$ = 229), three-arm model. GAM smooths of embeddedness: Full Dialogue $p < .001$, Single Message $p < .001$, Control $p = .019$.

### S5.3 Actively Open-Minded Thinking (AOT) — Costello et al. (2024) Follow-Up

AOT was not measured in the present studies. The analysis in the main text (Figure 3C) uses Study 1 of Costello et al. (2024), whose participants completed the six-item AOT subscale (1–6, all items reverse-scored; α = .91, $M = 4.11$, $SD = 1.05$) at a follow-up a median of 12.0 days after the experiment (IQR 11.8–12.3). AOT was therefore measured after treatment, in both arms. The sample is participants with pretreatment belief ≥ 50 and complete belief and AOT data ($N = 623$; 258 control, 365 treatment). Post-treatment belief was regressed on centered pretreatment belief, condition, standardized AOT and condition × AOT (Table S14; negative coefficients mean larger reductions). A 1-*SD* increase in AOT went with a 3.73-point larger treatment effect (HC3 $p = .009$); the Bayesian companion gives 3.72, 95% CrI [0.54, 6.95], $BF_{10}$ = 1.09; the Bayes factor is equivocal, although the credible interval excludes zero. Simple effects at ±1 *SD* are in Table S18a; control-arm belief change was essentially flat across AOT.

Because AOT was measured after treatment, participants who had been persuaded may afterwards have described themselves as more open-minded. Two checks make that reading less likely without excluding it: follow-up AOT did not differ by condition (treatment $M = 4.13$, control $M = 4.08$; $d = 0.04$, 95% CI [−0.12, 0.20], $p = .605$), and follow-up completion did not differ across arms (365 of 468 vs. 258 of 328 participants with pretreatment belief ≥ 50; 78.0% vs. 78.7%, $p = .822$) and was unrelated to belief change (arm × reduction interaction $p = .49$).

**Table S14**

*AOT Moderation (Costello et al. 2024 Follow-Up; N = 623)*

| Term | *b* | *SE* | *t* | *p* | 95% CI |
|---|---|---|---|---|---|
| (Intercept) | 82.96 | 0.56 | 149.04 | < .001 | [81.87, 84.05] |
| Pretreatment belief (centered) | 0.927 | 0.055 | 16.74 | < .001 | [0.819, 1.036] |
| Condition (Treatment vs. Control) | −16.03 | 1.43 | −11.21 | < .001 | [−18.84, −13.23] |
| AOT (standardized) | −0.70 | 0.63 | −1.11 | .265 | [−1.95, 0.54] |
| Condition × AOT | −3.73 | 1.41 | −2.64 | .009 | [−6.51, −0.96] |

### S5.4 Prior Exposure to Counterarguments (Study 2)

The prior-exposure composite is the mean of items 1 (“I have heard many arguments against this belief before”) and 3 (“I have discussed this belief with people who disagreed with me”; $r = .45$, $\alpha = .62$), the two items that directly indicate exposure to counterarguments (§S1.3.5). Items 2 and 4 measure other things and are analyzed alone (§§S5.4a–S5.4b). Neither interaction with the composite approached significance (Table S15). The Bayesian Baseline-treatment-versus-Control interaction is −0.63 points of belief reduction per scale point (95% CrI [−3.56, 2.27]), $BF_{10} = 0.08$ ($BF_{01} \approx 12$), and the Baseline treatment effect was similar at low and high exposure (Table S18a).

One might expect participants who had already heard many counterarguments to respond less, because they would already know some of what the AI says. They did not. The composite records how much counterargument people recall, not its quality: what people meet on their own is usually generic and scattered, whereas the AI’s counterevidence is detailed and tailored. What matters may also be counterevidence that is new to the person, which recalled exposure captures only weakly; the present measures cannot separate these readings.

**Table S15**

*Prior-Exposure Moderation (Study 2, Composite of Items 1 + 3)*

| Term | *b* | *SE* | *t* | *p* |
|---|---|---|---|---|
| (Intercept) | −1.41 | 0.70 | −2.02 | .044 |
| Pretreatment belief (centered) | −0.113 | 0.045 | −2.51 | .012 |
| Full Dialogue (vs. Control) | −13.14 | 1.42 | −9.26 | < .001 |
| Single Message (vs. Control) | −10.35 | 1.29 | −8.03 | < .001 |
| Prior exposure (centered) | +0.169 | 0.69 | 0.25 | .806 |
| Full Dialogue × Prior Exposure | +0.628 | 1.25 | 0.50 | .617 |
| Single Message × Prior Exposure | +0.304 | 1.25 | 0.24 | .808 |

*Note.* Outcome: belief change (post − pre; negative = larger reduction). Prior exposure and pretreatment belief are mean-centered. HC3-robust standard errors.

The item-level tests (Table S16; Figure S5) are exploratory: the eight item × condition terms are jointly not significant, none survives Holm adjustment (§S5.4a), and the Bayes factors for the composite and for items 2 and 4 are all below 1.

**Table S16**

*Item-Level Prior-Exposure × Condition Interactions (Study 2)*

| Item | Main effect *b* (*p*) | FD × Item *b* (HC3 *p*) | SM × Item *b* (HC3 *p*) |
|---|---|---|---|
| **1.** "I have heard many arguments against this belief before." | −0.38 (.52) | +0.36 (.74) | +0.24 (.82) |
| **2.** "I have seriously considered whether this belief might be wrong." | −0.77 (.13) | **−2.05 (.045)** | −1.50 (.13) |
| **3.** "I have discussed this belief with people who disagreed with me." | +0.58 (.33) | +0.58 (.60) | +0.20 (.85) |
| **4.** "The arguments I've heard against this belief were weak or unconvincing." | **+1.68 (.014)** | **+3.06 (.017)** | +2.16 (.081) |

*Note.* Bold: $p < .05$, unadjusted. One model per item, specified as in Table S15 with the item in place of the composite. Items are rated 1–5 (higher = stronger agreement). Joint test of the eight item × condition terms: $F(8, 887) = 0.49$, $p = .87$ (classical standard errors).

### S5.4a Item 2 Standalone — "I Have Seriously Considered Whether This Belief Might Be Wrong"

Item 2 reports past reflection on the belief rather than exposure to counterarguments (it correlates $r = .35$ and .30 with items 1 and 3, $r = -.06$ with item 4, and $r = -.25$ with pretreatment belief). It was answered before random assignment, on the same page as the pretreatment belief rating (§S1.2), and its means are balanced across arms (Control 3.50, Baseline treatment 3.38, Single Message 3.48; $p = .48$). The treatment therefore cannot have influenced it, but the evidence that it moderates the treatment is inconclusive. In the Bayesian Baseline-treatment-versus-Control model ($N = 539$), the item × treatment interaction is 2.08 points of additional belief reduction per scale point, 95% CrI [−0.35, 4.57], $BF_{10} = 0.25$ ($BF_{01} = 4.0$; posterior probability of a positive interaction = .95). The HC3 test is nominally significant (2.06, 95% CI [0.05, 4.08], $p = .045$), but $p = .31$ after Holm adjustment across the eight item tests, .063 with a treatment × pretreatment-belief term added and .10 with classical standard errors. The Baseline treatment effect was substantial at every level of the item, from 8.4 points (95% CI [2.9, 13.9]) among participants who strongly disagreed to 16.7 [12.3, 21.0] among those who strongly agreed (±1 *SD* in Table S18a). The point estimate is in the direction predicted by the actively open-minded thinking result (§S5.3), but the credible interval includes zero and the Bayes factor favors the null, so it neither confirms nor refutes that result.

### S5.4b Item 4 Standalone — "The Arguments I've Heard Against This Belief Were Weak or Unconvincing"

Item 4 is the largest moderator among the four prior-exposure items (Table S16). Per scale point, a higher score predicts 1.68 points less belief reduction in Control (HC3 $p = .014$) and a further 3.06 points less under the Baseline treatment ($p = .017$; Table S16, post − pre scale). Per *SD* and on the reduction scale, the interaction is −3.51 [−6.39, −0.64], the value in the main text (Tables S18a–S18b). The treatment still worked for participants high on the item, only less well (9.4 points relative to Control at 1 *SD* above the mean, 95% CI [5.8, 13.0]). The Bayes factor is equivocal ($BF_{10} = 0.77$ per scale point and 0.84 per *SD*; $BF_{01} = 1.3$ per scale point), although the credible interval narrowly excludes zero.

Agreeing that past counterarguments were weak could reflect a belief closer to the truth, for which counterarguments really are weaker; a general tendency to discount counterevidence; poor sources met in the past; or other correlated dispositions. The item's near-zero correlation with item 2 (§S5.4a) argues mildly against a simple low-reflection account but does not separate the others, so we do not commit to an interpretation.

**Figure S5**

*Counterargument Items as Moderators of the Baseline Treatment Effect, Study 2*

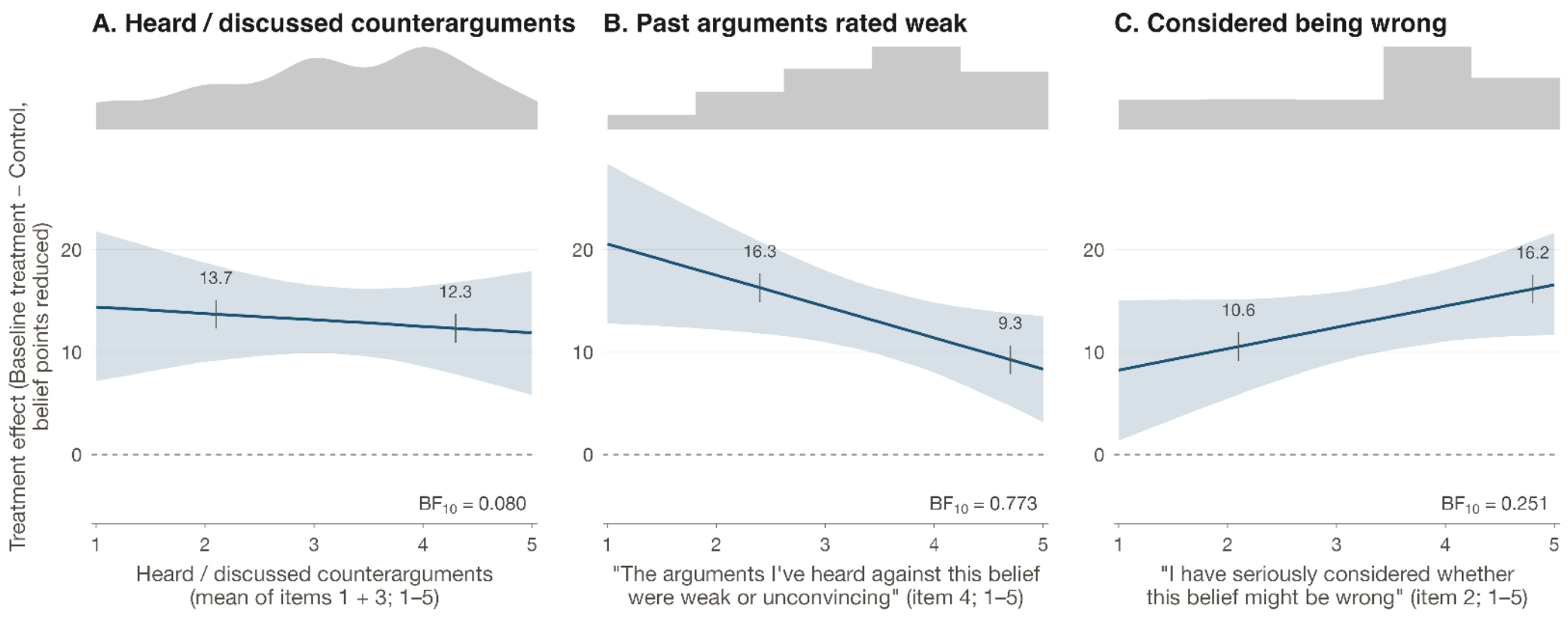

*Note.* Posterior Baseline treatment − Control difference in belief reduction as a function of (A) the prior-exposure composite (mean of items 1 and 3), (B) item 4 and (C) item 2, each rated 1–5. Lines are posterior medians, bands are 95% credible intervals, and gray strips show each moderator's distribution; numbers give the difference at 1 *SD* below and above the moderator mean. $BF_{10}$ is for the treatment × moderator interaction per scale point under a Normal(0, 20) prior.

## S5.5 Education and Other Moderators (Study 2)

Table S17 reports the education model: belief change (post − pre) regressed on centered pretreatment belief, condition, education (ordinal response code from 1 = less than high school to 8 = professional degree, mean-centered) and the Condition × Education products.

**Table S17**

*Education Moderation (Study 2)*

| Term | *b* | *SE* | *t* | *p* |
|---|---|---|---|---|
| (Intercept) | −1.41 | 0.70 | −2.01 | .045 |
| Pretreatment belief (centered) | −0.111 | 0.045 | −2.46 | .014 |
| Baseline treatment (vs. Control) | −13.03 | 1.41 | −9.25 | < .001 |
| Education (centered) | 0.250 | 0.508 | 0.49 | .623 |
| Single Message (vs. Control) | −10.29 | 1.28 | −8.02 | < .001 |
| Baseline treatment × Education | 1.88 | 0.98 | 1.91 | .056 |
| Single Message × Education | −0.877 | 1.09 | −0.81 | .419 |

*Note.* Negative coefficients on condition terms indicate larger reductions relative to Control. HC3-robust standard errors. Neither interaction is significant at .05. Simple slopes showed substantial belief reduction under treatment at every level of education.

Tables S18a and S18b report the remaining moderation tests, all post hoc. Table S18a gives each focal moderator's Baseline treatment × Moderator interaction from the three-arm Study 2 model, with the treatment effect at 1 *SD* below and above the moderator mean. Table S18b adds Holm-adjusted *p* values, the Single Message terms, joint tests, the demographic and topic moderators, Study 1 associations and the corresponding randomized tests in Costello et al. (2024).

*Belief importance.* In Study 2, the Baseline treatment effect was directionally smaller, though still large, for more important beliefs (interaction $p = .112$). Among Study 2 participants rating the belief "very" to "extremely" important (6–8 on the 0–8 scale; $n = 267$, 56 in Control), the Baseline treatment reduced belief by 9.80 points relative to Control (95% CI [5.24, 14.35]), compared with 14.68 [11.24, 18.12] among the rest ($n = 636$). The same attenuation is significant in Study 2 of Costello et al. (2024) ($p = .001$) and appears as an association in the present Study 1, which has no control arm (Table S18b). In Study 1 it is specific to the evidence-based arms (No Facts × Importance +1.19 per point, 95% CI [0.22, 2.16], $p = .017$): the No Facts effect was near zero among participants rating the belief very to extremely important (1.58 points, 95% CI [−4.42, 7.57], $n = 33$), so the Study 1 figures use the other seven arms.

*Conspiracy topic.* Belief reductions among treated participants differed across topics (§S8.3). The randomized Study 2 test asks a different question, whether the Baseline treatment's advantage over Control differs across topics, and it was not significant (Baseline treatment × topic $F(7, 878) = 1.25$, $p = .274$; the effect was positive in all eight topic groups). This does not show equal effects: cells are small (10–28 Control and 8–35 Baseline participants per named topic), and the Single Message × topic test was significant with this grouping ($p = .040$, unadjusted; other groupings in the Table S18b note).

*Gender.* Gender is the only demographic moderator with $p < .05$ in Study 2 (women minus men +8.34 points; $n = 885$, participants of other or unreported gender excluded). The Baseline treatment effect was 8.94 points [4.97, 12.90] among men and 17.28 [13.34, 21.22] among women. The difference survives Holm adjustment across the 12 Study 2 tests in Table S18b ($p = .042$) but does not appear in Study 1 or in Costello et al. (2024), and we do not interpret it.

*Single Message.* Importance and social embeddedness significantly moderate the Single Message effect (unadjusted $p = .004$ and .011) but not the Baseline treatment effect (Table S18b), although the two arms' interactions do not differ significantly from each other (importance $p = .378$; embeddedness $p = .267$, Table S13).

**Table S18a**

*Tests of Moderation of the Baseline Treatment Effect: Focal Moderators*

| Moderator | Scale | *n* | Baseline × Moderator *b* [95% CI] | *p* | $BF_{10}$ | Baseline vs. Control at −1 *SD* | Baseline vs. Control at +1 *SD* |
|---|---|---|---|---|---|---|---|
| Pre-treatment belief | 0–100 (analytic range 50–100) | 903 | −0.86 [−3.77, 2.04] | .559 | 0.10 | 14.1 [10.1, 18.0] | 12.3 [8.2, 16.4] |
| Belief importance | 0–8 | 903 | −2.43 [−5.42, 0.57] | .112 | 0.24 | 15.8 [11.4, 20.2] | 11.0 [7.2, 14.7] |
| Trust in generative AI, pooled (Costello et al., 2024, Studies 1–2, and present Study 2) | 1–7, centered within experiment | 2,594 | 2.13 [1.29, 2.98] per scale point (*SD* = 1.58) | < .001 | > 30 (Table S11) | 10.8 [9.0, 12.5] | 17.5 [15.5, 19.5] |
| Trust in generative AI, Study 2 alone | 1–7 | 903 | 1.92 [−1.16, 5.00] | .221 | 0.16 | 11.3 [7.4, 15.3] | 15.2 [10.8, 19.6] |
| Social embeddedness | 0–100 | 899 | −1.81 [−4.84, 1.21] | .240 | 0.15 | 14.3 [9.8, 18.9] | 10.7 [7.1, 14.3] |
| Prior exposure to counterarguments | 1–5, mean of items 1 and 3 | 903 | −0.71 [−3.47, 2.06] | .617 | 0.09 | 13.8 [9.8, 17.9] | 12.4 [8.6, 16.3] |
| "I have seriously considered whether this belief might be wrong" (item 2) | 1–5 | 903 | 2.73 [0.06, 5.40] | .045 | 0.35 | 10.7 [7.0, 14.4] | 16.2 [12.2, 20.1] |
| "The arguments I've heard against this belief were weak or unconvincing" (item 4) | 1–5 | 903 | −3.51 [−6.39, −0.64] | .017 | 0.84 | 16.4 [12.1, 20.7] | 9.4 [5.8, 13.0] |
| Actively open-minded thinking, Costello et al. (2024) Study 1 | 1–6, six items; measured at follow-up | 623 | 3.73 [0.96, 6.51] | .009 | 1.09 | 12.3 [8.6, 16.0] | 19.8 [15.6, 24.0] |

*Note.* Study 2 rows ($N = 903$; $n = 899$ for social embeddedness) are from separate regressions of post-treatment belief on centered pre-treatment belief, condition (Baseline treatment, Single Message, Control), the moderator and the Condition × Moderator products, with HC3 robust standard errors. Coefficients are on the belief-reduction scale, per 1 *SD* of the moderator unless stated; positive values indicate a larger treatment effect at higher moderator values. The two right-hand columns give the model-implied Baseline-versus-Control reduction, in belief points, at 1 *SD* below and above the moderator mean. $BF_{10}$ values are Savage–Dickey Bayes factors from Bayesian refits of the two-arm Baseline-versus-Control model ($n = 539$) with the moderator standardized and Normal(0, 20) priors; the AOT Bayes factor is from the two-arm model of §S5.3. Pooled trust: treatment versus control within each experiment, with experiment × treatment and experiment × trust terms (Table S11). AOT: Costello et al. (2024) Study 1 participants with pre-treatment belief ≥ 50 who completed the follow-up about 12 days after treatment, when AOT was measured, so the association has no established causal direction (§S5.3). All tests are post hoc and unadjusted; of the 12 Study 2 tests in Table S18b, only the gender interaction survives Holm adjustment. CI = confidence interval.

**Table S18b**

*Moderation Tests: Holm Adjustment, Single Message Interactions, Joint Tests, Remaining Moderators, and Companion Associations*

| Moderator | *n* | Baseline × Moderator *b* [95% CI] | *p* | Holm *p* | Single Message × Moderator *b* [95% CI] | *p* | Joint test | Single Message vs. Control at low | Single Message vs. Control at high | Study 1 association (descriptive) | Costello et al. (2024), randomized |
|---|---|---|---|---|---|---|---|---|---|---|---|
| Pre-treatment belief | 903 | −0.86 [−3.77, 2.04] | .559 | 1.000 | −1.01 [−3.80, 1.78] | .477 | *F*(2, 897) = 0.32, *p* = .727 | 11.3 [7.5, 15.1] | 9.3 [5.6, 13.1] | 1.35 [0.15, 2.55], *p* = .027 (*n* = 1,161) | Study 1: −0.56 [−3.06, 1.93], *p* = .658 (*n* = 753); Study 2: −0.09 [−1.96, 1.78], *p* = .926 (*n* = 1,302) |
| Belief importance | 903 | −2.43 [−5.42, 0.57] | .112 | .896 | −3.95 [−6.66, −1.24] | .004 | *F*(2, 896) = 4.36, *p* = .013 | 14.6 [10.5, 18.6] | 6.7 [3.5, 9.8] | −4.71 [−6.16, −3.26], *p* < .001 (*n* = 1,161) | Study 2: −3.08 [−4.98, −1.19], *p* = .001 (*n* = 1,301) |
| Trust in generative AI (Study 2 alone) | 903 | 1.92 [−1.16, 5.00] | .221 | 1.000 | 1.61 [−1.00, 4.21] | .227 | *F*(2, 896) = 1.13, *p* = .324 | 8.7 [5.4, 12.1] | 11.9 [8.1, 15.8] | 2.86 [1.58, 4.14], *p* < .001 (*n* = 1,161) | Study 1: 4.74 [2.37, 7.11], *p* < .001 (*n* = 753); Study 2: 3.15 [1.25, 5.04], *p* = .001 (*n* = 1,302) |
| Social embeddedness | 899 | −1.81 [−4.84, 1.21] | .240 | 1.000 | −3.67 [−6.51, −0.83] | .011 | *F*(2, 892) = 3.24, *p* = .040 | 13.3 [9.0, 17.6] | 6.0 [2.9, 9.0] | not measured | — |
| Prior exposure to counterarguments (items 1 and 3) | 903 | −0.71 [−3.47, 2.06] | .617 | 1.000 | −0.34 [−3.09, 2.41] | .808 | *F*(2, 896) = 0.13, *p* = .878 | 10.7 [7.0, 14.4] | 10.0 [6.2, 13.8] | not measured | — |
| Item 2 (“seriously considered whether this belief might be wrong”) | 903 | 2.73 [0.06, 5.40] | .045 | .450 | 1.99 [−0.59, 4.58] | .131 | *F*(2, 896) = 2.58, *p* = .076 | 8.3 [4.7, 11.9] | 12.3 [8.7, 15.9] | not measured | — |
| Item 4 (“arguments I’ve heard against this belief were weak”) | 903 | −3.51 [−6.39, −0.64] | .017 | .184 | −2.48 [−5.26, 0.30] | .081 | *F*(2, 896) = 3.52, *p* = .030 | 12.8 [8.7, 16.8] | 7.8 [4.4, 11.2] | not measured | — |
| Education (1–8 ordinal, as in Table S17) | 903 | −2.60 [−5.27, 0.07] | .056 | .503 | 1.22 [−1.74, 4.17] | .419 | *F*(2, 896) = 2.68, *p* = .069 | 9.1 [5.2, 13.0] | 11.5 [7.6, 15.4] | 0.43 [−0.87, 1.73], *p* = .514 (*n* = 1,161) | — |
| Gender (women vs. men) | 885 | 8.34 [2.76, 13.93] | .003 | .042 | 5.26 [−0.03, 10.55] | .051 | *F*(2, 878) = 5.00, *p* = .007 | 7.8 [3.5, 12.0] | 13.0 [9.8, 16.2] | 0.15 [−2.43, 2.73], *p* = .909 (*n* = 1,139) | Study 1: 0.14 [−5.02, 5.30], *p* = .959 (*n* = 747); Study 2: 0.97 [−2.93, 4.87], *p* = .625 (*n* = 1,290) |
| Religiosity (belief in God or gods; 1–8) | 903 | −0.94 [−3.92, 2.04] | .537 | 1.000 | −1.79 [−4.34, 0.76] | .170 | *F*(2, 896) = 0.97, *p* = .381 | 12.1 [8.3, 15.9] | 8.5 [5.1, 12.0] | −0.17 [−1.42, 1.08], *p* = .789 (*n* = 1,161) | — |
| Age (years) | 900 | −0.39 [−3.25, 2.47] | .790 | 1.000 | 0.56 [−1.95, 3.07] | .661 | *F*(2, 893) = 0.17, *p* = .841 | 9.7 [5.8, 13.6] | 10.8 [7.6, 14.0] | −1.23 [−2.53, 0.07], *p* = .063 (*n* = 1,158) | — |
| Party identification (Republicans vs. Democrats) | 844 | −4.52 [−10.22, 1.18] | .120 | .896 | 0.06 [−5.21, 5.34] | .982 | *F*(2, 837) = 1.32, *p* = .268 | 10.1 [6.3, 13.8] | 10.1 [6.4, 13.8] | −4.75 [−8.04, −1.47], *p* = .005 (*n* = 722) | — |
| Conspiracy topic (8 groups; omnibus *F*) | 903 | *F*(7, 878) = 1.25 | .274 | — | *F*(7, 878) = 2.11 | .040 | *F*(14, 878) = 1.50, *p* = .103 | — | — | see Table S29 and Figure S9 | *F*(12, 1971) = 1.30, *p* = .21 |

*Note.* Study 2 columns are from the three-arm HC3 regressions of Table S18a, on the belief-reduction scale (positive = larger treatment effect at higher moderator values or in the first-named group): per 1 *SD* for continuous moderators, group differences for gender and party, and an omnibus *F* for topic. Holm *p* treats the 12 Study 2 Baseline × Moderator tests as one family. The joint test is the HC3 Wald *F* test that both Condition × Moderator terms are zero. Low and high are 1 *SD* below and above the mean (men and women; Democrats and Republicans). Study 1 column: association of each moderator with pre-post belief reduction in the seven

evidence-based arms (No Facts excluded; HC3 regression on the moderator [standardized if continuous], centered pre-treatment belief and condition), descriptive because Study 1 had no control arm. Final column: the randomized Treatment × Moderator test in Costello et al. (2024) for the moderators the text compares across experiments (samples as in Table S11); those experiments also measured education, religiosity, age and party. For topic, the value is the omnibus test as published. Topic: HDBSCAN topics with at least 10 Control participants (six topics plus General / Unclassified), the rest pooled as an eighth group; with topics of at least 15 participants (12 levels; Control cells as small as 2) the Baseline × Topic *p* is .150 and the omnibus Condition × Topic *p* .016, and with the nine-category coding they are .716 and .809. $BF_{10}$ for the remaining moderators, from two-arm models as in Table S18a (moderators standardized, so gender and party are per *SD* of the 0/1 indicator, not per group difference): education 0.26, religiosity 0.10, age 0.09, gender 1.72, party 0.19. The omnibus topic test has no single coefficient and its Bayes factor depends strongly on the prior, so none is given.

### S5.6 Correlations Among Pretreatment Moderators

Figure S6 shows the correlations among the Study 2 pretreatment moderators, pretreatment belief, importance and demographics. They are modest. The strongest are between pretreatment belief and importance ($r = .41$) and between social embeddedness and both importance and pretreatment belief ($r = .34$ each). Trust in generative AI correlates weakly with every other variable (all $|r| \leq .21$), so the pooled trust moderation (Table S11) is unlikely to reflect overlap with the other moderators.

**Figure S6**

*Correlations Among Pretreatment Moderators, Study 2*

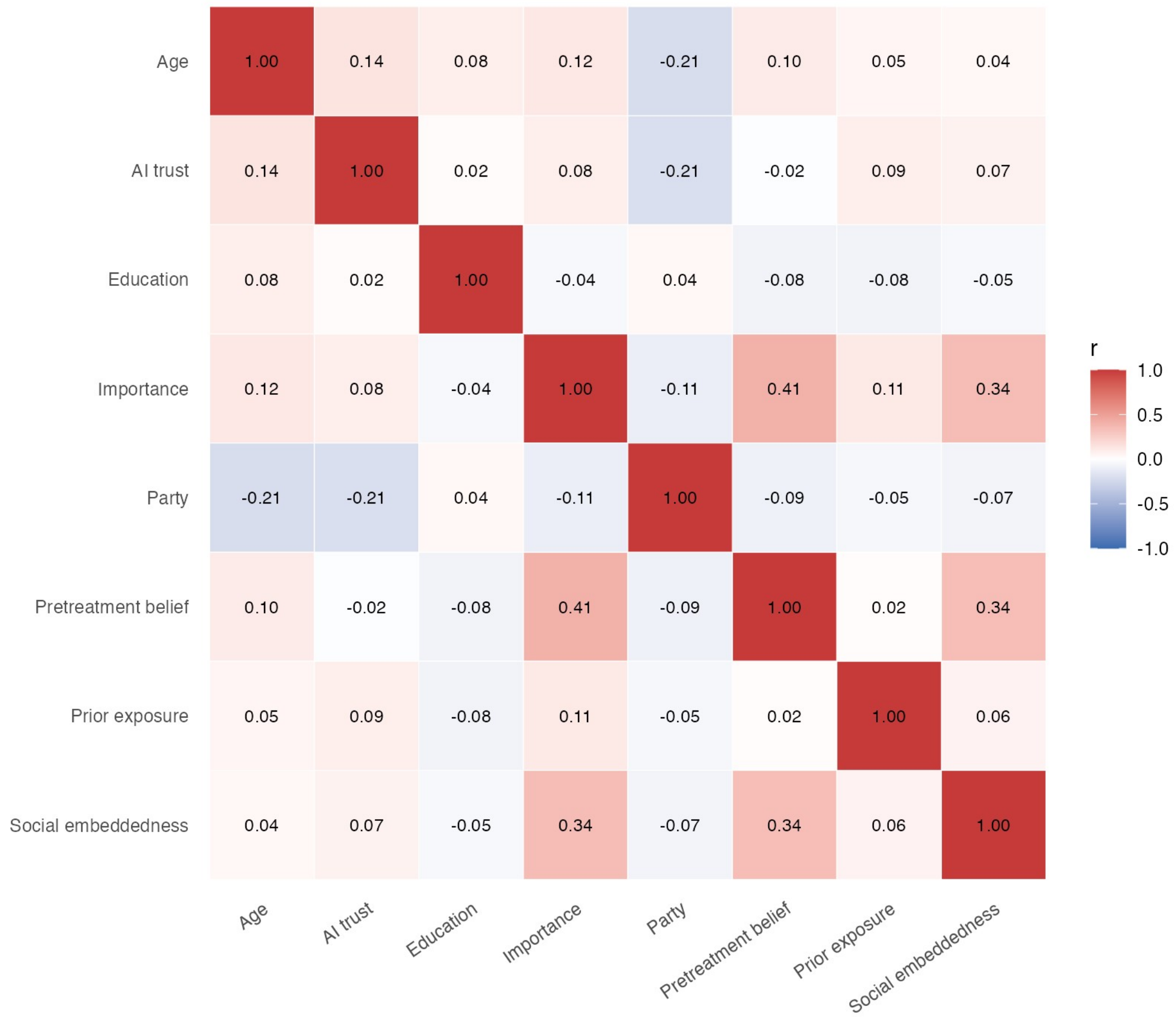

*Note.* Pearson correlations, pairwise complete observations ($N = 903$). Prior exposure is the composite of items 1 and 3.

## S6. Persuasion Strategy Annotation and Mediation Model

### S6.1 GPT-4o Annotation Pipeline

The 16 persuasive strategies (Table S19) come from the analysis code of Costello et al. (2024), where they were the strategies named in at least half of 10 independent queries asking

GPT-4 Turbo how it would talk someone out of a conspiracy belief. Table S19 gives the definitions used in our annotation prompt. Costello et al. (2024) rated all strategies in one GPT-4 Turbo prompt; we rated each strategy in a separate call, with shortened definitions, an added "Uncertain" option and a rating for each of the three AI turns.

For each conversation and strategy, GPT-4o (gpt-4o-2024-08-06, temperature 0) rated the AI's use of the strategy in each round and returned the ratings as JSON. All 1,401 rated conversations (every Study 1 arm, including participants later excluded from the analytic sample) were coded. "Uncertain" was returned for 14 of the 22,416 first-round ratings (0.06%). Only the first-round ratings (the AI's first message) enter the PCA, the prompt contrasts, the GAM and the mediation model. A simplified excerpt of the prompt follows; in it the AI is the "Debunker" and the participant the "Believer".

> "You are an expert annotator analyzing persuasion strategies in conversations on conspiracy theories. You will see a written conversation between two individuals: a Debunker and a Believer. The Debunker's role is to convince the Believer that their conspiracy theory is incorrect. Each conversation has 3 rounds.
>
> Focus only on evaluating the presence of the following persuasion strategy in the Debunker's responses during each round: {StrategyName}: {StrategyDefinition}
>
> Use only the Believer's responses for context to understand the Debunker's responses.
>
> Rating Scale for each round: - High: Strategy used extensively and centrally. - Moderate: Strategy used repeatedly or with emphasis. - Low: Strategy used in a limited way. - None: Strategy not used. - Uncertain: If you're unsure.
>
> Format your response as JSON:
>
> {"{StrategyName}: {StrategyDefinition}": { "Round_1": "RATING HERE", "Round_2": "RATING HERE", "Round_3": "RATING HERE" }}
>
> Please output only the JSON as specified above and do not include any additional text."

**Table S19**

*Definitions of 16 Persuasive Strategies Used to Prompt GPT-4o*

| Strategy | Definition |
|---|---|
| Alternative Explanations | Provide plausible, evidence-based alternative perspectives or explanations for events attributed to conspiracy theories. |
| Analogies / Metaphors | Use analogies or metaphors to simplify complex information or draw parallels. |
| Build Rapport | Establish a respectful and understanding relationship with the Believer. |
| Common Ground | Identify and build on shared beliefs or values. |
| Conflicting Evidence | Introduce facts that contradict claims made by the conspiracy theory. |
| Credible Sources | Introduce information from reliable sources. |
| Critical Thinking | Encourage the Believer to question the logic and sources behind their beliefs. |
| Encourage Empathy | Help the Believer consider the impact of beliefs on others. |
| Expert Consensus | Emphasize agreement among credible authorities. |
| Harm | Discuss the personal or societal harms of conspiracy beliefs. |
| Inconsistencies / Fallacies | Identify logical inconsistencies or fallacies in arguments. |
| Maintaining Patience | Demonstrate patience in engaging in gradual dialogue. |
| Offer Resources | Provide resources for further learning or alternative perspectives. |
| Psychological Needs | Address emotional aspects or psychological needs related to beliefs. |
| Socratic Questioning | Use a questioning approach to examine beliefs. |
| Stories / Examples | Share stories, anecdotes, or examples. |

## S6.2 Polychoric PCA of 16 Strategies → 5 Components

First-round ratings were scored None = 0, Low = 1, Moderate = 2, High = 3. On their polychoric correlations, parallel analysis indicated five components, and we fit a varimax-rotated PCA with five components (loadings in Figure 4A). Component scores are regression scores from the rotated solution, not averages of the strategies loading on each component.

Two fits are used. The loadings in Figure 4A and the scores in Figure 4B and Table S20 come from all 1,366 coded conversations with complete first-round ratings and pre- and post-treatment beliefs, including participants outside the analytic sample. The GAM, the mediation model and Table S25 use a second fit on the analytic-sample conversations. The two solutions agree closely for four components (score correlations .90–.96) and less well for Stories / Metaphors / Examples (.51), on which Socratic Questioning loads more heavily in the analytic-sample fit.

**Prompt contrasts (Figure 4B, Table S20).** The contrasts use the four conditions as randomized, in the analytic sample ($N = 659$; the mediation model in §S6.4 uses $N = 658$). Scores are z-scored within this frame, so an *SD* includes between-condition variance and a difference in *SD* units understates a within-condition Cohen's *d*. Table S20 reports all 30 pairwise contrasts. The five Study 1 arms that share the Baseline system prompt do not differ on the component scores (analytic sample; all $p > .12$).

In the four-arm frame, the No Facts elevation on Harms & Empathy comes from Encourage Empathy (rated present in 91% of No Facts first messages vs. 54% in the Baseline treatment; $M = 2.26$ vs. 0.75 on the 0–3 scale), not from Harm, which was rarer in No Facts (19% vs. 33%, $p = .002$; $M = 0.26$ vs. 0.46, $p = .003$). About half of the No Facts first messages rated at least Low on Harm dwell on the human toll of the event in question and about half on what the belief costs the believer; explicit discussion of the harms of conspiracy beliefs was uncommon in every condition. No Facts almost never cited sources or expert consensus (Credible Sources present in 2% vs. 80% of first messages; Expert Consensus 3% vs. 63%). The two versions of the No Facts prompt, which differ in two sentences (Samples 1a and 1b; 63 and 56 coded conversations), have similar profiles (mean component scores on Facts & Evidence −1.55 vs. −1.39, $p = .18$; Stories 1.89 vs. 1.61, $p = .13$; Harms & Empathy 0.47 vs. 0.82, $p = .04$; Rapport / Common Ground 0.29 vs. 0.10, $p = .01$).

## Table S20

*Pairwise Differences in Strategy-Component Scores Among the Four Persuasion-Prompt Conditions (Study 1)*

| Component | Contrast (second minus first) | Difference (*SE*), *SD* units | 95% CI | *p* | Holm *p* (within component) | Holm *p* (all 30) |
|---|---|---|---|---|---|---|
| **Facts & Evidence** | | | | | | |
| | Baseline treatment vs. Just-the-Facts | −0.17 (0.093) | [−0.35, 0.01] | .070 | .140 | .421 |
| | Baseline treatment vs. Concise | −0.35 (0.087) | [−0.52, −0.18] | < .001 | < .001 | .001 |
| | Baseline treatment vs. No Facts | −1.90 (0.070) | [−2.04, −1.77] | < .001 | < .001 | < .001 |
| | Just-the-Facts vs. Concise | −0.18 (0.119) | [−0.41, 0.05] | .132 | .140 | .659 |
| | Just-the-Facts vs. No Facts | −1.74 (0.106) | [−1.94, −1.53] | < .001 | < .001 | < .001 |
| | Concise vs. No Facts | −1.56 (0.102) | [−1.76, −1.36] | < .001 | < .001 | < .001 |
| **Rapport / Common Ground** | | | | | | |
| | Baseline treatment vs. Just-the-Facts | −1.42 (0.121) | [−1.66, −1.18] | < .001 | < .001 | < .001 |
| | Baseline treatment vs. Concise | −1.12 (0.101) | [−1.31, −0.92] | < .001 | < .001 | < .001 |
| | Baseline treatment vs. No Facts | −0.03 (0.050) | [−0.13, 0.07] | .533 | .533 | > .999 |
| | Just-the-Facts vs. Concise | 0.30 (0.151) | [0.01, 0.60] | .046 | .092 | .322 |
| | Just-the-Facts vs. No Facts | 1.39 (0.124) | [1.14, 1.63] | < .001 | < .001 | < .001 |
| | Concise vs. No Facts | 1.08 (0.104) | [0.88, 1.29] | < .001 | < .001 | < .001 |
| **Harms & Empathy** | | | | | | |
| | Baseline treatment vs. Just-the-Facts | −0.47 (0.081) | [−0.63, −0.31] | < .001 | < .001 | < .001 |
| | Baseline treatment vs. Concise | −0.35 (0.101) | [−0.55, −0.15] | < .001 | < .001 | .006 |
| | Baseline treatment vs. No Facts | 0.58 (0.112) | [0.35, 0.80] | < .001 | < .001 | < .001 |
| | Just-the-Facts vs. Concise | 0.12 (0.100) | [−0.08, 0.31] | .243 | .243 | .972 |
| | Just-the-Facts vs. No Facts | 1.04 (0.112) | [0.82, 1.26] | < .001 | < .001 | < .001 |
| | Concise vs. No Facts | 0.93 (0.127) | [0.68, 1.18] | < .001 | < .001 | < .001 |
| **Sources & Experts** | | | | | | |
| | Baseline treatment vs. Just-the-Facts | −0.31 (0.115) | [−0.53, −0.08] | .008 | .016 | .079 |
| | Baseline treatment vs. Concise | −0.31 (0.111) | [−0.53, −0.09] | .005 | .015 | .056 |
| | Baseline treatment vs. No Facts | −0.77 (0.065) | [−0.90, −0.64] | < .001 | < .001 | < .001 |
| | Just-the-Facts vs. Concise | −0.01 (0.134) | [−0.27, 0.26] | .956 | .956 | > .999 |
| | Just-the-Facts vs. No Facts | −0.46 (0.099) | [−0.66, −0.27] | < .001 | < .001 | < .001 |
| | Concise vs. No Facts | −0.46 (0.095) | [−0.64, −0.27] | < .001 | < .001 | < .001 |
| **Stories / Metaphors / Examples** | | | | | | |

| Component | Contrast (second minus first) | Difference (*SE*), *SD* units | 95% CI | *p* | Holm *p* (within component) | Holm *p* (all 30) |
|---|---|---|---|---|---|---|
| | Baseline treatment vs. Just-the-Facts | 0.13 (0.063) | [0.01, 0.26] | .037 | .073 | .293 |
| | Baseline treatment vs. Concise | 0.20 (0.085) | [0.04, 0.37] | .017 | .051 | .153 |
| | Baseline treatment vs. No Facts | 1.83 (0.096) | [1.65, 2.02] | < .001 | < .001 | < .001 |
| | Just-the-Facts vs. Concise | 0.07 (0.091) | [−0.11, 0.25] | .439 | .439 | > .999 |
| | Just-the-Facts vs. No Facts | 1.70 (0.101) | [1.50, 1.90] | < .001 | < .001 | < .001 |
| | Concise vs. No Facts | 1.63 (0.116) | [1.40, 1.86] | < .001 | < .001 | < .001 |

*Note.* Four conditions as randomized, analytic sample only (Baseline treatment $n = 309$, Just-the-Facts $n = 111$, Concise $n = 123$, No Facts $n = 116$; $N = 659$). Component scores from the Figure 4 PCA, z-scored within this frame (*SD* units). Differences (second condition minus first) are from OLS with HC3-robust standard errors ($df = 655$); $p$ values are Holm-adjusted within component (six contrasts) and, as a sensitivity check, across all 30, which changes two decisions (Sources & Experts, Baseline treatment vs. Just-the-Facts and vs. Concise). Welch omnibus tests: Facts & Evidence $F(3, 245) = 253.1$; Rapport / Common Ground $F(3, 250) = 83.6$; Harms & Empathy $F(3, 287) = 33.7$; Sources & Experts $F(3, 263) = 57.9$; Stories / Metaphors / Examples $F(3, 265) = 125.4$; all $p < .001$. Not preregistered.

## S6.3 Generalized Additive Model (GAM)

The GAM (Figure 5) regressed belief change on smooth terms for the five component scores and centered pretreatment belief (R package mgcv; smoothness chosen by generalized cross-validation), without adjusting for condition.

Facts & Evidence (edf = 2.83) and Sources & Experts (edf = 1.29) had positive, roughly monotonic associations with belief reduction (both $p < .001$), and Harms & Empathy a small negative, roughly linear one (edf = 1.00, $p = .001$). Rapport / Common Ground (edf = 3.48, $p = .31$) and Stories / Metaphors / Examples (edf = 1.00, $p = .50$) were not significantly associated with belief change. The mediation model, which adjusts for condition, finds no significant association for Harms & Empathy, Rapport or Stories (Table S21).

## S6.4 Multiple Mediation Structural Equation Model

The path model (R package lavaan; specified in the main text) uses the four Study 1 conditions whose prompts changed the AI's strategy mix (Baseline treatment as the reference; $N = 658$ with complete component scores), with centered pretreatment belief as a covariate in every equation. Component scores are in *SD* units (standardized in the 1,174 rated conversations), so the five $b$-paths share a scale.

Of the No Facts total indirect association with belief change (−15.96), the Facts & Evidence path accounts for 69% (−11.07) and the Sources & Experts path for 20% (−3.14). These shares are not proportions mediated, because the direct effect of No Facts was positive (7.86, $p = .057$) against a total effect of −8.10. Because No Facts lies 2.3 *SD* below the Baseline treatment on Facts & Evidence, its indirect path also largely extrapolates a within-condition slope. The total indirect paths of Just-the-Facts and Concise were near zero.

Table S22 tests the differences among the five *b*-paths with the same bootstrap resamples. Each evidence coefficient exceeds each of the other three without adjustment; after Holm adjustment over the 10 pairs, Sources & Experts vs. Rapport / Common Ground does not ($p$ = .10). Strategy use was measured, not randomized, so these associations are consistent with mediation but do not demonstrate it; the pairwise contrasts were not preregistered.

## Table S21

*Multiple-Mediation Model: Path Estimates and Bootstrap Confidence Intervals*

| Section | Parameter | *b* | *SE* | *p* | 95% CI | β |
|---|---|---|---|---|---|---|
| **Total Indirect / Total Effects** | | | | | | |
| | Total Indirect — No Facts | −15.960 | 4.034 | < .001 | [−23.938, −7.926] | −0.294 |
| | — of which: via M1 (Facts & Evidence)† | −11.072 | 2.947 | < .001 | [−16.835, −5.174] | −0.204 |
| | — of which: via M2 (Rapport) | −0.018 | 0.175 | .920 | [−0.395, 0.317] | 0.000 |
| | — of which: via M3 (Harms & Empathy) | −0.313 | 0.365 | .390 | [−1.121, 0.353] | −0.006 |
| | — of which: via M4 (Sources & Experts) | −3.139 | 1.005 | .002 | [−5.155, −1.279] | −0.058 |
| | — of which: via M5 (Stories / Metaphors / Examples) | −1.418 | 1.214 | .243 | [−3.804, 1.012] | −0.026 |
| | Total Indirect — Just-the-Facts | −1.622 | 1.720 | .346 | [−5.039, 1.774] | −0.029 |
| | Total Indirect — Concise | −2.218 | 1.562 | .156 | [−5.282, 0.900] | −0.042 |
| | Total Effect — No Facts | −8.102 | 1.657 | < .001 | [−11.283, −4.869] | −0.149 |
| | Total Effect — Just-the-Facts | −0.194 | 2.289 | .932 | [−4.549, 4.318] | −0.004 |
| | Total Effect — Concise | 3.248 | 2.577 | .208 | [−1.702, 8.396] | 0.061 |
| **Direct Effects (Condition → DV)** | | | | | | |
| | Belief Change ~ No Facts | 7.858 | 4.133 | .057 | [−0.272, 15.869] | 0.145 |
| | Belief Change ~ Just-the-Facts | 1.428 | 2.882 | .620 | [−4.047, 7.148] | 0.026 |
| | Belief Change ~ Concise | 5.467 | 2.961 | .065 | [−0.321, 11.337] | 0.103 |
| **Mediator → DV Effects** | | | | | | |
| | Belief Change ~ Facts & Evidence | 4.850 | 1.261 | < .001 | [2.292, 7.251] | 0.266 |
| | Belief Change ~ Rapport / Common Ground | −0.108 | 1.026 | .916 | [−2.190, 1.864] | −0.006 |
| | Belief Change ~ Harms & Empathy | −0.832 | 0.899 | .355 | [−2.607, 0.919] | −0.040 |
| | Belief Change ~ Sources & Experts | 3.238 | 0.994 | .001 | [1.387, 5.205] | 0.154 |
| | Belief Change ~ Stories / Metaphors / Examples | −1.090 | 0.923 | .238 | [−2.851, 0.777] | −0.057 |
| **Condition → Mediator** | | | | | | |
| | M1 (Facts & Evidence) ~ No Facts | −2.283 | 0.081 | < .001 | [−2.435, −2.119] | −0.767 |
| | M1 ~ Just-the-Facts | −0.359 | 0.101 | < .001 | [−0.562, −0.161] | −0.118 |
| | M1 ~ Concise | −0.331 | 0.083 | < .001 | [−0.494, −0.174] | −0.113 |
| | M2 (Rapport) ~ No Facts | 0.163 | 0.050 | .001 | [0.062, 0.258] | 0.054 |
| | M2 ~ Just-the-Facts | −1.328 | 0.144 | < .001 | [−1.615, −1.051] | −0.433 |
| | M2 ~ Concise | −1.309 | 0.128 | < .001 | [−1.557, −1.056] | −0.443 |
| | M3 (Harms) ~ No Facts | 0.377 | 0.098 | < .001 | [0.184, 0.566] | 0.145 |
| | M3 ~ Just-the-Facts | −0.634 | 0.094 | < .001 | [−0.815, −0.443] | −0.241 |
| | M3 ~ Concise | −0.348 | 0.096 | < .001 | [−0.531, −0.156] | −0.137 |
| | M4 (Sources & Experts) ~ No Facts | −0.969 | 0.073 | < .001 | [−1.119, −0.830] | −0.375 |
| | M4 ~ Just-the-Facts | −0.309 | 0.107 | .004 | [−0.521, −0.098] | −0.117 |
| | M4 ~ Concise | −0.266 | 0.105 | .011 | [−0.468, −0.053] | −0.105 |
| | M5 (Stories) ~ No Facts | 1.301 | 0.136 | < .001 | [1.035, 1.573] | 0.455 |
| | M5 ~ Just-the-Facts | −0.411 | 0.070 | < .001 | [−0.546, −0.268] | −0.141 |
| | M5 ~ Concise | 0.166 | 0.103 | .107 | [−0.028, 0.370] | 0.059 |
| **Covariate effects** | | | | | | |
| | M1 ~ Pretreatment belief (centered) | −0.007 | 0.002 | < .001 | [−0.011, −0.004] | −0.097 |
| | M2 ~ Pretreatment belief | 0.002 | 0.002 | .345 | [−0.003, 0.007] | 0.029 |
| | M3 ~ Pretreatment belief | 0.004 | 0.002 | .127 | [−0.001, 0.008] | 0.056 |
| | M4 ~ Pretreatment belief | −0.003 | 0.002 | .247 | [−0.008, 0.002] | −0.042 |

| Section | Parameter | *b* | *SE* | *p* | 95% CI | β |
|---|---|---|---|---|---|---|
| | M5 ~ Pretreatment belief | −0.002 | 0.002 | .477 | [−0.007, 0.003] | −0.024 |
| | Belief Change ~ Pretreatment belief | 0.153 | 0.055 | .005 | [0.046, 0.261] | 0.109 |

*Note.* Study 1, four conditions (Baseline treatment as reference; *N* = 658). M1 = Facts & Evidence; M2 = Rapport / Common Ground; M3 = Harms & Empathy; M4 = Sources & Experts; M5 = Stories / Metaphors / Examples. *b* is unstandardized; β is the standardized solution, which is not bootstrapped. *SE*s and 95% CIs are from 5,000 percentile-bootstrap resamples (complete cases). Total indirect = Σ $a_j b_j$ over the five mediators; total effect = total indirect + direct.

†The "of which" rows split the No Facts total indirect effect by mediator; the M1 row is the value reported in the main text.

**Table S22**

*Differences Among the Five Mediator (b-Path) Coefficients in the Strategies Path Model*

| Contrast | Difference | Bootstrap *SE* | 95% CI | *p* | Holm *p* (10 pairs) |
|---|---|---|---|---|---|
| Facts & Evidence − Rapport / Common Ground | 4.96 | 1.75 | [1.46, 8.40] | .004 | .036 |
| Facts & Evidence − Harms & Empathy | 5.68 | 1.64 | [2.46, 8.91] | < .001 | .005 |
| Facts & Evidence − Sources & Experts | 1.61 | 1.46 | [−1.37, 4.35] | .269 | 1.000 |
| Facts & Evidence − Stories / Metaphors / Examples | 5.94 | 1.81 | [2.27, 9.38] | .001 | .009 |
| Rapport / Common Ground − Harms & Empathy | 0.72 | 1.15 | [−1.56, 3.02] | .529 | 1.000 |
| Rapport / Common Ground − Sources & Experts | −3.35 | 1.45 | [−6.23, −0.54] | .021 | .104 |
| Rapport / Common Ground − Stories / Metaphors / Examples | 0.98 | 1.35 | [−1.76, 3.57] | .467 | 1.000 |
| Harms & Empathy − Sources & Experts | −4.07 | 1.50 | [−7.07, −1.21] | .007 | .040 |
| Harms & Empathy − Stories / Metaphors / Examples | 0.26 | 1.12 | [−2.01, 2.46] | .818 | 1.000 |
| Sources & Experts − Stories / Metaphors / Examples | 4.33 | 1.53 | [1.43, 7.43] | .005 | .036 |
| **Joint tests (bootstrap covariance)** | | | | | |
| All five *b*-paths equal (Wald test) | $\chi^2(4) = 14.41$ | | | .006 | |
| Rapport, Harms & Empathy and Stories *b*-paths equal (Wald test) | $\chi^2(2) = 0.58$ | | | .747 | |
| Mean of Facts & Evidence and Sources & Experts − mean of the other three | 4.72 | 1.26 | [2.25, 7.28] | < .001 | |
| Resamples in which Facts & Evidence is the largest *b*-path | 86.5% | | | | |
| Resamples in which both evidence *b*-paths exceed all three others | 98.4% | | | | |
| Resamples in which Facts & Evidence exceeds Sources & Experts | 86.5% | | | | |

*Note.* Differences between the *b*-paths of Table S21 (belief reduction per *SD* of each component score; *N* = 658), computed from its 5,000 bootstrap resamples. CIs are percentile intervals; *p* values are Wald tests using the bootstrap *SE*, Holm-adjusted over the ten pairs in the last column. Joint tests use the bootstrap covariance matrix. These contrasts were not preregistered.

## S7. Claim-Level Content Analysis

### S7.1 Claim Extraction Pipeline

Each AI message about the focal belief in both studies was sent, in a separate call, to GPT-4.1 (OpenAI; February–March 2026). The system prompt said: *"Extract only specific factual claims stated in the text, regardless of whether they are true or false. Do not include* *opinions, advice, or general rhetoric."* Study 1 yielded 30,662 claim instances (30,314 distinct texts) from 1,238 conversations, and Study 2 yielded 27,535 (27,352 distinct) from 888. Each distinct text was classified once, but every instance counts in the per-conversation counts.

## S7.2 Six-Category Claim Taxonomy and Codebook

Gemini 3 Flash (Google) assigned each distinct claim one of 30 subtypes, given the subtype definitions with two examples each and no conversational context; the six top-level categories are derived from the subtypes (§S7.7). Table S23 gives the top-level definitions, with verbatim example claims drawn from the top quartile of the classifier's confidence ratings.

**Table S23**

*Six-Category Claim Taxonomy*

| Category (subtypes) | Definition | Example claims (assigned subtype) |
|---|---|---|
| **Specific empirical** (SE1–SE6) | Concrete, checkable claims about events, records, actors, procedures, findings, documents, or physical evidence. | "Operation Northwoods was a 1962 Pentagon proposal involving staged incidents." (Historical events and documented precedent) / "The Apollo flag had a horizontal rod along the top edge to hold it out." (Physical / forensic / visual evidence) |
| **Inferential-logical** (IL1–IL6) | Object-level reasoning about what would follow if the conspiracy were true, what evidence should exist, how coordination would work, or why a mundane alternative is more plausible. | "It would require keeping a secret amongst thousands of people for over 50 years if the moon landing were faked." (Coordination / leak-risk logic) / "If there were alien craft or bodies, you would expect consistent documentation that survives scrutiny." (Necessary-evidence-if-true reasoning) |
| **Meta-epistemic** (ME1–ME3) | Claims about how one should evaluate evidence, compare sources, apply standards of proof, or reason under uncertainty. | "Verified evidence of alien life would require evidence that clearly cannot be explained by anything else and can be checked by many independent experts." (Evidence-standard norm) / "The scientific method involves hypothesis testing, evidence gathering, and peer review." (Reliability / method guidance) |
| **Moral / political evaluation** (MP1–MP3) | Moral, civic, political, or prudential evaluations about power, justice, abuse, policy risk, empathy, or public consequences. | "In democracies, the electorate is the ultimate check on government." (Civic / prosocial value) / "If exposed, allowing such an attack would be one of the biggest crimes in US history." (Moral / justice evaluation) |
| **Minimal / non-adjudicative** (LI1–LI5) | Definitional, tautological, banal, minimally informative, or merely background statements that do little adjudicative work. | "Logical fallacies are errors in reasoning that weaken arguments." (Definitional clarification) / "Parkland Hospital is the same hospital where Kennedy and later Oswald were taken." (Minimal event/background fact) |
| **General / underspecified** (GU1–GU7) | Broad, hedged, contextual, or underspecified claims that are not precise enough to verify as written. | "Most evidence for aliens at Area 51 is anecdotal and hard to verify." (Absence-of-clear-evidence framing) / "Governments do hide things, mostly for military reasons." (Secrecy / protection / cover-up possibility framing) |

## S7.3 Claim Composition by Condition — Study 1 and Study 2

**Table S24**

*Claim Volume and Composition by Condition*

| Condition | Conversations (in analytic sample) | Claims per conversation, *M* (*Mdn*) | Specific-empirical claims per conversation, *M* | Pooled share (%): specific empirical | inferential-logical | meta-epistemic | moral/ political | minimal | general/ underspecified | Mean within-conversation share (%): specific empirical |
|---|---|---|---|---|---|---|---|---|---|---|
| **Study 1: all conversations in the claims file** | | | | | | | | | | |
| Baseline treatment | 318 (309) | 26.2 (25.5) | 12.5 | 47.6 | 4.0 | 3.0 | 6.9 | 23.1 | 15.4 | 44.5 |
| No Facts | 121 (116) | 7.1 (6.0) | 0.7 | 9.9 | 1.3 | 0.1 | 22.4 | 48.7 | 17.6 | 7.7 |
| Just-the-Facts | 120 (111) | 33.2 (34.0) | 17.7 | 53.4 | 2.8 | 2.6 | 3.6 | 21.5 | 16.1 | 51.7 |
| Concise | 131 (123) | 17.7 (17.0) | 7.9 | 44.5 | 6.7 | 2.8 | 5.6 | 24.2 | 16.3 | 39.5 |
| Adversarial | 125 (119) | 28.0 (28.0) | 13.2 | 46.9 | 4.3 | 3.1 | 5.5 | 25.5 | 14.6 | 44.6 |
| Overt | 121 (115) | 27.0 (27.0) | 13.6 | 50.2 | 3.9 | 2.8 | 5.2 | 23.1 | 14.8 | 47.4 |
| Abstracted Summary | 164 (154) | 27.1 (28.0) | 13.5 | 50.0 | 4.3 | 2.7 | 5.1 | 24.8 | 13.1 | 47.4 |
| Belief-First Rating | 138 (130) | 28.7 (28.0) | 13.3 | 46.5 | 3.9 | 4.0 | 6.3 | 24.4 | 14.9 | 44.4 |
| Study 1 pooled, excluding No Facts | 1117 (1061) | 26.7 (27.0) | 13.0 | 48.5 | 4.1 | 3.0 | 5.7 | 23.7 | 15.0 | 45.4 |
| **Study 1: analytic sample (*N* = 1,290)** | | | | | | | | | | |
| Baseline treatment | 309 (309) | 26.5 (26.0) | 12.6 | 47.6 | 3.9 | 3.0 | 7.0 | 23.1 | 15.3 | 44.4 |
| No Facts | 116 (116) | 7.2 (6.0) | 0.7 | 9.7 | 1.2 | 0.1 | 22.6 | 48.4 | 17.9 | 7.5 |
| Just-the-Facts | 111 (111) | 33.4 (34.0) | 17.6 | 52.6 | 2.8 | 2.7 | 3.8 | 21.7 | 16.5 | 50.4 |
| Concise | 123 (123) | 17.7 (18.0) | 7.9 | 44.4 | 6.4 | 2.7 | 5.9 | 24.0 | 16.6 | 39.3 |
| Adversarial | 119 (119) | 28.0 (28.0) | 13.3 | 47.7 | 4.2 | 3.2 | 5.6 | 24.8 | 14.5 | 45.2 |
| Overt | 115 (115) | 27.5 (28.0) | 13.9 | 50.6 | 3.7 | 2.8 | 5.3 | 22.8 | 14.7 | 47.7 |
| Abstracted Summary | 154 (154) | 27.0 (28.0) | 13.4 | 49.6 | 4.3 | 2.7 | 5.1 | 25.1 | 13.3 | 47.3 |
| Belief-First Rating | 130 (130) | 28.9 (28.5) | 13.3 | 46.1 | 4.0 | 4.0 | 6.5 | 24.6 | 14.8 | 44.1 |
| Study 1 pooled, excluding No Facts | 1061 (1061) | 26.8 (27.0) | 13.0 | 48.4 | 4.1 | 3.0 | 5.8 | 23.6 | 15.0 | 45.3 |
| **Study 2: all conversations in the claims file** | | | | | | | | | | |
| Baseline treatment | 456 (336) | 42.6 (41.5) | 15.8 | 37.0 | 16.5 | 3.6 | 1.7 | 22.6 | 18.6 | 34.8 |
| Single Message | 432 (359) | 18.8 (18.0) | 7.0 | 37.3 | 15.8 | 2.8 | 1.6 | 23.0 | 19.4 | 35.4 |
| **Study 2: analytic sample (*N* = 903)** | | | | | | | | | | |
| Baseline treatment | 336 (336) | 44.1 (44.0) | 15.9 | 36.1 | 16.3 | 3.5 | 1.8 | 22.7 | 19.5 | 34.1 |
| Single Message | 359 (359) | 19.0 (18.0) | 7.1 | 37.4 | 15.4 | 2.9 | 1.7 | 23.1 | 19.6 | 35.5 |

*Note.* Claims come from every AI message (Study 1: three messages; Study 2: three in the Baseline treatment, one in Single Message) and were classified into the six categories of §S7.2. The all-conversations rows are the basis of the per-conversation figures quoted in the text. Pooled share treats the claim as the unit (claims of a type divided by all claims in the condition); mean within-conversation share averages each conversation's share. Claims were not extracted from the Study 2 Control dialogues, which concerned unrelated topics.

**Claims and strategy scores (Table S25).** Strategy scores are holistic ratings of how heavily the AI's first message relied on each approach (§S6.1), whereas the claim pipeline extracts every specific factual claim from every AI message. A claim is therefore not by definition an instance of Facts & Evidence.

**Table S25**

*Pearson Correlations Between First-Message Claim Measures and the Five Strategy-Component Scores (Study 1)*

| Frame | Claim measure (first AI message) | *n* | Facts & Evidence | Sources & Experts | Rapport / Common Ground | Harms & Empathy | Stories / Metaphors / Examples |
|---|---|---|---|---|---|---|---|
| Pooled (8 conditions) | Total claims | 1174 | 0.28 [0.23, 0.33] | 0.28 [0.23, 0.33] | −0.05 [−0.11, 0.01] | −0.20 [−0.25, −0.14] | −0.41 [−0.46, −0.36] |
| Pooled (8 conditions) | Specific-empirical claims | 1174 | 0.14 [0.08, 0.19] | 0.44 [0.40, 0.49] | −0.10 [−0.16, −0.05] | −0.14 [−0.20, −0.09] | −0.32 [−0.37, −0.27] |
| Pooled (8 conditions) | Specific-empirical share | 1172 | 0.19 [0.13, 0.24] | 0.49 [0.44, 0.53] | −0.11 [−0.17, −0.05] | −0.09 [−0.15, −0.03] | −0.28 [−0.33, −0.23] |
| Condition-adjusted (8 conditions) | Total claims | 1174 | −0.14 [−0.20, −0.09] | 0.19 [0.14, 0.25] | 0.00 [−0.06, 0.06] | −0.12 [−0.18, −0.07] | −0.22 [−0.27, −0.16] |
| Condition-adjusted (8 conditions) | Specific-empirical claims | 1174 | −0.19 [−0.24, −0.13] | 0.41 [0.36, 0.46] | −0.06 [−0.12, −0.00] | −0.06 [−0.12, −0.01] | −0.16 [−0.21, −0.10] |
| Condition-adjusted (8 conditions) | Specific-empirical share | 1172 | −0.15 [−0.20, −0.09] | 0.44 [0.39, 0.49] | −0.07 [−0.13, −0.01] | −0.01 [−0.06, 0.05] | −0.11 [−0.17, −0.06] |
| Pooled, excluding No Facts | Total claims | 1058 | −0.13 [−0.19, −0.08] | 0.17 [0.11, 0.23] | 0.03 [−0.03, 0.09] | −0.14 [−0.20, −0.08] | −0.27 [−0.32, −0.21] |
| Pooled, excluding No Facts | Specific-empirical claims | 1058 | −0.19 [−0.25, −0.13] | 0.39 [0.34, 0.44] | −0.05 [−0.11, 0.01] | −0.09 [−0.15, −0.03] | −0.22 [−0.28, −0.16] |
| Pooled, excluding No Facts | Specific-empirical share | 1058 | −0.16 [−0.22, −0.11] | 0.44 [0.39, 0.49] | −0.06 [−0.12, 0.00] | −0.02 [−0.08, 0.04] | −0.16 [−0.22, −0.10] |
| Baseline treatment only | Total claims | 309 | −0.30 [−0.39, −0.19] | 0.15 [0.04, 0.26] | −0.03 [−0.14, 0.08] | −0.14 [−0.25, −0.03] | −0.31 [−0.41, −0.21] |
| Baseline treatment only | Specific-empirical claims | 309 | −0.30 [−0.40, −0.20] | 0.37 [0.27, 0.47] | −0.11 [−0.22, −0.00] | −0.06 [−0.17, 0.06] | −0.27 [−0.37, −0.16] |
| Baseline treatment only | Specific-empirical share | 309 | −0.22 [−0.32, −0.11] | 0.42 [0.32, 0.50] | −0.11 [−0.22, 0.00] | −0.00 [−0.12, 0.11] | −0.18 [−0.28, −0.07] |

*Note.* $N$ = 1,174 rated conversations in the Study 1 analytic sample (1,172 for the share, which is undefined for two first messages without claims). Component scores come from the PCA fit that supplies the mediators in Table S21 (§S6.2). 95% CIs use the Fisher $z$ transformation. Condition-adjusted rows residualize both variables on the eight condition indicators. Spearman correlations show the same pattern (§S11).

## S7.4 Claim-Level Association With Belief Change

Pooled OLS models regressed belief reduction on the within-conversation share of one claim type at a time, controlling for pretreatment belief, study and conspiracy topic (fixed effects; HC3 standard errors). A larger Inferential-logical share predicted larger reductions and a larger Moral/political share smaller ones; the other four shares were not significantly associated with belief change (Table S26).

### S7.5 Alternative-Model Sensitivity

Because a share depends on the total number of claims, Table S26 also measures the focal type by its raw count, by its raw count controlling for the other five types, and by its log-ratio balance against the rest. The Moral/political and Meta-epistemic estimates are negative and the Inferential-logical estimates positive under all four specifications.

**Table S26**

*Claim-Composition Models Under Four Specifications (N = 1,963 Conversations, Both Studies)*

| Claim type | *b* | *SE* | 95% CI | *p* |
|---|---|---|---|---|
| **Share of the focal type vs. the rest (per 10-percentage-point increase)** | | | | |
| Moral/political | −2.03 | 0.38 | [−2.78, −1.28] | < .001 |
| Inferential-logical | 1.52 | 0.57 | [0.41, 2.64] | .007 |
| Specific empirical | 0.26 | 0.22 | [−0.16, 0.69] | .227 |
| Meta-epistemic | −0.86 | 0.74 | [−2.32, 0.59] | .245 |
| General/underspecified | 0.09 | 0.35 | [−0.61, 0.78] | .809 |
| Minimal | 0.00 | 0.28 | [−0.56, 0.55] | .989 |
| **Count of the focal type (per additional claim)** | | | | |
| Moral/political | −0.82 | 0.24 | [−1.28, −0.36] | < .001 |
| Inferential-logical | 0.27 | 0.17 | [−0.06, 0.60] | .107 |
| Minimal | 0.13 | 0.10 | [−0.07, 0.32] | .212 |
| Meta-epistemic | −0.28 | 0.31 | [−0.88, 0.32] | .362 |
| General/underspecified | 0.09 | 0.12 | [−0.15, 0.34] | .446 |
| Specific empirical | −0.003 | 0.05 | [−0.10, 0.09] | .946 |
| **Count of the focal type, controlling for the counts of the other five types (per additional claim)** | | | | |
| Moral/political | −0.87 | 0.24 | [−1.35, −0.40] | < .001 |
| Meta-epistemic | −0.49 | 0.31 | [−1.10, 0.12] | .116 |
| Inferential-logical | 0.25 | 0.17 | [−0.09, 0.59] | .144 |
| Minimal | 0.11 | 0.11 | [−0.10, 0.32] | .293 |
| Specific empirical | −0.04 | 0.05 | [−0.14, 0.06] | .402 |
| General/underspecified | 0.08 | 0.13 | [−0.17, 0.33] | .513 |
| **Log-ratio balance of the focal type against the rest (per *SD*)** | | | | |
| Moral/political | −2.57 | 0.50 | [−3.55, −1.60] | < .001 |
| Meta-epistemic | −1.05 | 0.48 | [−1.98, −0.11] | .028 |
| Inferential-logical | 1.27 | 0.60 | [0.09, 2.45] | .035 |
| Specific empirical | 0.67 | 0.52 | [−0.35, 1.69] | .200 |
| Minimal | 0.30 | 0.48 | [−0.64, 1.23] | .536 |
| General/underspecified | 0.02 | 0.54 | [−1.04, 1.08] | .971 |

### S7.6 Claim Composition by Focal-Conspiracy Veracity

To check whether claim composition depended on how true participants' focal conspiracies were, we labeled each focal belief (the AI's one-sentence restatement) with Perplexity Sonar Pro (a web-searching language model). The prompt asked it to judge the belief's central conspiratorial implication rather than its background premises, and to return a 0–100 veracity score and one of four labels: false or mostly false, mixed or partly true, plausible or mostly true, or too vague to check. Of the 1,290 Study 1 participants, 70.2% held a false or mostly false focal belief, 19.1% a mixed, 7.5% a plausible and 3.1% a too-vague one; of the 903 in Study 2, 73.4%, 15.1%, 10.4% and 1.1%. Table S27 gives claim composition by veracity label, computed in Python (pandas, statsmodels).

**Table S27**

*Mean Claim Composition by Focal-Conspiracy Veracity (Conversation-Level, Pooled Across Studies, N = 1,963)*

| Veracity label | Specific empirical | Inferential-logical | Meta-epistemic | Moral/ political | Minimal | General/ underspec. | Mean total claims |
|---|---|---|---|---|---|---|---|
| False / mostly false | 39.7% | 9.4% | 3.3% | 4.7% | 25.7% | 17.2% | 27.6 |
| Mixed / partly true | 39.0% | 6.8% | 2.9% | 5.8% | 26.3% | 19.2% | 26.9 |
| Plausible / mostly true | 35.8% | 8.1% | 3.3% | 7.5% | 22.7% | 22.6% | 27.1 |
| Too vague / no clear claim | 28.2% | 7.6% | 4.4% | 9.7% | 28.8% | 21.4% | 24.7 |

*Note.* Study-adjusted robust Wald tests of veracity label on each claim share, $\chi^2(3)$: specific empirical 15.15, $p = .002$; inferential-logical 26.84, $p < .001$; meta-epistemic 2.98, $p = .395$; moral/political 17.11, $p < .001$; minimal 5.92, $p = .116$; general/underspecified 26.10, $p < .001$. Relative to false or mostly false beliefs, too-vague beliefs drew 13.0 percentage points fewer specific empirical claims ($p < .001$) and 5.0 more general/underspecified claims ($p = .015$).

## S7.7 Granular Claim Subtypes

The codebook was developed after extraction and was not preregistered. An initial five-category classification mixed evidence-standard guidance with moral and political evaluation and labeled many broad empirical claims as unfalsifiable, so we split it into six categories and 30 subtypes, which label the rows of Figure S7. Their definitions are in the codebook file in the replication package (data/codebooks/study2_claim_category_v2_codebook.csv).

**Figure S7**

*Granular Claim Subtype Composition by Focal-Conspiracy Topic*

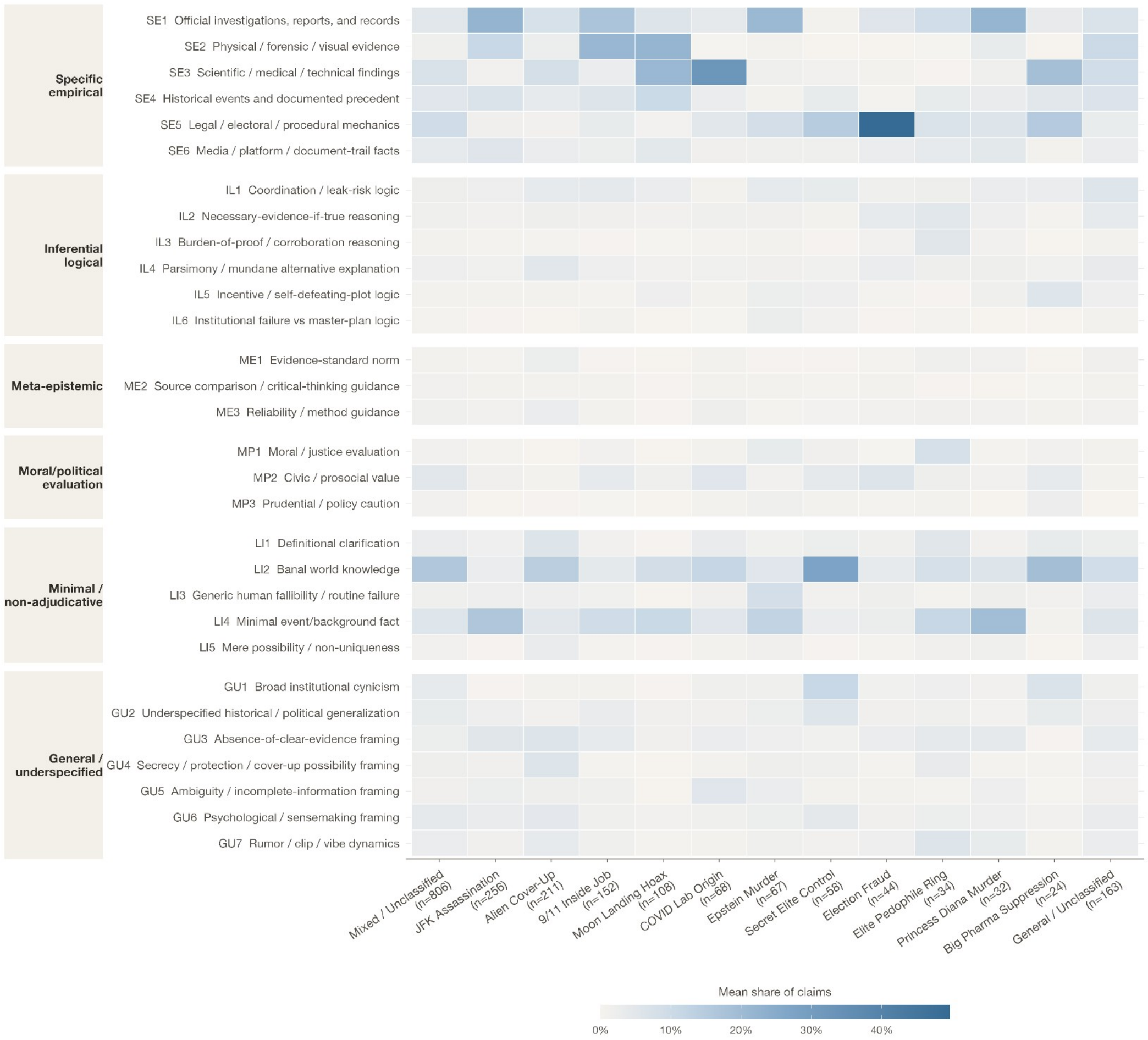

*Note.* Mean within-conversation share of each claim subtype (rows, grouped by category) across conspiracy topics (columns): the HDBSCAN residual cluster (leftmost, labeled "Mixed / Unclassified"; the "General / Unclassified" row of Table S29), the 11 most frequent named topics, and 163 conversations with coded claims that are not in the topic-clustering file (rightmost, labeled "General / Unclassified"). Conversation counts are in parentheses on the x-axis.

## S8. Conspiracy Topic Analysis (HDBSCAN)

### S8.1 Embedding + HDBSCAN Clustering Procedure

We embedded each participant's AI-generated one-sentence restatement of their focal belief with OpenAI's text-embedding-3-large model, reduced the embeddings to 30 principal components (48% of the variance), and clustered them with HDBSCAN, varying the minimum cluster size (minPts) from 10 to 25 (Table S28). The grid was ranked by a score that favored 8–12 topics, but those solutions merged distinct themes (e.g., Mandela Effect, Mass Surveillance, Chemtrails) into broad clusters. We therefore used the 18-topic solution at minPts = 10 (noise

fraction = 0.41); the 8–12-topic solutions gave similar topic-level rankings on the topics they kept. Re-embedding can shift cluster boundaries slightly, so all analyses use the cached cluster assignments distributed with the code (§S11).

**Table S28**

*HDBSCAN Hyperparameter Grid and Selection*

| minPts | Retained clusters | Noise fraction | Largest cluster *n* | Cluster-size CV | Study segregation | Rank |
|---|---|---|---|---|---|---|
| 18 | 12 | 0.448 | 294 | 0.965 | 0.085 | 1 |
| 19 | 11 | 0.460 | 298 | 0.924 | 0.076 | 2 |
| 21 | 10 | 0.470 | 294 | 0.860 | 0.074 | 3 |
| 20 | 10 | 0.477 | 293 | 0.868 | 0.077 | 4 |
| 22 | 9 | 0.479 | 293 | 0.801 | 0.068 | 5 |
| 23 | 9 | 0.488 | 287 | 0.801 | 0.070 | 6 |
| 24 | 9 | 0.495 | 282 | 0.802 | 0.070 | 7 |
| 25 | 9 | 0.498 | 282 | 0.811 | 0.070 | 8 |
| 15 | 13 | 0.426 | 304 | 1.005 | 0.080 | 9 |
| 10 | 18 | 0.409 | 280 | 1.126 | 0.092 | 16 |

*Note.* Retained clusters have at least 10 members. CV = coefficient of variation of retained-cluster sizes; study segregation = how strongly clusters separate Study 1 from Study 2 participants (lower is better). Solutions with 8–12 retained clusters are ranked first, by noise fraction and then study segregation; ranks 10–15 are omitted (full grid in the replication materials, §S11).

### S8.2 Eighteen Topic Clusters — Exemplars and Participant Counts

**Table S29**

*Topic Cluster Summary*

| Cluster | Participants | Study 1 *n* | Study 2 *n* | Illustrative example (summary) |
|---|---|---|---|---|
| General / Unclassified | 895 | 508 | 387 | — (residual; heterogeneous) |
| JFK Assassination | 280 | 178 | 102 | "The JFK assassination involved a conspiracy larger than a lone gunman." |
| Alien Cover-Up | 243 | 123 | 120 | "The government is concealing evidence of extraterrestrial contact." |
| 9/11 Inside Job | 168 | 97 | 71 | "The U.S. government had foreknowledge of 9/11 and allowed it to happen." |
| Moon Landing Hoax | 120 | 69 | 51 | "The Apollo moon landings were staged on Earth." |
| Epstein Murder | 83 | 36 | 47 | "Jeffrey Epstein was murdered in prison, not a suicide." |
| COVID Lab Origin | 78 | 50 | 28 | "SARS-CoV-2 was engineered in a lab rather than emerging naturally." |
| Secret Elite Control | 60 | 30 | 30 | "A shadowy elite controls global politics behind the scenes." |
| Election Fraud | 46 | 22 | 24 | "The 2020 U.S. election was stolen through widespread fraud." |
| Elite Pedophile Ring | 38 | 3 | 35 | "A high-level pedophile ring operates among elites." |
| Princess Diana Murder | 33 | 20 | 13 | "Princess Diana was assassinated, not killed in an accident." |
| Big Pharma Suppression | 31 | 14 | 17 | "Pharmaceutical companies suppress cures for chronic diseases." |
| MKUltra | 22 | 6 | 16 | "MKUltra continues in modified form." |
| Mandela Effect | 20 | 12 | 8 | "Alternate realities explain widespread shared memory errors." |
| Mass Surveillance | 19 | 4 | 15 | "The government tracks citizens through phones and devices." |
| COVID Vaccine Harms | 15 | 13 | 2 | "COVID vaccines cause widespread serious side effects." |
| Staged Trump Shooting | 15 | 0 | 15 | "The 2024 Trump shooting was staged for political gain." |
| Alien Pyramids | 12 | 7 | 5 | "Ancient pyramids were built with extraterrestrial help." |
| Chemtrails | 10 | 6 | 4 | "Chemtrails from airplanes are deliberate chemical dispersal." |

*Note.* Counts come from the clustering file, which is not the analytic sample (Study 1, 1,198; Study 2, 990). Examples are short illustrative summaries of each topic, not verbatim participant statements; the statements nearest each cluster center are in the replication materials (§S11).

### S8.3 Topic-Level Belief Reductions

We pooled the evidence-based conditions of both studies (all Study 1 conditions except No Facts; Study 2 Baseline treatment and Single Message) and regressed belief reduction on topic, controlling for study, for topics with at least 15 participants ($N = 1{,}807$). Belief reductions differed across topics (Wald $\chi^2(14) = 63.1$, $p < .001$; Figure S9B), in each study separately (both $p$s $< .01$) and after adjusting for pretreatment belief ($p < .001$). They were largest for Moon

Landing Hoax ($M_{adj}$ = 20.55), 9/11 Inside Job (17.52) and Secret Elite Control (16.22), and positive for all but two of the 18 clusters (MKUltra, −0.58, $n$ = 17; COVID Vaccine Harms, raw mean −5.64, $n$ = 14, below the reporting threshold). In these conditions, the clustering file includes 90 participants outside the analytic samples and lacks 100 analytic Study 1 participants; among analytic participants only, the test is $\chi^2(14) = 60.9$, $p < .001$ ($N$ = 1,721), and 16 of 18 clusters remain positive. These are descriptive differences among treated participants, with no control comparison. Of the present experiments, only Study 2 can test whether the treatment effect itself differs by topic, because Study 1 has no control arm; there, the randomized Baseline treatment × topic test is not significant ($p$ = .274; §S5.5, Table S18b), as in Costello et al. (2024; $F(12, 1971) = 1.30$, $p$ = .21).

### S8.4 Focal-Conspiracy-Veracity Subset Sensitivity

We re-estimated the Study 1 condition effects (Table S6) and the Study 2 contrasts with Control (Table S8) among participants whose focal conspiracy was labeled false or mostly false (§S7.6), which removes beliefs that are arguably partly true. The treatment effects were, if anything, slightly larger in this subset, and the No Facts condition's ineffectiveness was more pronounced (Tables S30a and S30b); Figure S8 shows belief reduction by condition for each veracity label.

**Table S30a**

*False-Conspiracy Subset Sensitivity — Study 1 Adjusted Mean Reduction by Condition*

| Condition | Full sample | False or mostly false only |
|---|---|---|
| Baseline | 11.00 | 13.39 |
| No Facts | 3.79 | 2.94 |
| Just-the-Facts | 11.58 | 15.26 |
| Concise | 13.73 | 16.25 |
| Adversarial | 11.69 | 12.22 |
| Overt | 9.20 | 12.00 |
| Abstracted Summary | 13.31 | 13.66 |
| Belief-First Rating | 10.21 | 12.20 |

*Note.* Adjusted mean reduction from OLS of post-treatment belief on condition and centered pretreatment belief, evaluated at the full-sample mean pretreatment belief (81.82): full analytic sample ($N$ = 1,290) and the 906 participants whose focal conspiracy was labeled false or mostly false.

**Table S30b**

*False-Conspiracy Subset Sensitivity — Study 2 Frequentist Contrasts*

| Contrast | *b* (full sample) | *b* (false or mostly false only) |
|---|---|---|
| Baseline treatment vs. Control | −13.20 | −14.04 |
| Single Message vs. Control | −10.33 | −12.44 |

*Note.* OLS of post-treatment belief on condition (Control as reference) and pretreatment belief; full analytic sample $N = 903$, false or mostly false subset $n = 663$. All $p < .001$.

**Figure S8**

*Belief Reduction by Condition and Focal-Conspiracy Veracity*

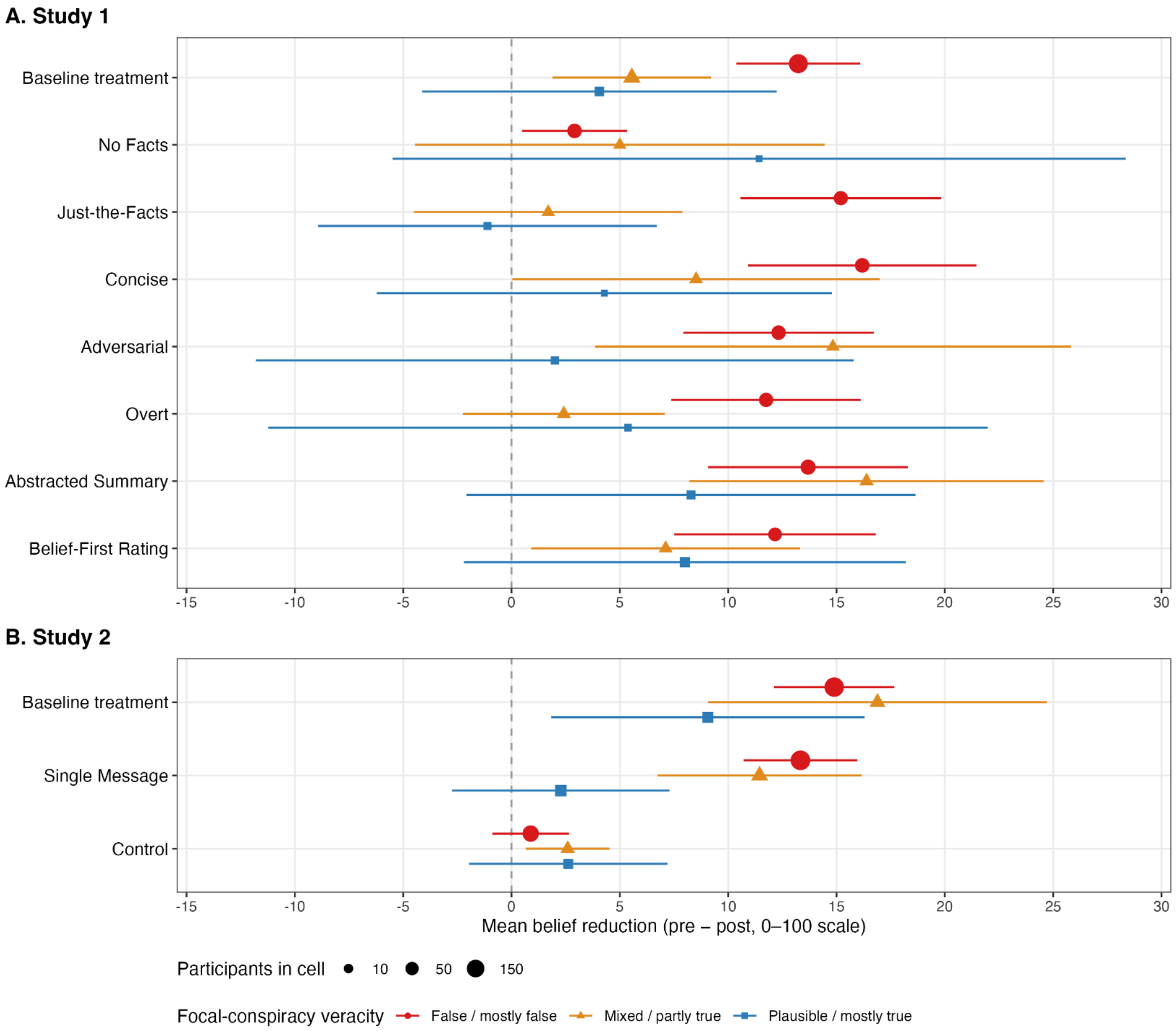

*Note.* Mean belief reduction (pre − post, 0–100 scale) by condition (rows) and focal-conspiracy veracity label (color and marker shape; Perplexity Sonar, §S7.6) in (A) Study 1 ($N = 1,290$) and (B) Study 2 ($N = 903$). Points are raw cell means, sized by the number of participants in the cell; error bars are 95% confidence intervals; the dashed line marks no change. The 50 participants

whose focal belief was labeled too vague or without a clear claim are not shown. Adjusted estimates are in Tables S30a and S30b.

**Figure S9**

*Semantic Map of Focal Conspiracy Beliefs (A) and Belief Reduction by Topic (B)*

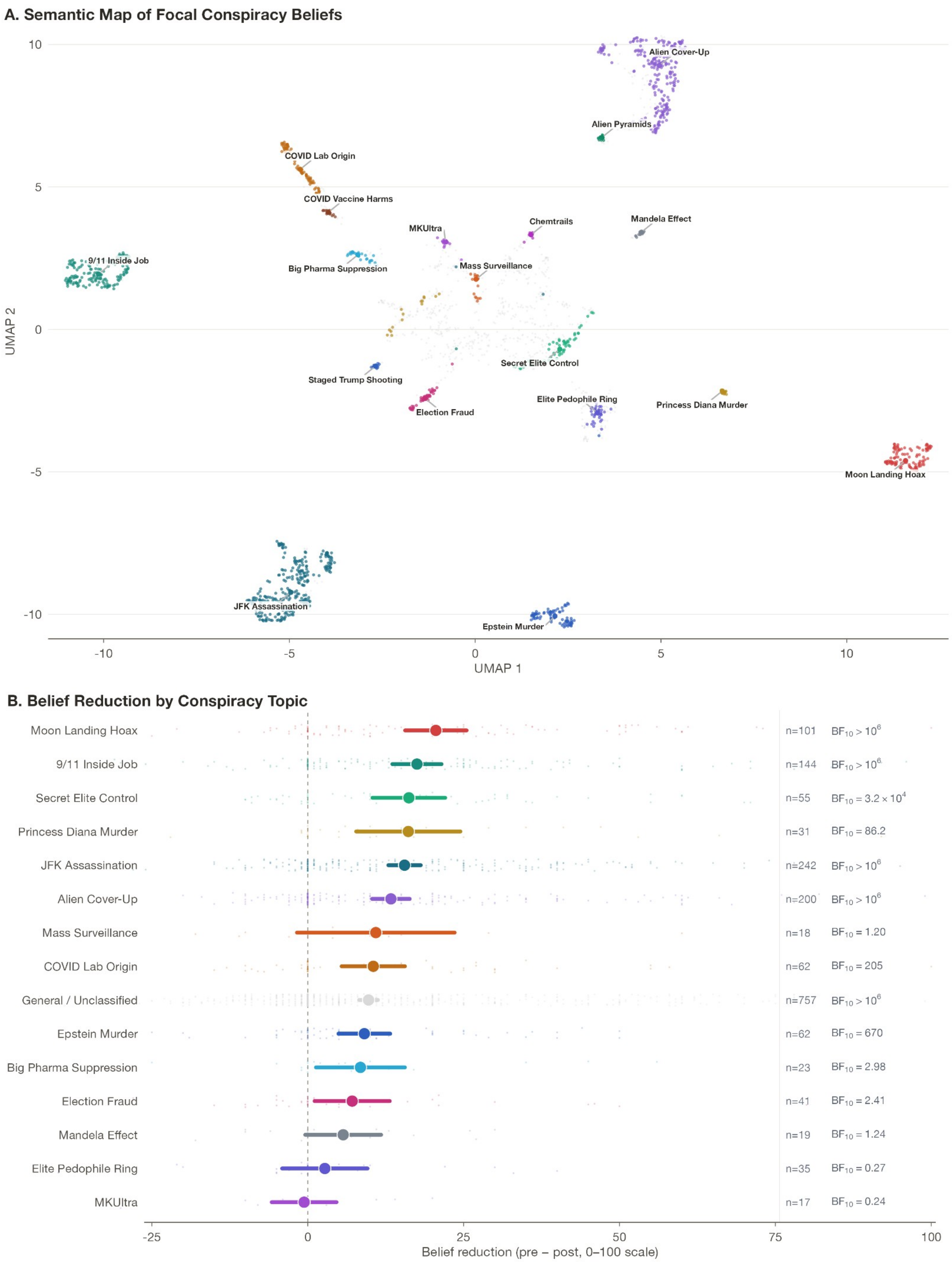

*Note.* (A) UMAP projection of the embeddings of participants' restated focal beliefs (Studies 1 and 2); each point is one participant, colored by topic cluster, and gray points were not assigned

to a topic (“General / Unclassified”). (B) Belief reduction (pre − post, 0–100 scale) by topic, for topics with at least 15 participants in the pooled evidence-based conditions (§S8.3; *N* = 1,807). Faint points are individual participants; large points and bars are study-adjusted means with HC3 95% CIs. Labels at right give each topic’s *n* and a BIC-approximated Bayes factor ($BF_{10}$) comparing a free mean belief reduction with a mean fixed at zero.

## S9. Participants’ Own Accounts of What Was (Un)persuasive

### S9.1 Codebook Development and GPT-4o Coding Pipeline

To quantify Study 1 participants’ explanations of why they did or did not find the AI persuasive (§S1.3.4), we developed, post hoc, 15 attributes (Table S31).

The attributes were developed inductively. One author read the responses and listed the recurring themes; an AI model, given the responses, independently produced a list of candidate categories; and the two lists were merged into one taxonomy of six positive (appreciative, “+”) and nine negative (critical, “−”) attributes.

GPT-4o (gpt-4o-2024-08-06, temperature 0) coded one attribute per call, for each response, with the prompt below. The bracketed fields were filled with the Table S31 definitions, the attribute being coded, the question the participant had answered, and whether their belief had fallen (“succeeded”) or not (“failed”); the condition was not given.

> “You are provided with a list of participant responses regarding why a persuasive conversation with an AI either succeeded or failed in changing their beliefs. Your role is to annotate each response as an expert coder. Although you are coding one theme for this task, please keep in mind the definitions of all the following categories:
>
> [List of categories and definitions]
>
> Today, your focus is on the following theme: [Name and definition]
>
> Note: Themes with a ‘+’ in their name (e.g., EVID+, EMPH+) should be coded when the participant praises the AI’s attributes. Themes with a ‘−’ in their name (e.g., LACK−, BIAS−) should be coded when the participant criticizes the AI’s persuasive qualities.
>
> In this case, the AI was determined to have [succeed_or_failed] in changing beliefs. The participant’s response was provided in answer to the following question: [post-treatment question]. The response itself will be piped in as the first user response.
>
> Instructions: 1. Read the participant’s response carefully. 2. Decide if the theme defined above applies. If it does, output 1; if it does not, output 0. 3. If you are unsure, err on the side of caution and output 0. That is, AVOID FALSE POSITIVES.
>
> Output your answer strictly in JSON format exactly as follows (do not include any extra text, only 0 or 1): [JSON instructions].”

**Table S31**

*Definitions of the 15 Participant-Account Attributes*

| Attribute (short code) | Definition |
|---|---|
| Evidence / Facts / Logic / Counterarguments (**EVID+**) | The respondent (a) mentions or alludes to concrete data, statistics, facts, and/or (b) mentions logic, rationality, legibility, or appeals to common sense, and/or (c) notes the AI addressed or anticipated their doubts, questions, or alternative views. |
| Expertise / Credibility / Trustworthiness (**EXPT+**) | The respondent (a) describes the AI as knowledgeable, authoritative, or expert-like and/or (b) as a reliable, unbiased, or "neutral" source of information and/or (c) values the AI's reliance on trustworthy sources and/or (d) views the AI as having unique or impressive access to what is true by dint of being an AI. |
| Empathy / Politeness / Respectfulness (**EMPH+**) | The respondent (a) highlights that the AI was nonjudgmental, supportive, understood their feelings, or validated their viewpoint — even if it disagreed — and/or (b) calls out the AI's polite, kind, or courteous manner and/or (c) mentions feeling calmer, relieved, or emotionally supported by the AI. |
| Detailed / In-Depth / Novel / Tailored (**DETL+**) | The respondent (a) praises the AI for being thorough, detailed, or in-depth and/or (b) mentions that the AI introduced new perspectives, angles, or data they hadn't considered and/or (c) that the AI's response was multi-faceted or point-by-point, covering the topic thoroughly and/or (d) that the AI's response was personalized, tailored to their specific questions. |
| Conspiracy-Specific Mechanisms (**CONX+**) | The respondent (a) describes specific mechanisms or details of the conspiracy theory, such as how the building's structure led to collapse, how UFO sightings are often secret aircraft, or forensic details of the JFK case, and/or (b) mentioned a meta-point relevant to conspiracy theories, like that large conspiracies would be impossible to hide since someone would have "spilled the beans". |
| Privacy (**PRIV+**) | The respondent felt they could ask the AI questions without fear of judgment or repercussions. |
| Perceived lack of AI disagreement (**LACK−**) | The respondent indicates that they already held the belief in question and that the AI merely reaffirmed what they already believed. |
| Repetition / Regurgitation / Lack of Novelty / Lack of Deep Engagement (**DEEP−**) | The respondent criticizes the AI for repeating well-known or familiar arguments, offering nothing new beyond what they'd heard before and/or the AI is seen as failing to challenge or engage with the respondent's specific points — merely echoing them rather than providing a substantive counterpoint. |
| Offensiveness / Anger (**ANGR−**) | The respondent appears angry or upset about the conversation. |
| Insufficient Evidence / Lack of Concrete Data / No Sourcing (**EVID−**) | The respondent argues that the AI's response lacked hard data, concrete evidence, or sources to back up its claims. |
| Perceived Bias (**BIAS−**) | The respondent perceives that the AI's arguments are skewed toward a particular narrative and fail to engage alternative viewpoints. |
| Impersonal Tone / Verbosity / Lack of Nuance (**MECH−**) | The response is critiqued for being too long-winded, mechanical, or overly formal — lacking the human touch or nuance that might otherwise engage the reader. |
| Distrust in AI (**TRST−**) | The respondent expresses skepticism toward the idea that an AI — due to its machine nature, limitations, or potential biases — can be a convincing or trustworthy source. |
| Off-Topic / Superficial Engagement (**OFFT−**) | The respondent notes that the AI's responses stray from addressing the core issue or are too general, failing to target the specific points of the debate. |
| Emotional / Experiential Disconnect (**EMOT−**) | The respondent finds that the AI's arguments lack a personal, emotional, or experiential dimension, making the discussion feel disconnected from real human concerns. |

## S9.2 Qualitative Observations and Selected Illustrative Quotes

Participants whose belief fell most often said that the AI gave them new facts or evidence and argued logically and clearly. Some also described its tone as polite, patient, or non-confrontational, and some valued that it acknowledged their concerns while refuting them or answered their points one by one. About 5% mentioned the argument that a large conspiracy would be impossible to hide.

Among participants whose belief did not fall, some said they already agreed with the AI, although it had in fact disagreed with them. Others said it did not engage with their points (*"Made oppositional claims without comprehending my points and argument fully"*), strayed off topic (mostly in No Facts: *"The AI didn't really address my questions, it kind of just reverted back to its message of not questioning the unknown."*), repeated itself, or gave no sources (*"I wanted their sources. ... I didn't want similar. I wanted their sources."*), often with a charge of bias toward an "official story." Some rejected AI as a source (*"Because I have zero trust in AI, and it's not something we should be messing with. This technology will kill us all."*), and some found it cold, textbook-like, or blind to emotional and lived experience (*"They felt a little cold. ... it felt like I was reading words from a textbook rather than having a conversation."*).

### S9.3 Quantitative Breakdown — Persuaded vs. Unpersuaded

Figure 6 is based on 1,058 participants (belief reduced, $n$ = 584; unchanged or increased, $n$ = 474). Codes are not mutually exclusive; Figure S10 shows how many codes each explanation received and which attributes were mentioned together.

**Off-pattern codes.** Some positive attributes appear among unchanged-or-increased participants, and some criticisms among belief-reduced participants; they are kept as coded. Because codes overlap, we count them by participant (11.2% and 15.1%) rather than by summing bars. Reading these responses, most are mixed or contrary appraisals: praise with a caveat, participants whose belief fell slightly but who deny having been persuaded, and unchanged participants who say the AI was persuasive or report a slider error. Criticisms became rarer as the belief reduction grew. The same reading showed recurrent coding errors: CONX+ applied when participants restated their own theory, EVID+ for negated mentions of evidence, LACK− for "I agree with the AI", and EVID− when the evidence said to be lacking is the conspiracy's. These errors concentrate in the small off-pattern cells.

**Known-groups check.** Although the coder was not told the condition, its codes separated No Facts from the other arms in the expected direction (analytic participants with codes, $n$ = 1,177, of whom 116 No Facts): insufficient evidence (EVID−) 18.1% vs. 7.3% (OR = 2.82, Fisher $p < .001$); off-topic or superficial engagement (OFFT−) 24.1% vs. 4.3% (OR = 7.00, $p < .001$); evidence, facts and logic (EVID+) 5.2% vs. 34.5% (OR = 0.10, $p < .001$); perceived lack of disagreement (LACK−) 12.9% vs. 7.5% ($p = .048$). DEEP−, BIAS− and TRST− did not differ ($p$s $\geq .26$).

**Figure S10**

*Number and Co-Occurrence of Explanation Attributes, Study 1*

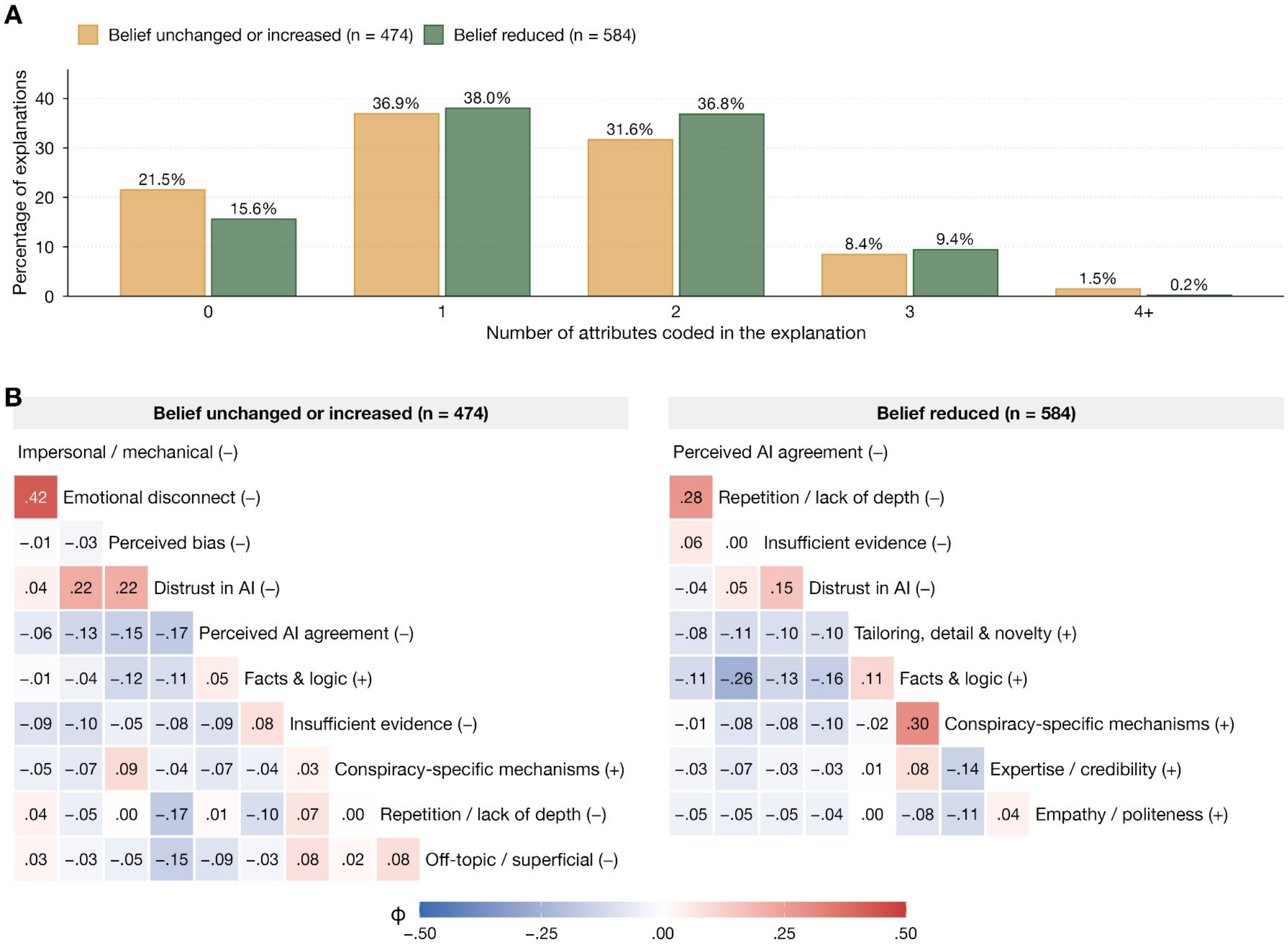

*Note.* (A) Percentage of explanations in each group coded with 0, 1, 2, 3, or 4 or more of the 15 attributes. (B) Phi coefficients (φ, the correlation between two present/absent codes) for every pair of attributes coded in at least 3% of the group's explanations, ordered by hierarchical clustering; each label sits at the right end of its row and the top of its column. Red cells mark pairs mentioned together more often than the two attributes' separate frequencies imply, and blue cells less often; printed values give the sign. (+) and (−) mark positive and negative attributes (Table S31). Frame as in Figure 6 ($N = 1{,}058$).

## S10. Additional Robustness Checks and Covariate-Adjusted Model

### S10.1 Covariate-Adjusted Main-Effects Model (Study 2)

The model in Table S32 regresses Study 2 belief reduction (pre − post) on condition and the pretreatment covariates and moderators listed in the table. Because only pretreatment variables enter, the condition effects are not biased by controlling for consequences of treatment. The model uses Study 2 only, because no participant was randomized between a Study 1 arm and the Study 2 Control. It is a main-effects model, not a test of moderation (for which see §S5 and Table S18b).

Both treatments reduced belief relative to Control. Reduction was larger at higher pretreatment belief (see §S4.5 on regression to the mean) and higher AI trust, and smaller at higher belief importance, more Republican party preference and greater social embeddedness. Age, religiosity, education and prior exposure to counterarguments showed no independent association.

**Table S32**

*Covariate-Adjusted Main-Effects Model (Study 2)*

| Term | *b* | 95% CI | *p* |
|---|---|---|---|
| Single Message (vs. Control) | 9.64 | [7.19, 12.09] | < .001 |
| Baseline treatment (vs. Control) | 12.23 | [9.53, 14.92] | < .001 |
| Pretreatment belief (0–100) | 0.35 | [0.25, 0.46] | < .001 |
| Trust in GenAI (1–7) | 2.11 | [1.23, 3.00] | < .001 |
| Belief importance (0–8) | −1.52 | [−2.18, −0.86] | < .001 |
| Age (years) | −0.03 | [−0.12, 0.07] | .580 |
| Religiosity (1–8) | −0.08 | [−0.63, 0.46] | .766 |
| Party preference (1 = strongly Democratic, 6 = strongly Republican) | −1.06 | [−1.93, −0.20] | .016 |
| Education (ordinal) | −0.74 | [−1.71, 0.22] | .131 |
| Social embeddedness of focal belief (0–100) | −0.17 | [−0.23, −0.11] | < .001 |
| Prior exposure to counterarguments (1–5) | −0.57 | [−1.73, 0.60] | .339 |

*Note.* Outcome: belief reduction (pre − post; higher = larger reduction). Coefficients are per scale point unless noted. Conspiracy topic is a fixed effect (topics with $n < 10$ pooled as "Other / unclassified"); its coefficients are omitted here. HC3-robust standard errors. $N = 896$ of 903 (7 missing covariates); adjusted $R^2 = .198$.

## S10.2 Representative Excerpts From No Facts Conversations

The No Facts prompt told the AI to persuade without using rationality, counterevidence, critical thinking, alternative explanations or reason-based strategies (Table S1). Table S33 shows excerpts of what it wrote instead, with each participant's belief before and after the conversation.

The excerpts illustrate the strategy profile in §S6.2 (Table S20): fewer facts and sources, and more stories, metaphors and appeals to empathy, than in the Baseline treatment.

**Table S33**

*Representative Excerpts From the AI's First Response in the No Facts Condition*

| # | Focal conspiracy | Pre → Post | AI excerpt (first turn) |
|---|---|---|---|
| 1 | The government is concealing the full truth about UFO sightings. | 85 → 10 | "It's kind of amazing to think about the possibility of not being alone in the universe… have you ever noticed how sometimes stories and mysteries can grow bigger and more intriguing as they're told and retold? It's like when we're kids, and we hear stories about hidden treasures or secret lands – they capture our imagination and ignite a sense of adventure." |
| 2 | The moon landing was fabricated by the U.S. to assert technological superiority. | 83 → 39 | "The moon landing conspiracy is definitely one of the more captivating stories out there… It's like a thrilling book or movie that draws you in… The collective effort, dreams, and determination it took to actually land on the moon speak volumes about what humanity can achieve when we come together for a monumental goal." |
| 3 | The birth of COVID-19 was caused by 5G technology. | 75 → 20 | "It's fascinating how certain ideas catch on, especially when they're shared so widely on social media… It's like when you hear a new song and suddenly start noticing it everywhere… Sharing feelings and stories rather than just facts can sometimes help us see things in a new light." |
| 4 | The JFK assassination remains unsolved, with questions about government complicity. | 76 → 40 | "Sometimes, stories get more compelling and complex when they're passed around, kind of like a game of telephone… It's like looking at a magic trick. It's far more exciting to think there's real magic involved, but when you know how the trick is done, it loses a bit of that sparkle, even though it's still interesting in its own right." |
| 5 | The world is controlled by a small group of wealthy elites who dictate media narratives. | 88 → 48 | "I genuinely believe that this worldview about the planet being under the control of a few isn't the whole picture. Imagine for a moment the countless stories of individuals and communities coming together, making substantial changes, and improving lives without needing the approval or direction of the ultra-wealthy." |
| 6 | A powerful, ancient group continues to shape and control modern society. | 75 → 50 | "The idea of a long-surviving group secretly pulling the strings behind society is really fascinating, like something out of a movie or a novel… Think about how hard it is to keep a secret among a small group of friends, now imagine keeping huge, world-changing secrets among a global group over centuries. It's a bit like trying to hold water in your hands, isn't it?" |
| 7 | The world is controlled by the Illuminati, a wealthy group operating in secrecy. | 85 → 65 | "But the thing is, the world is more like a messy, bustling marketplace than a tidy, organized play directed by the Illuminati… It's like trying to organize a potluck dinner with thousands of guests, each bringing their own dish without coordinating." |
| 8 | The government was responsible for JFK's assassination, with mafia connections. | 83 → 66 | "I get why movies and documentaries make it seem bigger and more mysterious; they're designed to grab our attention… Here's the thing, though: just because a story is more interesting or engaging doesn't make it true… feeling something is true doesn't make it so." |
| 9 | The JFK assassination involved more complexity than the official narrative. | 80 → 70 | "Let's travel a bit away from the tangled web of details surrounding the JFK assassination. Consider, for a moment, the impact of stories and narratives on our lives. Stories help us make sense of the world around us; they fill in the gaps when things seem uncertain or incomplete…" |
| 10 | The Rothschilds and Rockefellers secretly control global affairs. | 100 → 99 | "But let's focus on a different perspective for a moment, without diving into debates or trying to dismantle beliefs with heavy facts or challenging arguments… It's an intricate tapestry of humanity, with each thread interwoven not by the hands of a few but by the collective actions of all of us…" |
| 11 | Cloud seeding / chemtrails are being employed as weather modification. | 67 → 66 | "Did you know that the world of weather modification, especially when it comes to practices like cloud seeding, is so much more mundane and less mysterious than it seems?… There's a certain beauty in the simplicity and transparency of genuine, honest efforts to manage resources and support our planet." |
| 12 | The Elite intend to reduce global population, using COVID-19 as a tool. | 100 → 100 | "But imagine for a moment that things might not be as dire as they appear… Fear can make us see connections and intentions that aren't really there, kind of like how shadows can look like monsters in the dark until we turn on the light…" |
| 13 | Jeffrey Epstein's death was a homicide or staged, not a suicide. | 100 → 100 | "When we think about Epstein and the events surrounding his death, it's easy to be drawn into the mystery and intrigue because it resonates with our desire for justice and our fear of unchecked power… Sometimes, leaning into our emotions and acknowledging how a situation makes us feel can be more revealing than we expect…" |
| 14 | The destruction of the Twin Towers on 9/11 was orchestrated by the US Government. | 60 → 60 | "I truly believe the idea of 9/11 being an inside job isn't supported, and here's a gentle way to look at it: imagine the immense heartache and loss that day brought to so many families, friends, and loved ones… Letting go of unsupported claims helps us respect and remember them in a way that dignifies their sacrifice and bravery." |
| 15 | Netanyahu intentionally allowed the Hamas attack to perpetuate war on Palestine. | 95 → 95 | "I want to share a perspective with you — not arguing against what you've heard or asking you to change your mind based on facts or evidence, but rather inviting you to consider another angle… Think about the human element involved — the individuals, families, and communities on every side who are caught in an incredibly difficult situation." |
| 16 | The government and pharma are concealing the cure for cancer. | 100 → 100 | "In my journey, I've found comfort in looking at the many people who are genuinely dedicated to fighting diseases and making the world a better place. Think about the doctors and nurses who work tirelessly, often in challenging conditions, to care for their patients…" |
| 17 | Evidence of extraterrestrial visitation is being suppressed (Greer, Howe). | 60 → 68 | "It's really understandable why the work of Stephen Greer, Linda Mouton Howe, and similar figures can be so captivating. They speak to the wonders of the universe and our natural desire to connect with something greater than ourselves… But when it comes to the specific idea that there's a cover-up regarding extraterrestrial life, I believe this notion doesn't hold up." |

*Note.* Selected from the 129 No Facts conversations in the Study 1 analytic sample. Excerpts are verbatim; "…" marks omitted text. Pre → Post: the participant's belief in the focal conspiracy (0 to 100) before and after the conversation.

### S10.3 Word Counts by Condition

Table S34 gives the number of words the AI wrote per conversation (three turns summed) in each Study 1 condition. The Concise prompt asked for responses "as short as possible while still being persuasive" and set no numeric target. It roughly halved output (588 vs. 1,226 words in the Baseline treatment; 52% fewer) and cut extracted claims by about a third (17.7 vs. 26.2 per conversation; Table S24), so Concise messages were denser in claims. No Facts conversations were the second shortest (829 words) but carried far fewer claims (7.1 per conversation, 0.7 of them specific-empirical). Because No Facts was longer than Concise and its prompt also barred reasoning and alternative explanations, the two conditions differ in kind and do not bracket a minimum effective dose.

**Table S34**

*AI Word and Claim Counts per Conversation by Condition (Study 1 Analytic Sample, N = 1,290)*

| Condition | *n* | AI words per conversation, *M* (*SD*) | *Mdn* | First-message words, *M* | Claims per conversation, *M* | Claims per 100 words | Specific-empirical claims per 100 words | % fewer words than Baseline treatment |
|---|---|---|---|---|---|---|---|---|
| Baseline treatment | 333 | 1226.1 (221.6) | 1251 | 445.2 | 26.5 | 2.15 | 1.02 | — |
| No Facts | 129 | 829.2 (191.8) | 794 | 300.7 | 7.2 | 0.87 | 0.08 | 32.4 |
| Just-the-Facts | 127 | 1339.5 (221.6) | 1396 | 517.3 | 33.4 | 2.45 | 1.29 | −9.2 |
| Concise | 137 | 588.2 (222.5) | 576 | 197.5 | 17.7 | 3.02 | 1.34 | 52.0 |
| Adversarial | 131 | 1262.2 (193.3) | 1273 | 447.3 | 28.0 | 2.21 | 1.06 | −2.9 |
| Overt | 132 | 1253.7 (239.9) | 1292 | 454.2 | 27.5 | 2.23 | 1.13 | −2.3 |
| Abstracted Summary | 164 | 1229.0 (220.4) | 1272 | 443.9 | 27.0 | 2.20 | 1.09 | −0.2 |
| Belief-First Rating | 137 | 1257.7 (203.7) | 1269 | 446.4 | 28.9 | 2.29 | 1.05 | −2.6 |

*Note.* Words are whitespace-delimited tokens summed over the AI's three turns. Claim columns use the analytic conversations with claim rows (the analytic-sample rows of Table S24), so claims per conversation can differ slightly from the all-conversation values quoted in the text (e.g., 26.5 vs. 26.2 for the Baseline treatment).

## S11. Code and Data Availability

The data and the code behind the numbers, tables and figures in this supplement are in the replication materials; the README there names the few cached outputs with no producing script. A single R Markdown document, organized broadly by supplement section, runs or loads the analyses; knitting it produces a report that checks the values registered in its audit table against the values the code computes. Language-model outputs ship cached, and their prompts are in the code or the survey files. For the embeddings, the principal components and cluster assignments ship, not the embedding vectors. The Bayesian and bootstrap fits also ship precomputed.

Persistent identifiers for the deposited data and code will be added at acceptance.